\pdfoutput=1
\RequirePackage{latex/atlaslatexpath}
\documentclass[PAPER=true, atlasdraft=false, texlive=2016, UKenglish, tocdepth=2,atlaslogo=false,cernlogo=false, datetop=false, preprint=true]{atlasdoc}

\usepackage{atlaspackage}
\usepackage{atlasbiblatex}

\usepackage[other=true, jetetmiss=true]{atlasphysics}

\graphicspath{{logos/}{figures/}}

\usepackage{ANA-LARG-2019-01-PAPER-defs}
\usepackage{flafter}

\AtlasTitle{The Phase-I Trigger Readout Electronics Upgrade of the ATLAS Liquid Argon Calorimeters}
 
\AtlasAbstract{The Phase-I trigger readout electronics upgrade of the ATLAS 
Liquid Argon calorimeters enhances the
physics reach of the experiment during the upcoming operation at
increasing Large Hadron Collider luminosities.
The new system, installed during the second Large Hadron Collider Long Shutdown,
increases the trigger readout granularity by up to a factor of ten
as well as its precision and range.
Consequently, the background rejection at trigger level is improved
through enhanced filtering algorithms utilizing the additional information
for topological discrimination of electromagnetic and hadronic shower shapes.
This paper presents the final designs of the new electronic elements,
their custom electronic devices, the procedures
used to validate their proper functioning, and the performance achieved
during the commissioning of this system.
}

\author{\begin{flushleft}
\hypersetup{urlcolor=black}
\begin{small}
\AtlasOrcid[0000-0002-6665-4934]{G.~Aad}$^\textrm{\scriptsize 19}$,    
\AtlasOrcid[0000-0002-2846-2958]{A.V.~Akimov}$^\textrm{\scriptsize 6,~22}$,    
\AtlasOrcid[0000-0002-0547-8199]{K.~Al~Khoury}$^\textrm{\scriptsize 7a}$,
\AtlasOrcid[0000-0002-1936-9217]{M.~Aleksa}$^\textrm{\scriptsize 6}$,    
\AtlasOrcid[0000-0002-4413-871X]{T.~Andeen}$^\textrm{\scriptsize 3}$,    
\AtlasOrcid{C.~Anelli}$^\textrm{\scriptsize 32}$,
\AtlasOrcid[0000-0001-9013-2274]{N.~Aranzabal}$^\textrm{\scriptsize 6}$,    
\AtlasOrcid[XX]{C.~Armijo}$^\textrm{\scriptsize 2}$,
\AtlasOrcid{A.~Bagulia}$^\textrm{\scriptsize 22}$,
\AtlasOrcid{J.~Ban}$^\textrm{\scriptsize 7a}$
% ...
\\
\end{small}
\bigskip
\begin{footnotesize}
% ...
$^{2}$Department of Physics, University of Arizona, Tucson AZ; United States of America.\\
$^{3}$Department of Physics, University of Texas at Austin, Austin TX; United States of America.\\
% ...
$^{6}$CERN, Geneva; Switzerland.\\
${^7}$$^{(a)}$Nevis Laboratory, Columbia University, Irvington NY;$^{(b)}$Department of Electrical Engineering, Columbia University, New York NY; United States of America.\\
% ...
$^{19}$CPPM, Aix-Marseille Universit\'e, CNRS/IN2P3, Marseille; France.\\
% ...
$^{22}$P.N. Lebedev Physical Institute of the Russian Academy of Sciences, Moscow; Russia.\\
% ...
$^{32}$Department of Physics and Astronomy, University of Victoria, Victoria BC; Canada.\\
% ...
$^{a}$ Also at Institute of Particle Physics (IPP); Canada.\\
% ...
\end{footnotesize}
\end{flushleft}
 }

\AtlasJournal{JINST}

 \hypersetup{pdftitle={LAr document},pdfauthor={The ATLAS LAr Collaboration}}

\begin{document}

\maketitle
\tableofcontents

\section{Introduction}
\label{sec:intro}
During the second Large Hadron Collider (LHC) Long Shutdown (LS2), 
the Liquid Argon (LAr) calorimeters of the ATLAS experiment have been 
equipped with new trigger readout electronics~\cite{phase1tdr}, which 
provides digital information to the ATLAS trigger system~\cite{TRIG-2011-02}.
The purpose of this so-called \phaseone upgrade is to enhance the
physics reach of the experiment during the upcoming operation at
increasing LHC luminosities. ATLAS ran
in the \runtwo data taking period (years 2015--2018)
at a typical maximum instantaneous luminosity of $1.9\times$\highL
and with an average mean number of proton-proton interactions of
$\avgmu=33$~\cite{ATLAS-CONF-2019-021}.
During \runthree (years 2022--2025) the instantaneous luminosity was originally
expected to increase to a peak value of $3\times$\highL, with
$\avgmu\approx 80$~\cite{phase1tdr}. 
The updated LHC plan for \runthree targets a beam intensity increase of 50\% 
compared to \runtwo with a luminosity leveled to $2\times$\highL over periods
up to 12 hours. The center-of-mass energy will also be raised
from \sqs=\SI{13}{\tev} to \SI{13.6}{\tev}.
These changes will result in an overall detector occupancy increase,
while the first level trigger (L1) bandwidth will
remain at \SI{100}{kHz} during \runthree. If the currently used LAr trigger
readout system, referred to as \emph{legacy} trigger readout in the following,
were to remain unchanged, the transverse energy (\ET) trigger
thresholds would need to be raised, degrading the
physics performance. To avoid this efficiency loss and enhance the
physics reach of the experiment in \runthree and beyond, the new system
installed during LS2 increases the
readout granularity by up to a factor of ten: instead of summing the \ET of
calorimeter cells in areas as small as
$\Delta\eta\times\Delta\phi = 0.1\times 0.1$
to form the so-called \emph{Trigger Towers} of the legacy
readout, additional lateral and longitudinal segmentation is
introduced to form smaller clusters called \emph{Super Cells}.
Trigger Towers are typically split longitudinally into four layers,
and some layers are split laterally along $\eta$ into four strips each.
One Super Cell can thus cover an area as small as
$\Delta\eta\times\Delta\phi =0.025\times 0.1$, depending on which
longitudinal layer it is located in. 
In addition, the precision and range of the \ET measurement is also increased.
Consequently, the background
rejection at trigger level is improved due to better filtering
algorithms being possible already at L1, based on topological discrimination
between electromagnetic shower shapes and hadronic activity. In particular, the
resulting electron, photon and tau lepton identification is more
efficient. Furthermore, thanks to more advanced reconstruction
algorithms and better pile-up subtraction techniques, the energy
resolution for electromagnetic, jet and missing \ET
objects is improved, leading to a sharper rise of the per-event trigger efficiency 
toward its plateau value.

The \phaseone upgrade project is part of a broad upgrade program of the
LAr calorimeters through the lifetime of the LHC and is fully compatible
with the future \phasetwo upgrade program of the ATLAS experiment planned
for the third Long Shutdown (LS3) in 2026--2028.

This paper presents the new system, initially proposed in
Reference~\cite{phase1tdr} and installed during LS2, together with its
performance. The final designs of the
new electronic boards and their Application-Specific Integrated
Circuits (ASICs) are described as well as the procedures used to
validate their proper functioning. The basic performance achieved
during the commissioning of this system is also presented.
In \Sect{\ref{sec:overview}}, the legacy readout electronics is
briefly recapitulated and the new trigger readout electronics is introduced, as
are the new data paths. The new electronics installed on the
ATLAS detector (Front-End) and off the detector (Back-End) is
described in \Sects{\ref{sec:fe}}{\ref{sec:be}},
respectively. Finally, the integration tests performed to validate the
full chain of this new readout electronics are reported
in \Sect{\ref{sec:integration}} before the conclusions are presented in
\Sect{\ref{sec:conclusion}}.
 
\section{Overview of the LAr readout electronics}
\label{sec:overview}
\begin{figure}[tb]
  \centering
  \includegraphics[width=\textwidth]{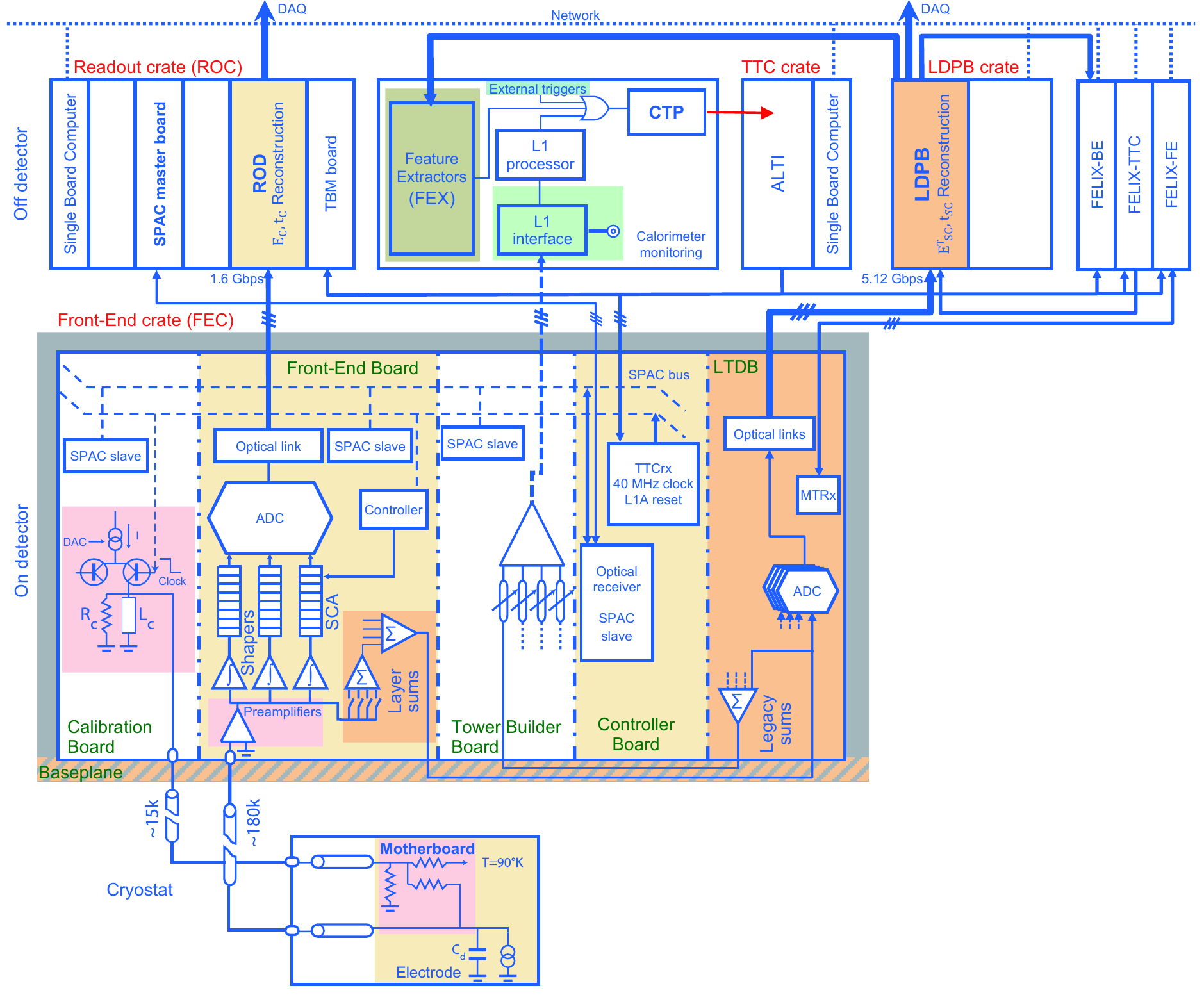}
  \caption{The ATLAS LAr calorimeter electronic architecture
    as of LHC \runthree. The new LAr boards are highlighted in orange. This
    diagram depicts the EM calorimeters; HEC and FCal electronics
    are slightly different.}
  \label{fig:run3arch}
\end{figure}

The readout electronics of the LAr calorimeters is designed to record
energies approximately ranging from \SI{50}{\MeV} to \SI{3}{\TeV},
measured in 182418 calorimeter cells.
The energy resolution of the calorimeter for electromagnetic showers 
can be written as
$\sigma_{E}/E = a / \sqrt{E} \oplus b \oplus c/E$, with an intrinsic sampling term
$a$ that is typically  10--11\%~$\sqrt{\GeV}$~\cite{TestBeam2006,TestBeam2007},
a constant term $b$ that is 1--2\%~\cite{PERF-2017-03},
and a noise term $c$ that is 10--\SI{600}{\mev} without pile-up contribution~\cite{LArElecPerf2010}
and is expected to be \SI{30}{\mev}--\SI{3}{\gev} in \runthree pile-up conditions~\cite{phase1tdr},
depending on the pseudorapidity region considered (up to $|\eta|<3.2$).
The triangular pulse coming from the ionization in a LAr cell is shaped
into a bipolar pulse that is sampled at \SI{40}{MHz}. The digitized
samples are read at the L1 trigger rate, that can go up
to \SI{100}{kHz}.  The LAr calorimeter system is composed of four different
components: the electromagnetic barrel (EMB) and end-caps (EMEC), both
referred to as EM calorimeters; the hadronic end-cap (HEC); and
the forward calorimeter (FCal). These components are divided in two sides, 
A and C, oriented respectively along the positive and negative $z$-axis of the experiment.
The LAr readout electronics is divided into a Front-End and a Back-End
system, both interconnected by cables and optical fibers as shown in
Figure~\ref{fig:run3arch}.

The Front-End system is composed of 58 {Front-End Crates} (FECs),
each hosting at most two baseplanes which interconnect the following legacy system 
boards:
\begin{description}
 \item[Front-End Boards (FEBs):] process typically 128 channels,
 each in a specific longitudinal
  layer of the calorimeter. They amplify, shape, sample and digitize
  the ionization signals and transmit them to the Back-End readout. 
  The analog sums needed for the trigger system are also
  prepared by the Linear Mixers and the Layer Sum Boards (LSBs),
  both hosted on the FEBs. The Linear Mixer electronics is part of the
  shaper chip located on the FEB motherboard, while the LSB is a plug-in
  card connected to the Linear Mixer.
 \item[Tower Builder Boards (TBBs):] form the legacy Trigger Towers from the
  analog sums provided by the LSBs over the FEC baseplanes. The 
  Trigger Tower analog signals are sent to the L1 trigger system via copper cables.
  In the HEC and FCal no further sums need to be produced after the LSB. 
  Therefore, {Tower Driver Boards} (TDBs) are used in place of
  TBBs.
 \item[Calibration Boards:] inject current calibration pulses whose shape is
  close to the pulse generated from the LAr ionization signal.
 \item[Controller Boards:] receive the configuration and monitoring
  commands from the Serial Protocol for Atlas Calorimeter~\cite{SPAC} (SPAC)
  masters installed in the Back-End Readout Crates. Controller boards
  receive also the Timing Trigger and Control~\cite{TTC} (TTC) signals
  coming from the Back-End TTC crates.  They distribute the SPAC and
  TTC commands to the other legacy boards in the FEC.
\end{description}

Some FECs also host monitor boards transmitting the information from the sensors
measuring possible mechanical stresses of the detector and the LAr temperature
and purity. 
Another function of some of the monitor boards is to measure the status of the 
FECs' low-voltage power supplies.
Each electronic board in a FEC is conductively cooled using two aluminium 
plates placed on each side of the board. These aluminium plates are part of a leak-less
water cooling system. 

On the Back-End side, two legacy systems are present:
\begin{description}
 \item[Readout Crates (ROCs):] Versa Module Eurocard (VME) crates hosting Readout
  Driver boards (RODs) that read the Analog-to-Digital Converter (ADC) data 
  sent by the FEBs and compute energy ($E_{\textrm{C}}$) and time ($t_{\textrm{C}}$) for each
  cell signal at the L1 trigger rate.
  The ROCs also house the SPAC master boards and
  the Trigger Busy Modules (TBMs). The latter collect the busy signals
  from the RODs and distribute the TTC signals.
 \item[TTC crates:] VME crates containing the modules that
  receive the TTC commands from the ATLAS Central Trigger Processor (CTP) system or
  generate locally these commands and transmit them to the other LAr
  boards.
\end{description}

The LAr electronics upgrade extends the legacy system with 
new Front-End components, sending the Super Cells' digital data to new Back-End 
components, which compute and transmit the Super Cells' \ET to the new trigger system. 
The architecture as of \runthree is illustrated in Figure~\ref{fig:run3arch}. 
This new LAr trigger readout system is
composed of new Layer Sum Boards providing the analog sums for the
higher trigger readout granularity; new baseplanes to route the
increased number of analog signals and host the new LAr Trigger
Digitizer Boards (LTDBs) that digitize the Super Cell analog signals
at \SI{40}{MHz} and provide the legacy sums for the TBBs; 
and new Back-End boards, the LAr Digital Processing
Blades (LDPBs), that read the Super Cell ADC signals from the LTDBs,
compute the Super Cells' \ET and send them to the new L1 trigger
system. This new trigger system comprises a set of
Feature EXtractors (FEX) with three subsystems targeting
electromagnetic (eFEX),
jet (jFEX) and global (gFEX) features~\cite{ATLAS-TDR-23}.
The choice to convert the Super Cells' energy to \ET at the LDPB level was
made to simplify the LTDB design.
The LDPB is built in Advanced Telecommunications Computing Architecture (ATCA) format 
and comprises one ATCA carrier blade named LAr Carrier
(LArC) and up to four Advanced Mezzanine Cards (AMCs) called LAr
Trigger prOcessing MEzzanines (LATOMEs). The hardware control and
monitoring of each blade proceeds via an {Intelligent Platform
Management Controller} (IPMC) plugged into the LArC. In total, 124
LTDBs are installed on the 114 new baseplanes. Furthermore, 30 LDPBs corresponding
to 30 LArCs and 116 LATOMEs, are installed in three ATCA shelves placed in one 19-inch
rack. This new LAr system uses also three separate Front-End LInk eXchange 
(FELIX)~\cite{felix} systems, comprising a total of 30 FELIX boards.

\begin{figure}[tb]
  \centering
  \includegraphics[width=\textwidth]{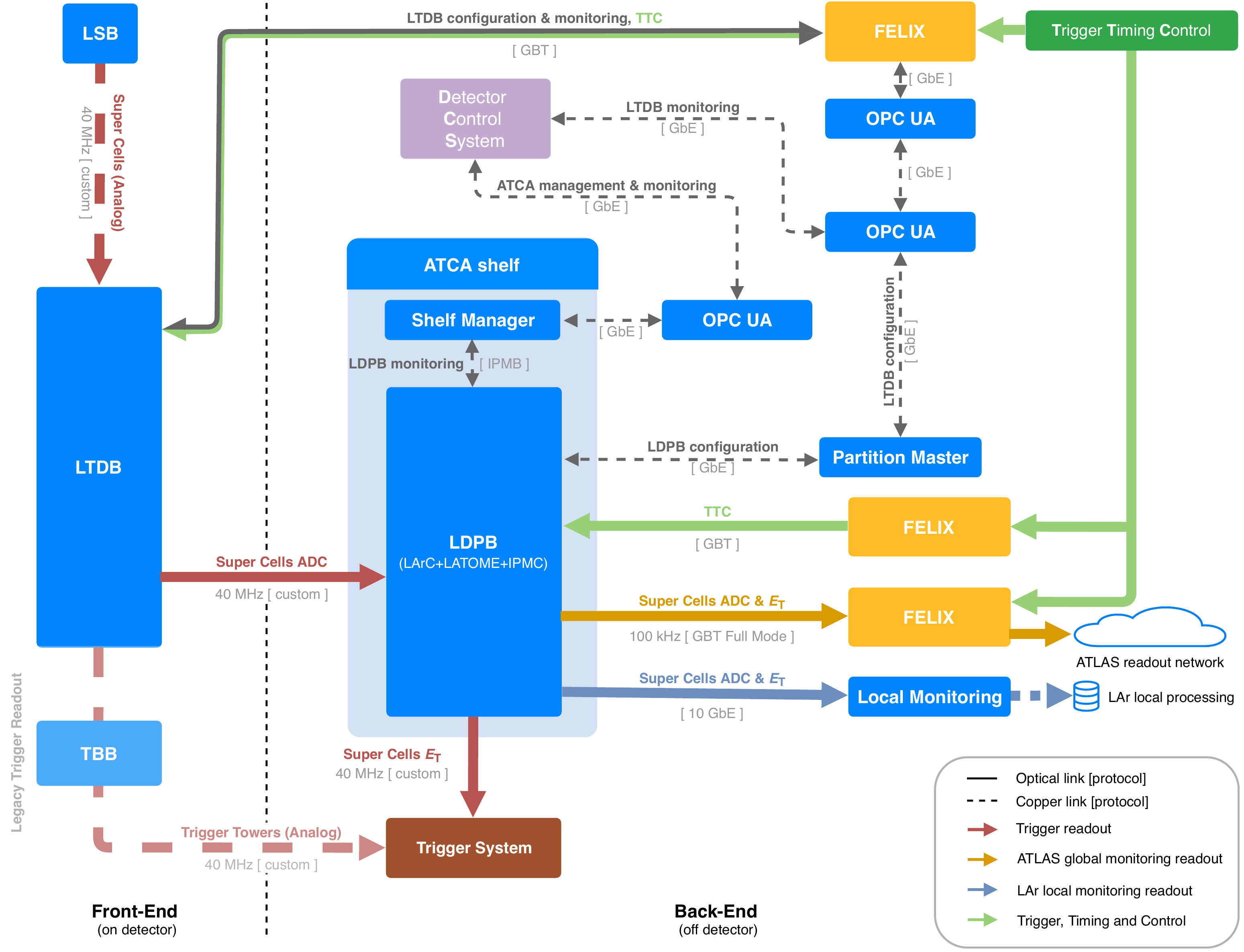}
  \caption{The ATLAS LAr calorimeter digital trigger readout
   system installed during the second LHC long shutdown.}
\label{fig:phaseIsys}
\end{figure}

The new trigger readout path is illustrated in
Figure~\ref{fig:phaseIsys} with solid red arrows. In order to ease the
new system commissioning and have a fallback solution in case of
unforeseen issues, the legacy trigger readout is kept functional since
the LTDB sends the re-summed analog signals to the
TBB, which will remain in operation during at least
the initial phase of the
LHC \runthree. The legacy trigger readout path is illustrated in
Figure~\ref{fig:phaseIsys} with dashed red arrows.

Several other data paths are present in the system. A global
monitoring path utilizes a Full Mode~\cite{felix} data link between 
each LATOME and the
FELIX system that connects to the ATLAS main data readout (path in
orange arrows in Figure~\ref{fig:phaseIsys}). Its purpose is to verify
that the \ET sent to the trigger system is correct, by reading the
Super Cell ADC data and the \ET values for all events selected by 
the first level trigger accept signal (L1A). Thus, this readout allows
a recomputation of the calculation performed in the LATOME boards.

A similar monitoring path utilizes the 10 Gigabit Ethernet (GbE) interface of the ATCA
shelf that hosts the LDPBs (blue arrow in Figure~\ref{fig:phaseIsys}).
This readout is not connected to the ATLAS main readout but is used by
a processing server specific to the LAr system. It enables a more flexible
and localized monitoring to readout Super Cell ADC data and \ET 
at the L1A signal rate or optionally at any other rate, independently of
the constraints present in the ATLAS main data readout in terms of
bandwidth and event rate.

The whole system is synchronized with the TTC signals
(green arrows in Figure~\ref{fig:phaseIsys}). These signals provide the
\SI{40}{MHz} LHC reference clock, the L1A, the Bunch Counter
Reset (BCR), the Event Counter Reset (ECR) and the
Trigger Type commands. The BCR corresponds to the LHC orbit
frequency and enables the determination of the collision time of an
event via its Bunch Crossing IDentifier (BCID). The ECR is used
to compute an event identifier and the Trigger Type records which
subsystem issued the L1A. The TTC signals are transmitted
to LTDB and LDPB through FELIX systems with Gigabit Transceiver
(GBT)~\cite{GBT1,GBT2} links.

The configuration of the LTDB and LDPB hardware proceeds via
a "Partition Master" PC. This computer is connected to LTDBs in
three stages: two Open Platform Communications Unified Architecture
(OPC UA) servers and a FELIX system, connected to LTDBs via GBT links.
The first OPC UA server takes care of the LTDB configuration command
generation, according to the run control commands received from the Partition
Master. The second OPC UA server packs these specific commands to follow
the GBT protocol.
The Partition Master connection to the LDPB is made directly via a 
\SI{1}{GbE} interface.

A Detector Control System~\cite{DCS} (DCS) handles the hardware
monitoring: it ensures that the hardware is operated under safe
conditions and provides basic safety mechanisms in case of failure.
The connection to the LTDB is made with a part of the same OPC UA and FELIX
chain used for the configuration. The LDPB is connected to the DCS via
an OPC UA server and the ATCA shelf manager, interfaced to the IPMC on 
the LDPB via an Intelligent Platform Management Bus (IPMB).

Some differences exist in the signal processing between the main and
trigger readout systems: the analog signal processing in the main readout
comprises pre-amplification followed by shaping performed in three
different gain scales, such that three overlapping gains are obtained
with a ratio of about 10. The bipolar shape is created with a time
constant of \SI{13}{ns}. The samples stored by the Switched
Capacitor Array (SCA) analog pipeline chip are digitized by a 12-bit
ADC and sent to the Back-End system via a 1.6 Gigabit per second (Gbps) optical
link. In the trigger readout, the analog processing has a similar
shaping with only one gain scale, but with a special conditioning described
in \Sect{\ref{sssec:ltdb-motherboard}}.
The Super Cells' digital samples
generated in the LTDBs are sent to the Back-End system with
a \SI{5.12}{Gbps} optical link.  While in the main readout the RODs
compute the cell energies and signal times at the L1A rate (up to \SI{100}{kHz}), 
the LDPBs of the trigger readout compute the
Super Cell transverse energies ($E^{\textrm{T}}_{\textrm{SC}}$) and signal times ($t_{\textrm{SC}}$) at the
LHC collision rate (\SI{40}{MHz}).

The new trigger readout aims to have an Integral Non-Linearity (INL) for each
Super Cell channel below \SI{1}{\%} with a Super Cell energy computation precision
below \SI{250}{\mev}. Even under high pile-up data taking conditions, this computation
must remain unbiased. 

In the following, the new hardware, as well as the procedures carried out to
validate its functionalities, are described in detail.
 
\section{New Front-End electronics}
\label{sec:fe}
\subsection{Baseplanes}
\label{ssec:baseplanes}

\begin{figure}[h]
  \centering
  \includegraphics[scale=0.5]{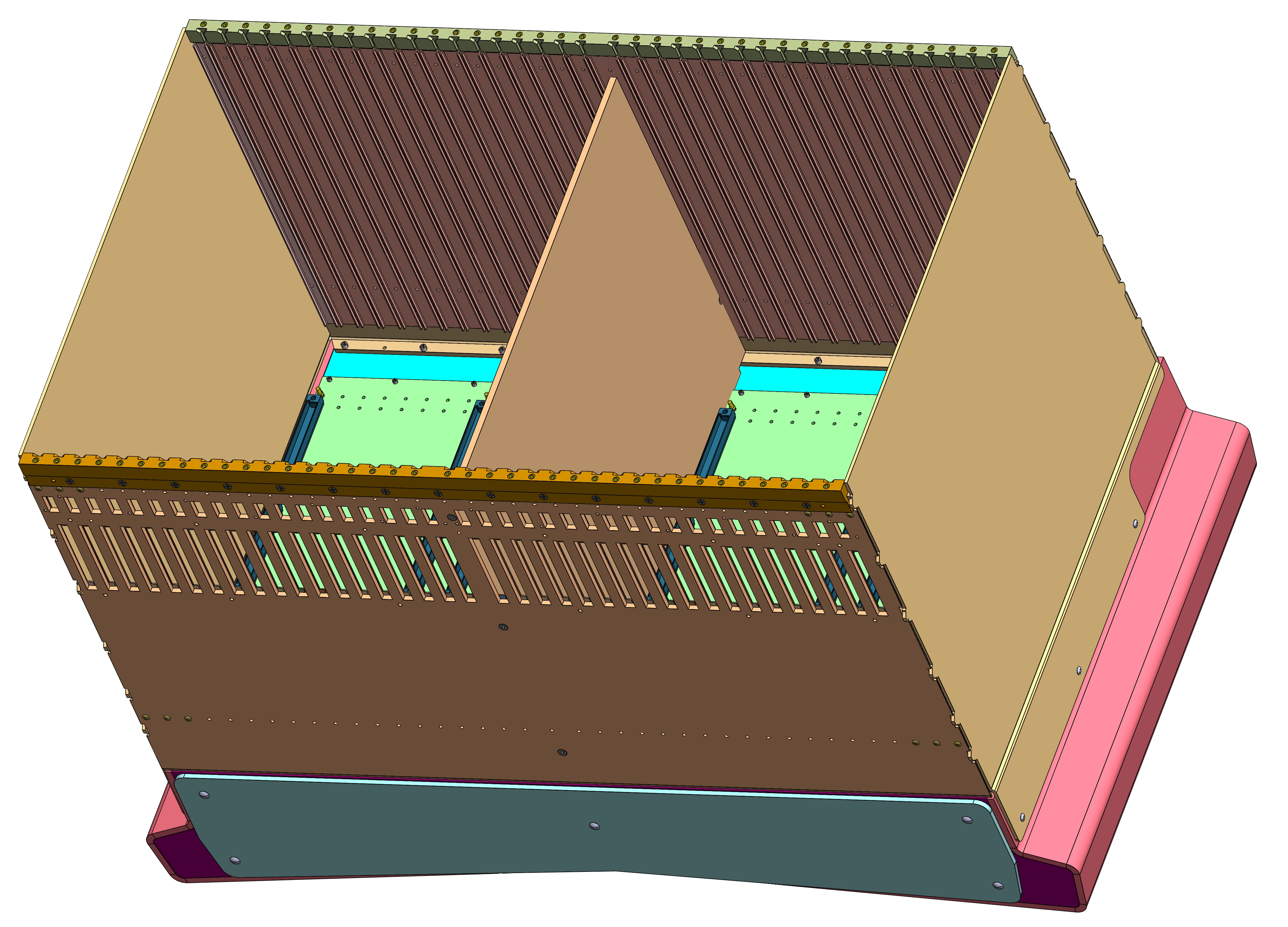}
  \caption{Perspective drawing of a FEC mounted on its
           pedestal. Simplified sketches of the two baseplanes can be 
	   seen mounted on the pedestal inside the crate.}
  \label{fig:BaseplaneinFEC}
\end{figure}

At each end of the barrel cryostat  
32 calorimeter signal feedthroughs~\cite{Axen:2005vb} are installed, 
distributed around the cylindrical periphery at nearly uniform 
angles. Each of the two end-cap cryostats has 25 signal 
feedthroughs distributed around the far end (from the collision 
point) of its periphery. Each feedthrough with cold and warm flanges connected via vacuum cables is made up of 
four {pin carriers}.  Each pin carrier has either seven or eight 
connectors and each connector has 64 signal pins.  A pedestal 
spans two adjacent feedthroughs and a FEC 
is mounted on each pedestal. Each FEC is divided 
into two halves with a baseplane installed in each half for almost all of 
the crates, see Figure~\ref{fig:BaseplaneinFEC}.  For each 
connector in a pin carrier a flex ribbon cable carries 64 signals 
from the feedthrough to a connector on the back side of the 
baseplane.  The pins in this connector extend through  
to the front side of the baseplane. A FEB has 
two 64-pin signal connectors and this board, mounted in a FEC, 
plugs into the baseplane.  Between the two signal connectors 
on each FEB is a third 64-pin connector, which returns analog 
trigger signals to the baseplane where they are routed to 
trigger-handling boards, also plugged into the baseplane.

For the Phase-I upgrade, new baseplanes are required (1) to
accommodate the finer trigger segmentation, (2) to make room for the
LTDBs (Section~\ref{ssec:ltdb}), and (3) to maintain the legacy Level-1
trigger system. The new mezzanine LSBs
(Section~\ref{ssec:lsb}), mounted on the existing FEBs,
provide more analog signals than before onto the baseplanes, which are
now routed to both the LTDBs and to the existing TBBs or TDBs
requiring additional signal layers on the baseplane Printed Circuit Boards (PCBs). Because the
LTDB baseplane connectors are wider than those for the other boards plugged
into the baseplane, there is no space next to an LTDB for the Radio Frequency (RF)
shields which protect the small-signal inputs to the FEBs from 
extraneous noise. This requires, in some cases, additional slots
and/or significant re-arrangement of the baseplane slot assignments,
as well as thicker baseplane PCBs, whose
thickness is highly constrained by the existing size of the FEC and 
FEBs. Table~\ref{tab:baseplanePCBs} gives an overview of the physical
dimensions of the various baseplane types.
\begin{table}[tb]
   \caption{Overview of the number of PCB layers, physical dimensions and numbers installed on the detector for the various baseplane types. Side A and side C are oriented along the positive and negative $z$-axis of the experiment.\label{tab:baseplanePCBs}}
      \begin{center}
        \begin{tabular}{rrrrrrr}
        \toprule

                  & \textbf{EMB} & \textbf{\begin{tabular}[c]{@{}c@{}}EMEC \\ Standard\end{tabular}} & \textbf{\begin{tabular}[c]{@{}c@{}}EMEC\\ Special\end{tabular}} & \textbf{HEC} & \textbf{\begin{tabular}[c]{@{}c@{}}FCAL \\ side A\end{tabular}} & \textbf{\begin{tabular}[c]{@{}c@{}}FCAL \\ side C\end{tabular}} \\
        \midrule
        PCB layers & 12 & 12 & 14 & 12 & 16 & 16\\
        Length [mm]     & 388.00 & 388.00 & 522.73 & 260.00 & 450.60 & 490.00\\
        Width [mm]      & 415.00 & 415.00 & 415.04 & 415.00 & 406.40 & 406.40\\
        Thickness [mm]  & 2.40 & 2.40 & 2.36 & 2.36 & 2.36 & 2.36\\
        \# on detector & 64 & 32 & 8 & 8 & 1 & 1 \\

        \bottomrule
        \end{tabular}
      \end{center}
\end{table}
                   
\begin{figure}[h]
  \begin{center}
  \includegraphics[width=0.53\linewidth]{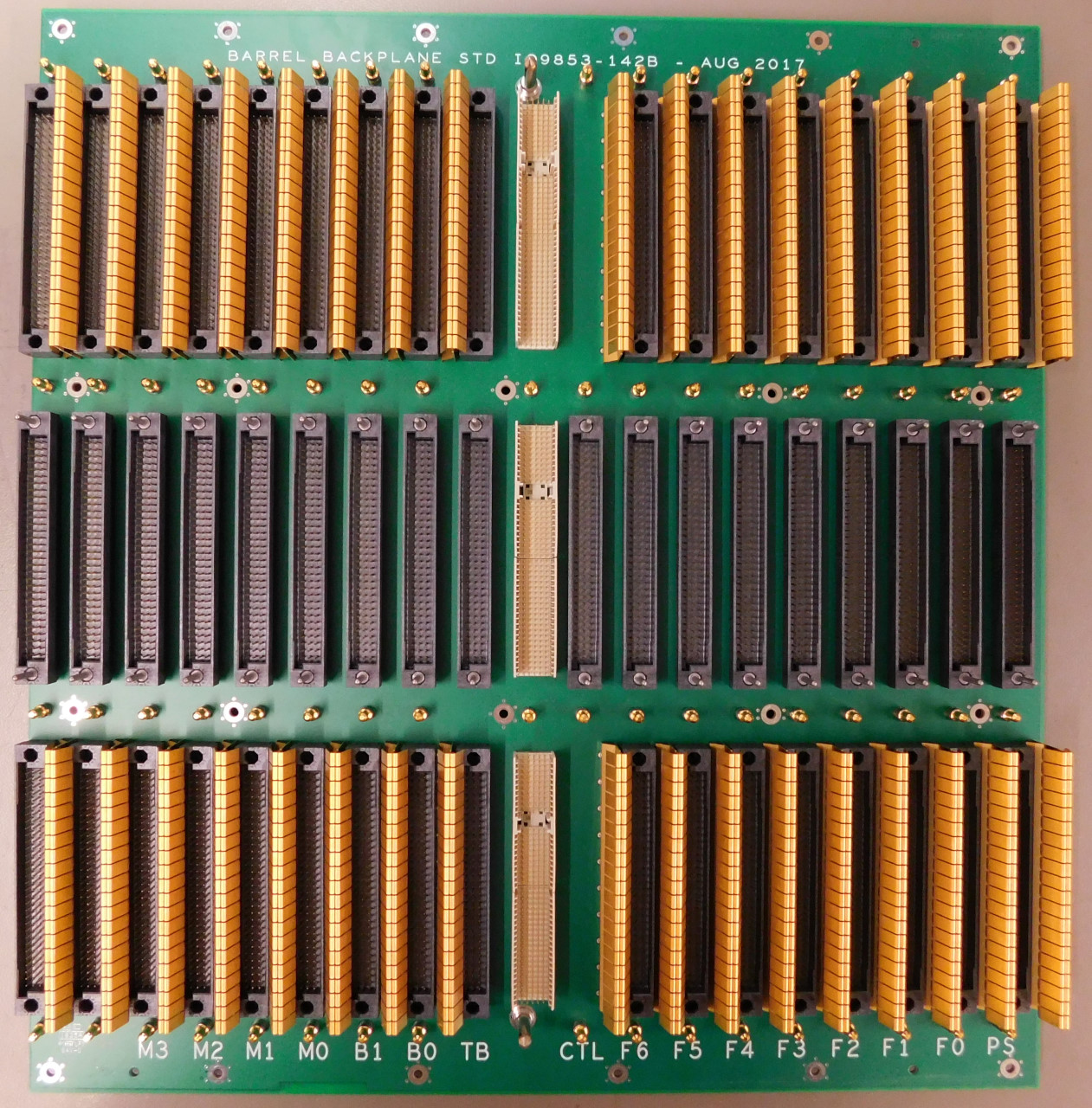}\quad\includegraphics[width=0.40\linewidth]{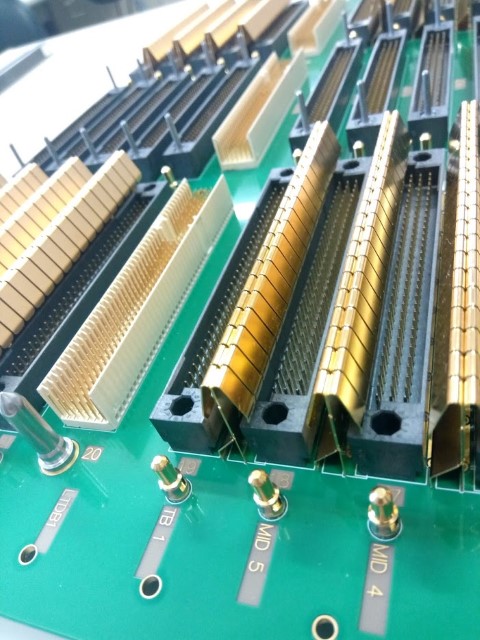}
  \caption{Left: Photograph of an EMB Baseplane. The slot with
           three wider baseplane connectors is for the LTDB. Just to
           the left of the LTDB is the slot for the TBB
           and to the right is the Controller board
           slot. The calibration board slot is on the far left and next to it is the monitor board slot. The
           gold-colored RF shields surrounding connectors in the top
           and bottom rows identify the FEB slots.  Right:
	   Close-up of an EMEC special baseplane showing the RF 
	   shields and the wider LTDB connectors.}
  \label{fig:EMBbaseplane}
  \end{center}
\end{figure}

The new EMB baseplane has the same number of slots (19) as the
original baseplane. Space for the single new LTDB comes from a
previously unallocated slot at one end of the FEC.  Some FEBs have
been shifted by one slot to allow the LTDB to sit in the middle slot
of the baseplane next to the TBB (see 
Figure~\ref{fig:EMBbaseplane}). The flex ribbon  cable connections 
underneath the baseplane are then also shifted by the same amount.

The EMEC crates house two different types of baseplanes. The EMEC
standard crate has baseplanes similar to the EMB but with
one fewer FEB, leaving a slot for an additional monitor board. The
EMEC special crate has one baseplane for the EMEC and
another for the HEC. The EMEC special baseplane has
two LTDBs and it extends well beyond the center of the crate. The HEC
baseplane in the other side of the crate consequently has room for
only 13 slots rather than the normal 19 slots of the EMB
baseplanes and the EMEC standard baseplanes. An overview of the 
different boards which plug into the various baseplane types can be found 
in Table~\ref{tab:baseplaneboards}.

\begin{table}[tb]
   \caption{Overview of the different board types which plug into each of the baseplane types.\label{tab:baseplaneboards}}
      \begin{center}
        \begin{tabular}{rrrrrr}
        \toprule

                  & \textbf{EMB} & \textbf{\begin{tabular}[c]{@{}c@{}}EMEC \\ Standard\end{tabular}} & \textbf{\begin{tabular}[c]{@{}c@{}}EMEC\\ Special\end{tabular}} & \textbf{HEC} & \textbf{FCAL} \\
        \midrule
        FEBs & 14 & 13 & 17 & 6 & 14 \\
        LTDBs & 1 & 1 & 2 & 1 & 2 \\
        TBBs & 1 & 1 & 3 & 0 & 0 \\
        TDBs & 0 & 0 & 0 & 2 & 2 \\
        Calibration & 1 & 1 & 2 & 1 & 1 \\
        Controller & 1 & 1 & 1 & 1 & 1 \\
        Monitor & 1 & 2 & 0 & 0 & 1 \\
        Low Voltage Power & 0 & 0 & 0 & 2 & 0 \\
        \midrule
        Total & 19 & 19 & 25 & 13 & 21 \\
        \bottomrule
        \end{tabular}
      \end{center}
\end{table}

The original FCal baseplanes had 19 slots, each fitting
within one-half of a FEC. The other half was used for devices which 
monitor the status of the cryostats. The new FCal baseplanes each
have two LTDBs and the baseplanes extend two (three) slots into the
now unused other half crate for the A-end (C-end). 
Integrated into each new FCal baseplane are traces carrying
pulses from the calibration board through attenuating injection resistors (surface mounted on the backside of the baseplane)
to those pins on the baseplane connectors which carry FEB signals.

Prototype and production baseplanes were subjected to a series of tests.
For those connector pins which are specified to be grounded, the impedance
to ground was confirmed to be consistent with zero. Likewise, connector pins which are not specified to be
grounded were also checked. The impedance of traces on the baseplanes 
was measured with a time-domain reflectometer.  Continuity between 
trace ends was compared with the net lists. These mapping tests  
were then compared with independent LTDB mapping requirements.  
Signal integrity on the traces and cross-talk between near-by traces 
were measured with a pulser and oscilloscope to verify that they met 
specifications.  In the case of the FCal baseplanes, the calibration 
injection resistors were all measured for correct values. 
All baseplanes were shown to meet the specifications.

\subsection{Layer Sum Boards}
\label{ssec:lsb}
\begin{figure}[!h]
        \centering
        \includegraphics[width=0.4\textwidth]{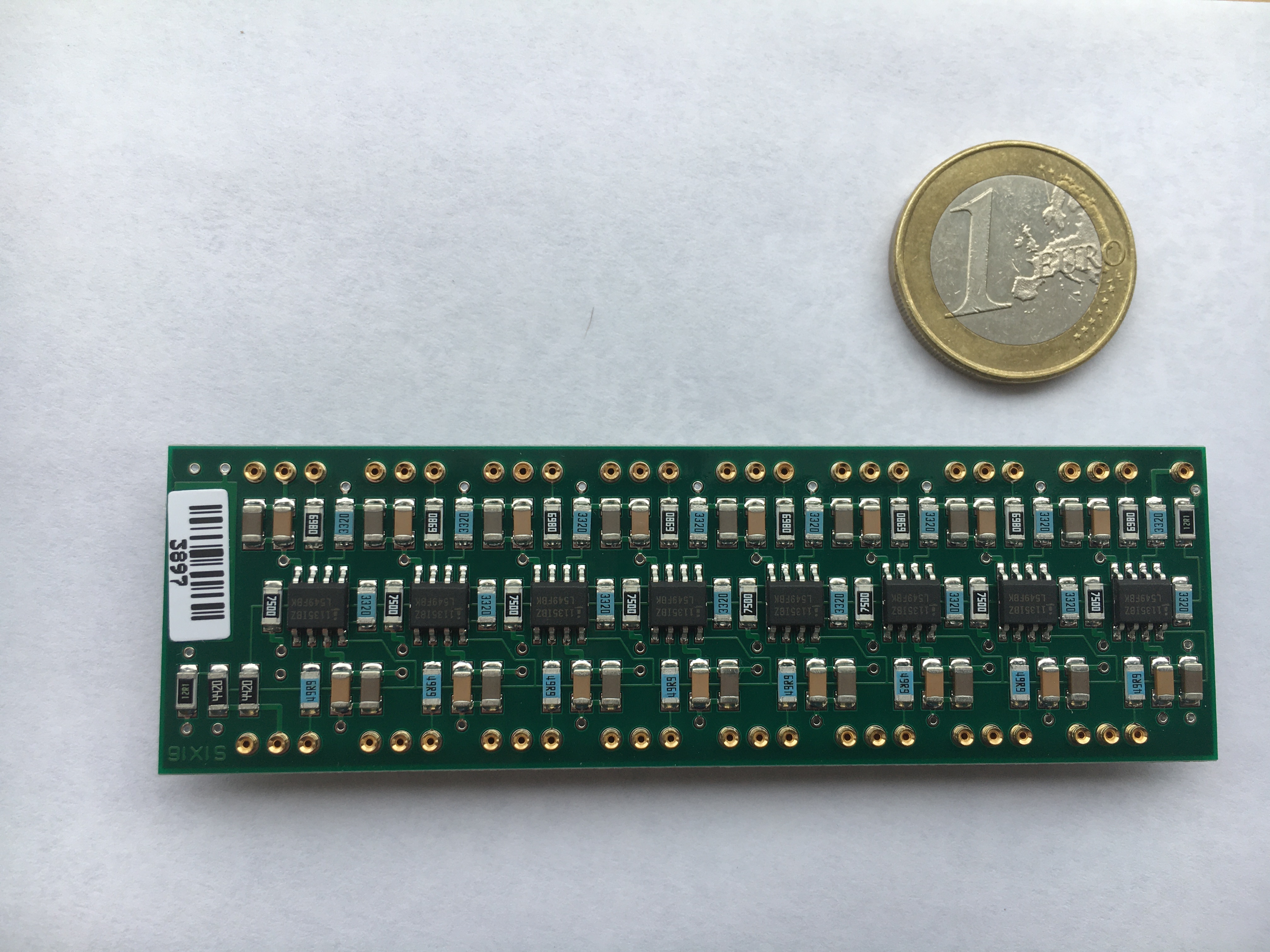}
        \caption[Photo of an LSB.]{Photo of an S1x16LN LSB.  Only one side is shown.  Half of the components are on the other side. A one-euro coin is shown for scale.}
        \label{fig:lsb_photo}
\end{figure}

In the LAr trigger readout, the first level of summing is provided by the Linear Mixer, which sums over four channels in azimuth with different gains in different regions of $\eta$.  
The upgrade to the trigger branch begins with the ouput of this signal.
The LSB is a plug-in card for the FEB
that performs a second level of summing of the analog signals~\cite{Buchanan:2008zza}.  
There are two LSBs mounted on each FEB.
For the \phaseone upgrade, the new LSB output provides signals for the finer 
granularity Super Cells in the front and middle layers, while 
retaining the signals needed for the legacy trigger path elsewhere. Six main types
of new LSBs have been produced:
\begin{description}
\item[ S2x8 ] The basic circuit on the board performs an analog sum of
  two input channels from the Linear Mixer. There are eight copies of
  this circuit on each LSB.
  There are 896 boards of this type required for
  the EMB and 192 for the EMEC, with 288 reused from the original construction.
\item[ S2x8D ] S2x8 LSBs were already used in the back layer of the EM
  calorimeters. For \phaseone, this output signal must be sent to both
  the new LTDB and the legacy TBB.  A simple splitter is added to the
  S2x8 in order to produce this dual output.
  There are 128 boards of this type required for
  the EMB and 160 for the EMEC.
  \item[ S1x16LN ] This LSB has no summing function and only serves as a
  line driver for the Linear Mixer signal.
  There are sixteen copies of this circuit on each LSB.
  There are 448 boards of this type required for
  the EMB and 544 for the EMEC.
\item[ S1x12+S6x2 ] For the front section of the EMEC in the range
  $1.8<|\eta|<2.0$, the output from the shapers covers the area
  $\Delta\eta\times\Delta\phi=0.0167\times 0.1$. These are sent
  individually to the LTDB, while sums of six inputs are sent to the
  TBB.
  There are 128 boards of this type required for the EMEC.
\item[ S2x6D+S1x4 ] At the end of the EM Barrel ($|\eta|>1.4$), each
  LSB must handle four signals originating from the middle layer and
  12 signals from the back layer. The signals from the back layer are
  summed and handled with the same circuitry as the S2x8D and sent to
  the LTDB and TBB, while the four signals from the middle are just
  passed directly to the LTDB as in the S1x16LN circuit.
  There are 128 boards of this type required for the EMB.
\item[ FCal ] All the LSBs above form an unweighted sum of cells in a
  narrow region of $\eta$, while in the FCal, one must perform a sum
  over a relatively wide range in pseudorapidity, over which the
  conversion factor from energy to transverse energy varies
  significantly. For this reason, the inputs to be summed on the FCal LSBs
  are weighted to perform the conversion to transverse energy before
  summing. This leads to a large variety of LSBs, which, within a
  given FCal module, differ only by the weighting resistor values.
  All 56 FCal LSBs were replaced by the new design.
\end{description}
The technical specifications for all the boards are that the gain should not vary by more than 2\% from nominal, the summing amplifier INL should be < 1\%, and the Direct Current (DC) offset should be < 2\% of the maximum output voltage.
All boards were tested before and after a week-long burn-in.  About 99\% passed the specification and were delivered to CERN.
As the replacement of the baseplanes required that all electronics in
the FECs be disconnected and removed, all FEBs were brought to a
surface laboratory. Here, the FEBs were ``opened'' (i.e. the cooling
plates dismounted) and the LSBs replaced by the new boards.
Figure~\ref{fig:lsb_photo} shows a photo of an example LSB. 
 
\subsection{LAr Trigger Digitizer Boards}
\label{ssec:ltdb}
An LTDB processes and digitizes up to 320 Super Cell signals (``channels'') and transmits them 
via optical links to the Back-End, while also providing summed analog legacy signals for the TBB
via the new baseplanes. The following sections describe the key ASICs for digitization and optical transmission, the motherboard that houses them, the mezzanine card providing the power, quality assurance and control for the assembled system as well as considerations on compatibility with the \phasetwo upgrade.

\subsubsection{Analog-to-Digital Converter}
\label{sssec:adc}
A custom, quad-channel, pipeline ADC was designed to process the shaped analog signals from the LAr calorimeter Super Cells. These signals are continuously sampled and digitized at 40~Megasamples per second. To digitize the energy of the trigger signals, the ADC has a dynamic range of approximately 12~bits (after calibration) with a required precision of at least 10~bits~\cite{phase1tdr}. Each ADC consumes \SI{45}{mW} per channel and has a latency of less than \SI{125}{ns},
meeting the power and latency budgets at the system level of less than \SI{145}{mW} per channel and \SI{200}{ns}, respectively.

Radiation tolerance is a key requirement for the ADC, as it is for all devices located on-detector and within the substantial radiation field. The device must operate reliably through the remaining LHC run, as well as the High Luminosity LHC (HL-LHC) runs, with a total expected integrated luminosity of \SI{3000}{\ifb}, possibly up to \SI{4000}{\ifb}. This places strict requirements on the ADC, particularly for the tolerance to total ionizing dose (TID) and single-event effects (SEE), which depend on the expected level of radiation at the readout electronics location~\cite{radiation_qualification}. The LAr on-detector ASICs are required to be radiation tolerant up to a total ionizing dose of \SI{180}{kRad}, non-ionising energy loss doses of up to \SI{4.9E12}{n~eq\per\cm\squared}, and be relatively insusceptible to (recoverable) SEEs for a total fluence of up to \SI{7.7E12}{hadrons\per\\cm\squared}~\cite{phase2tdr}.

\begin{figure}[!h]
	\centering
	\includegraphics[width=0.8\textwidth]{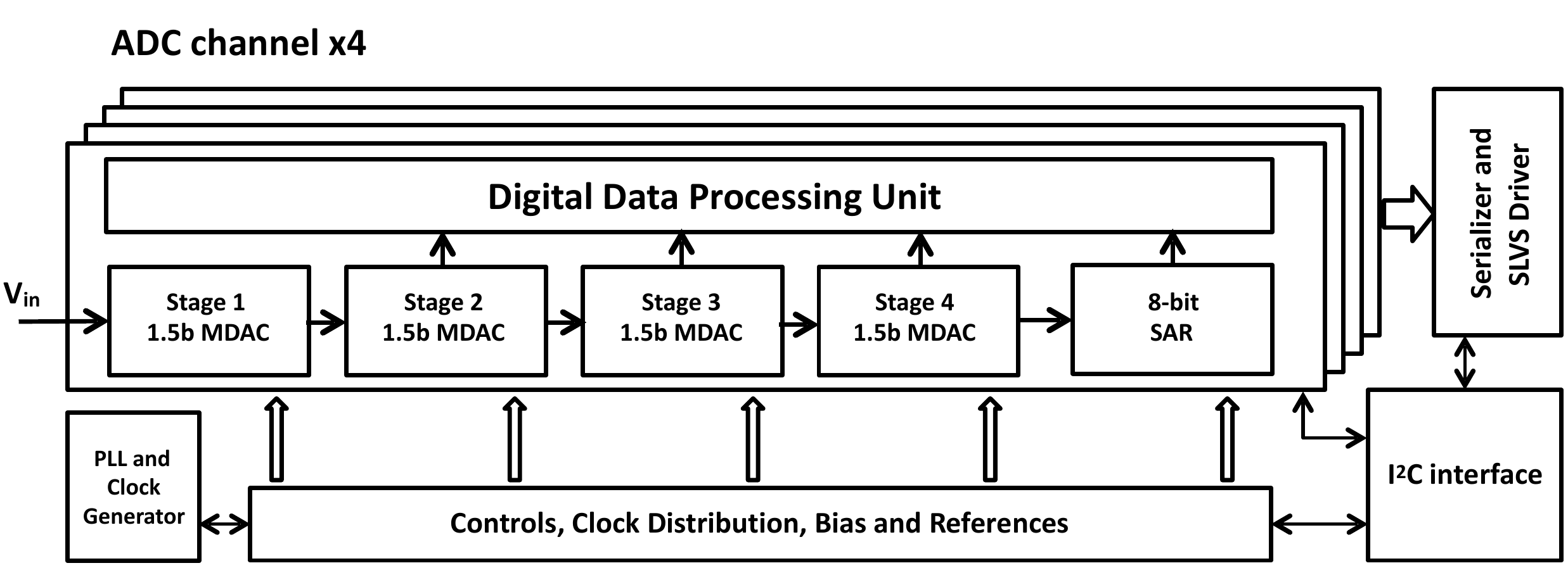}
	\caption[ADC block diagram.]{ADC block diagram.}
	\label{fig:adc_arch}
\end{figure}

\begin{figure}[!h]
	\centering
	\includegraphics[width=0.4\textwidth]{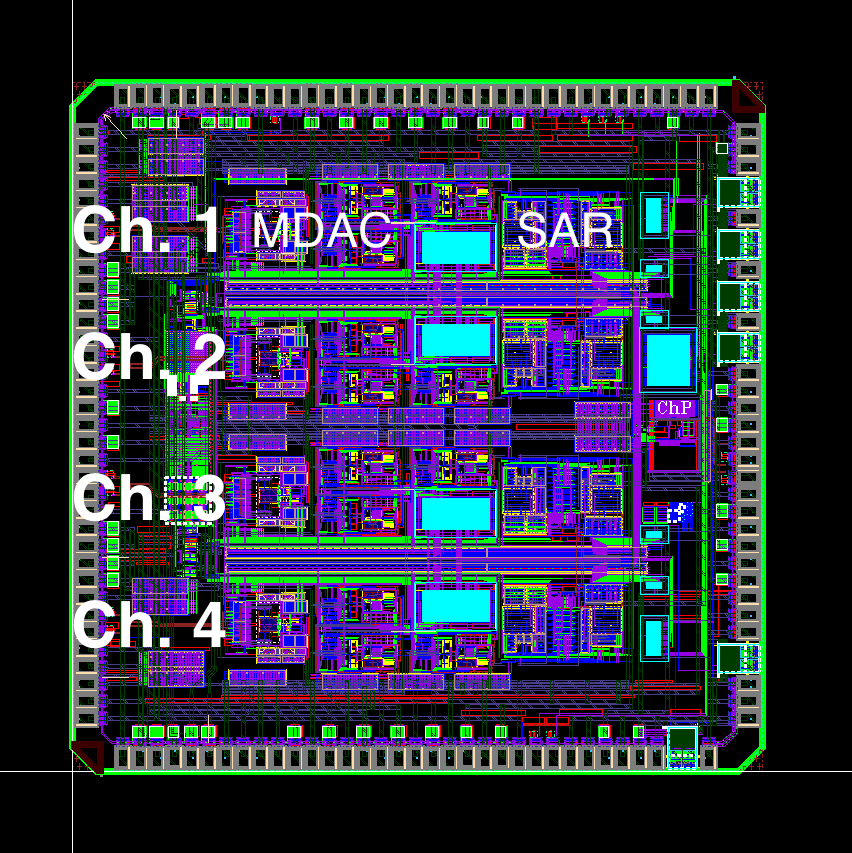}
	\caption[Image of ADC layout.]{Image of ADC layout ($ 3.6 \times 3.6$ mm).}
	\label{fig:adc_chip}
\end{figure}

The ADC chip consists of four identical ADC channels and is implemented in CMOS 8RF \SI{130}{nm} technology. Each channel consists of a multiplying digital-to-analog converter (MDAC) pipeline followed by a successive approximation (SAR) ADC, as shown in Figure~\ref{fig:adc_arch}. The MDAC pipeline contains four 1.5-bit stages, each with a nominal gain of two. The additional half-bit overlap allows for a digital on-chip calibration to correct capacitor mismatch. These stages resolve the four most significant bits \cite{nevis12}. The remaining eight least significant bits are resolved by the SAR stage. An internal Phase Lock Loop (PLL) generates a \SI{640}{MHz} clock from a \SI{40}{MHz} input clock (provided by the LTDB), of which both edges are used for the SAR operation. The MDAC stages operate with a supply voltage of \SI{2.5}{V} and a common-mode voltage of \SI{1.25}{V}, while for the SAR a supply voltage of \SI{1.2}{V} with a common-mode of \SI{0.6}{V} is used~\cite{nevis15}. A digital data processing unit~(DDPU) applies the calibration, forming the digital output, and the data are serialized and sent off the chip over a Scalable Low-Voltage Signaling (SLVS) driver at \SI{320}{MHz}. An Inter-Integrated Circuit (I$^2$C) interface is used for chip control. An image of the chip layout is shown in Figure~\ref{fig:adc_chip}.

The \SI{130}{nm} CMOS 8RF technology does not have sufficient fabrication accuracy to meet the 10~bit precision specification. Therefore, a digital calibration is performed to correct for capacitor mismatch in the MDAC stages. The algorithm measures the actual gain and compares it to the ideal MDAC gain of exactly two. The difference is then taken as a correction and applied to the digital output in the DDPU. By design, the gains are lower than two to avoid over-ranging the input to the following stage, which would lead to missing codes. This results in a reduction of up to 10\% of the dynamic range of the chip with respect to the nominal 12~bits (or 4096~counts), within the margin of acceptable performance. The calibration constants are calculated off-line but stored on-chip and must be reloaded to the chip after power cycling. 

The radiation tolerance of the ADC chip has been studied in its different prototyping stages. The TID and SEE tests performed with a proton beam at the Francis H. Burr Proton Therapy Center at Massachusetts General Hospital with prototypes are described in references~\cite{nevis12} and \cite{nevis10}. Radiation tolerance was established for a TID of up to \SI{10}{MRad} and for non-ionising energy loss up to a fluence of the order of \SI{E14}{n~eq\per\cm\squared}. Proton SEE cross-sections were measured to be of the order of \SI{E-12}{\cm\squared} per ADC channel. In addition to the proton radiation tests, SEE cross-sections were measured using four different species of heavy ions at the Cyclotron Resource Centre at Louvain-la-Neuve, Belgium, targeting a larger range of deposited energies and complementing the existing studies with protons \cite{nevis15}.

To equip the 124 LTDBs approximately 12000 fully functional ADC chips were required (80 chips per LTDB, plus spares). Allowing for yield factors 17200 chips were produced and packaged in molded QFN-72 packages. The individual ADCs were tested using a socketed test board 
which allowed for a sine wave input, low-jitter clock, power and readout via a Field Programmable Gate Array (FPGA). The functionality, dynamic range and precision, evaluated by the effective number of bits (ENOB), were measured. A total of 12838 chips were identified as having four channels with a dynamic range greater than 3600~counts and an ENOB larger than 9.9~bits (using the limited-precision socketed board) \cite{nevis15}, and thus were qualified to be used for LTDB production.

\subsubsection{Serializer and optical transmitter}
\label{sssec:locx2}
\begin{figure}[h]
  \centering
  \includegraphics[scale=0.5]{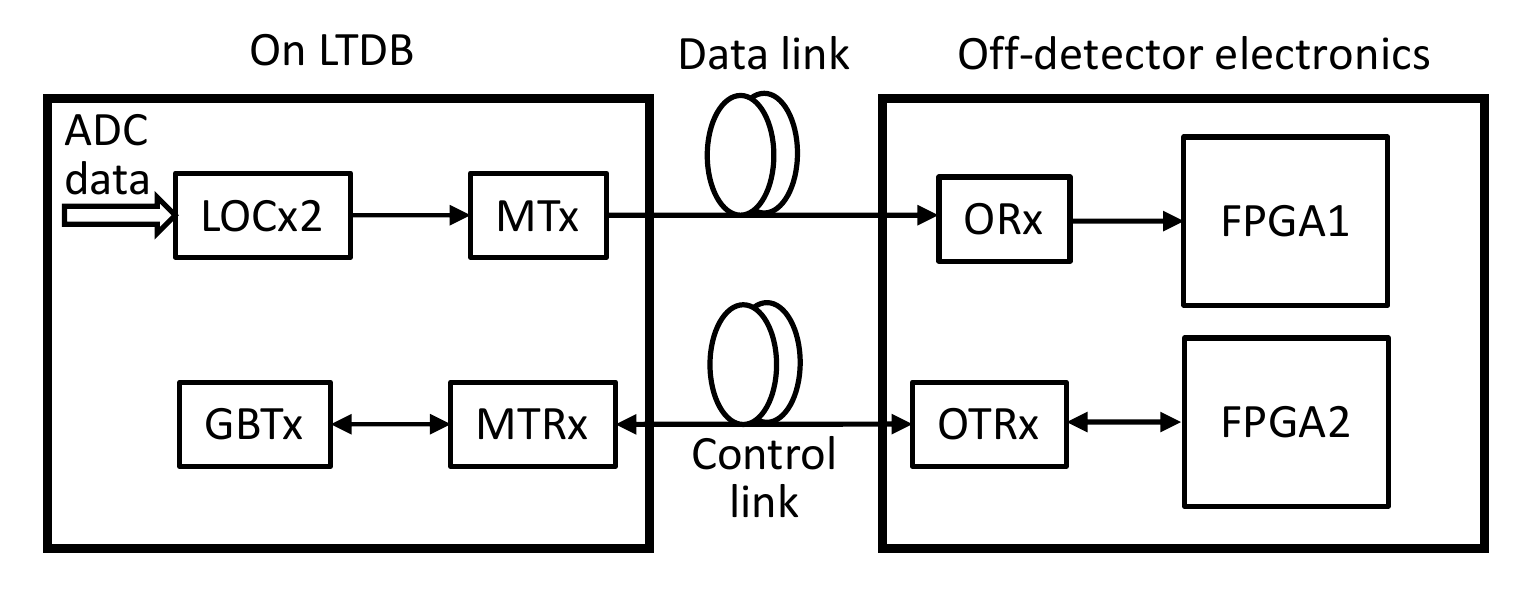}
  \caption{Block diagram of the LTDB optical links.}
  \label{fig:opticallinks}
\end{figure}

The digitized signals from up to 320 ADC channels per LTDB in the
format of serial bit streams at \SI{640}{Mbps} are prepared for
uni-directional, high-speed serial-data transmission over fiber optics
using the link-on-chip (LOC) serializer ASIC named
LOCx2~\cite{Xiao:2016dvu}. Each LOCx2 can process the data from four
ADC chips (a total of 16 ADC serial output channels) for transmission
of the data via two \SI{5.12}{Gbps} optical links to the Back-End. Five
bi-directional \SI{4.8}{Gbps} control links provide configuration, control
and monitoring based on the GBT
serializer-deserializer ASIC GBTx~\cite{Moreira:2010cjn}.
Consequently, each LTDB has up to 40 fibers for the data links and ten fibers 
for the control links, with all 124 LTDBs installed in ATLAS 
comprising a total of 5848 fibers and a throughput of more than 29 Tbps for 
the whole optical link system. The Miniature Transmitter (MTx) and
Miniature Transceiver (MTRx)~\cite{Zhao:2016czy} are the optical
transmitter and transceiver developed to fit between the LTDB PCB and
the cooling plate with a clearance in thickness of six millimeters. The MTx is
based on the LOCld~\cite{Li:2015hka} laser driver ASIC and an 850 nm
Vertical Cavity Surface Emitting Laser (VCSEL). The MTRx uses the
GBTIA~\cite{Menouni:2009ait} ASIC-loaded Receiver Optical Sub Assembly
(ROSA) from CERN for the receiving channel. All ASICs and the optical
components are tested to be radiation tolerant for applications in the
LAr FECs~\cite{Farthouat:1997tda, phase1tdr, radiation_qualification}. The design
of the two links, including the choice of the fiber, follows the
guidelines of the Versatile Link collaboration~\cite{VersatileLink}.
Figure~\ref{fig:opticallinks} shows a block diagram of the data and
control links.

As the data link transmits information for the ATLAS Level-1 trigger,
it is crucial that the transmission latency is fixed and is within the
time budget of the ATLAS trigger system. This requirement,
after subtracting the signal propagation in the optical fiber,
translates to a maximum latency of 75 ns for LOCx2~\cite{phase1tdr}.

\begin{figure}[h]
  \centering
  \includegraphics[scale=0.5]{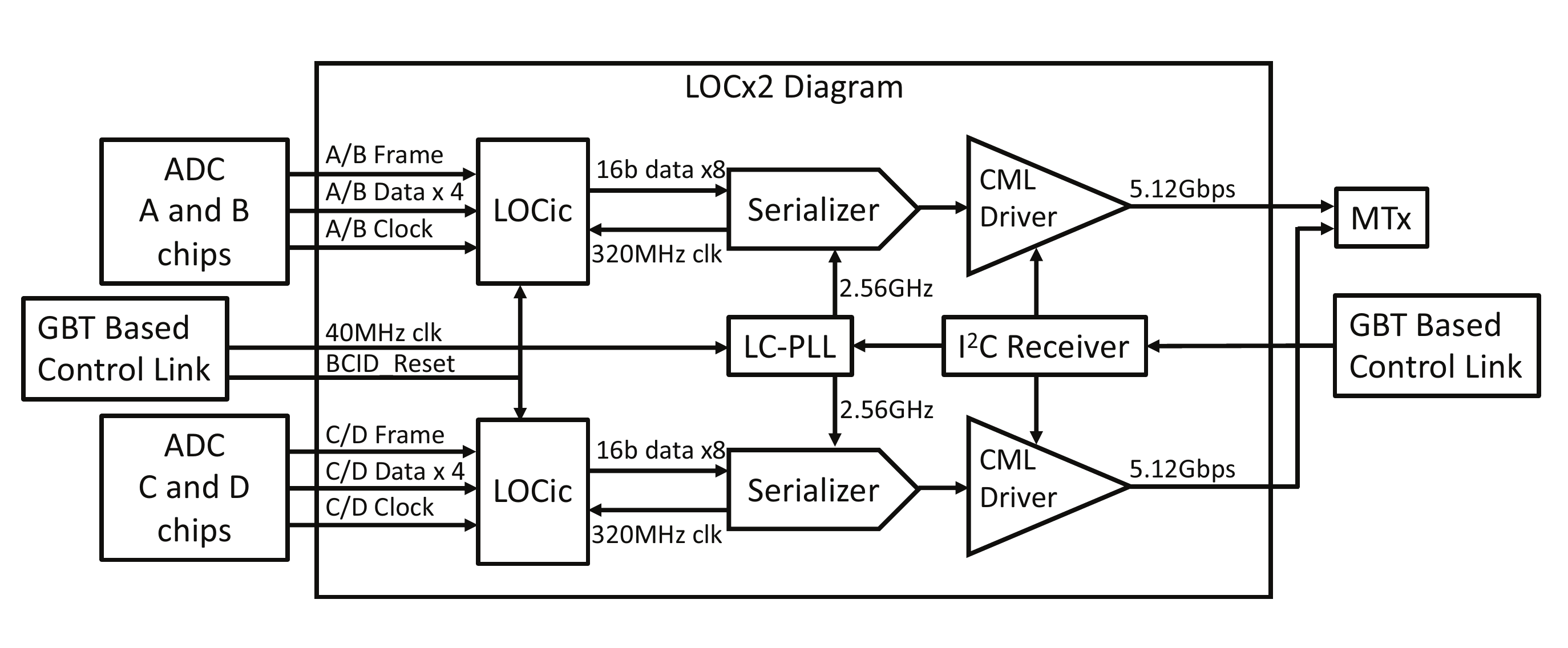}
  \caption{LOCx2 block diagram with connections to four ADCs (A,B,C and D), Control Link and MTx.}
  \label{fig:locx2diag}
\end{figure}

The LOCx2 and LOCld ASICs are specifically designed for the LTDB and
are based on commercial silicon-on-sapphire \SI{250}{nm} CMOS
technology. The two LOCx2 \SI{5.12}{Gbps} serializer channels share one
high-speed clock system based on an inductor-capacitor (LC) oscillator PLL. They
use a low-overhead custom encoding and framing transmission protocol
(LOCic~\cite{Xiao:2016gwm}) developed to meet the requirement of
transmission latency, and to achieve automatic link re-synchronization
should the receiving end of the link fail to recognize the
transmission frame due to single event upset (SEU) in the link
Front-End. The encoding protocol can detect up to three consecutive SEU
errors, enabling the link Back-End to discard data containing 
SEU errors. The LHC bunch-crossing ID is embedded in the protocol to
save transmission bandwidth. The LOCic also scrambles the data to achieve
DC balance in data transmission through fibers. The LOCld is the matching
dual-channel VCSEL driver for the LOCx2. Figure~\ref{fig:locx2diag} shows the block
diagram of the LOCx2.

\begin{figure}[h]
  \centering
  \includegraphics[scale=0.5]{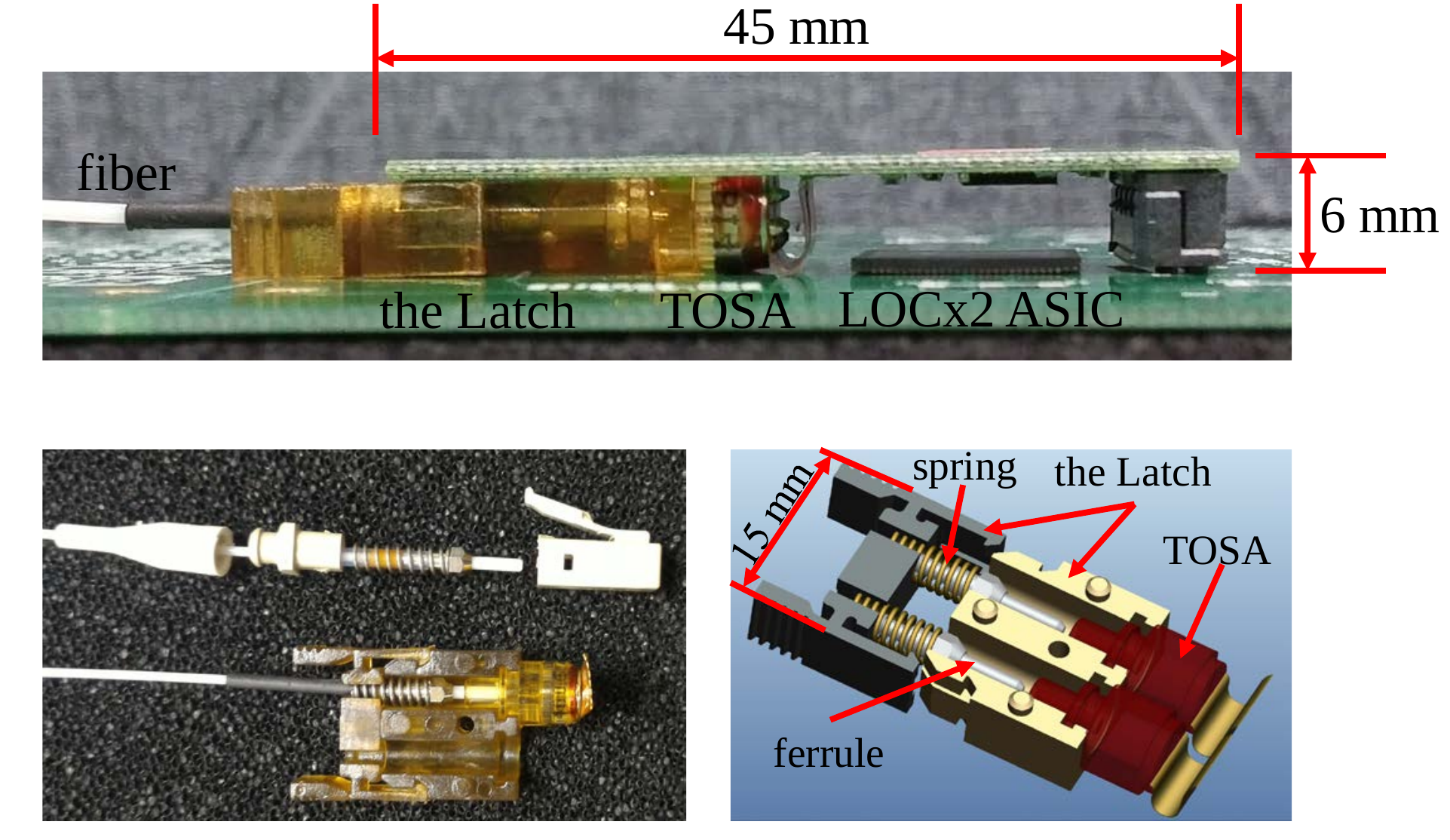}
  \caption{MTx and the Latch that couples the fiber with the TOSA or
           ROSA. Top: MTx on the LTDB with the LOCx2 under the module
           to save PCB space on the LTDB. Lower left: the LC ferrule
           taken from the industry standard LC fiber connector and the
           injection molded Latch. Lower right: a computer-aided design diagram
           of the Latch with the spring and ferrule to couple with the
           TOSA.}
  \label{fig:locx2}
\end{figure}

MTx and MTRx share the same mechanical design and electrical
connector. A custom mechanical coupler (the ``Latch'') holds the fiber and the
Transmitter Optical Subassembly (TOSA) or ROSA together. The fiber
couples with the TOSA or ROSA through an industry standard
ferrule. Figure~\ref{fig:locx2} shows a photo of an MTx with the custom
Latch to achieve the overall module height of 6 millimeters.

\subsubsection{Motherboard}
\label{sssec:ltdb-motherboard}
\begin{figure}[!h]
	\centering
	\includegraphics[width=0.8\textwidth]{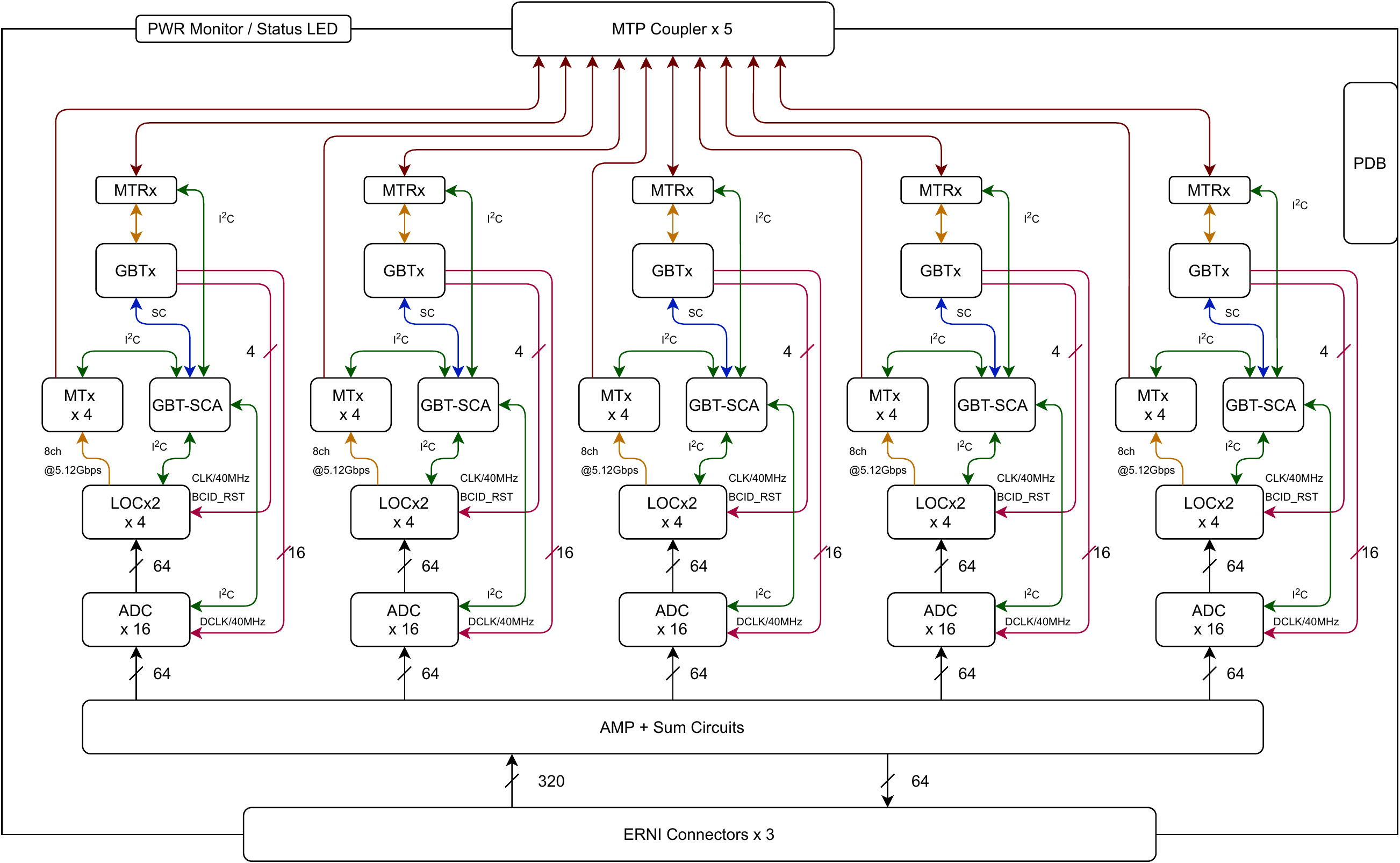}
	\caption{Block diagram of the LTDB: the board can be divided
	into five groups. Each group processes up to 64 Super Cell signals
	with 16 ADCs, four LOCx2 and four MTx. There is one GBTx set to
	implement timing, control and monitoring per group.}
	\label{fig:ltdb_blkdiagram}
\end{figure}

\begin{figure}[!h]
	\centering
	\includegraphics[width=0.7\textwidth]{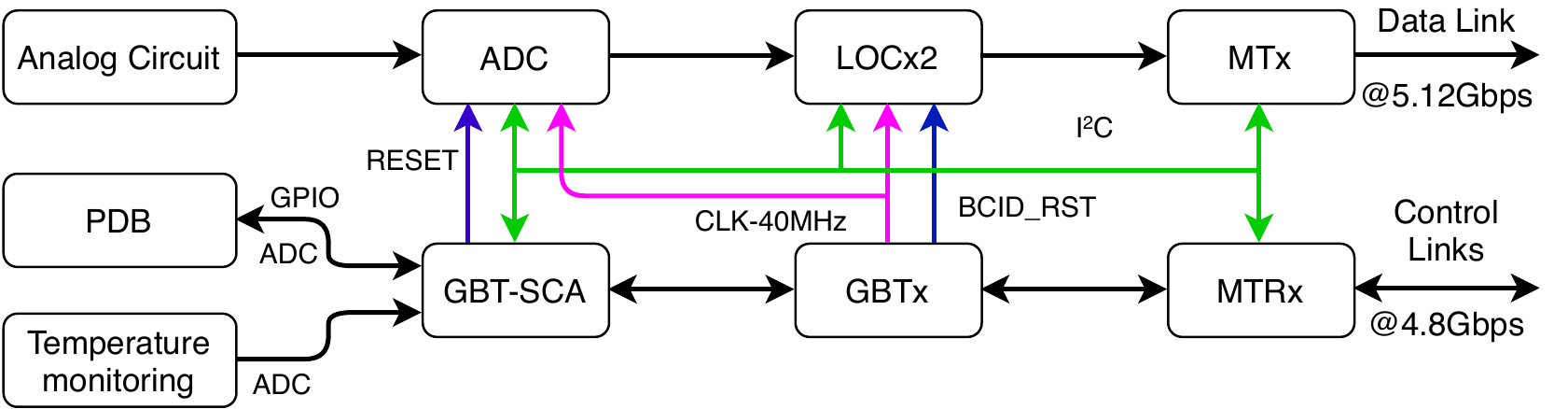}
	\caption{Block diagram of the LTDB data flow, showing the data
	link and the control link. The data link proceeds from the
	Super Cell signals injected into analog circuitry to
	ADC, LOCx2, and transmission through MTx. The control links
	send the clock and commands to GBTx, and collect monitoring
	information through MTRx.}
	\label{fig:ltdb_dataflow}
\end{figure}

The LTDB motherboard digitizes the Super Cell signals from the 
FEBs, and sends the data to the Back-End
electronics for trigger processing. Each LTDB is powered by a plug-in 
Power Distribution Board (PDB, see Section~\ref{sssec:pdb}) and
can receive up to 320 Super Cell signals which are amplified by
commercial amplifier chips and digitized by the custom ADCs. The
digitized data are then packaged and serialized in the LOCx2, and sent
out through 40 optical fibers, where each fiber transmits the data at
\SI{5.12}{Gbps}. For Super Cell signals from the front and the middle
layers, four neighboring signals are summed up, and the resulting 64
summed signals on the LTDB are sent to the legacy TBB
through the new baseplane.
Each LTDB comprises five independent identical logical groups, every one capable
of processing up to 64 Super Cell signals with 16 ADCs, four LOCx2 and 
four MTx. One GBTx set per group implements timing, control and monitoring.
A block diagram of the LTDB is
shown in Figure~\ref{fig:ltdb_blkdiagram}, and the data flow and
control links are shown in Figure~\ref{fig:ltdb_dataflow}.
While many elements of the LTDBs (such as powering) are common to all channels 
and consequently represent single points of failure for all channels, the 
independent logical groups enable normal operation of the other sections in 
case of a component failure in one section.

\begin{figure}[!h]
	\centering
	\includegraphics[width=0.5\textwidth]{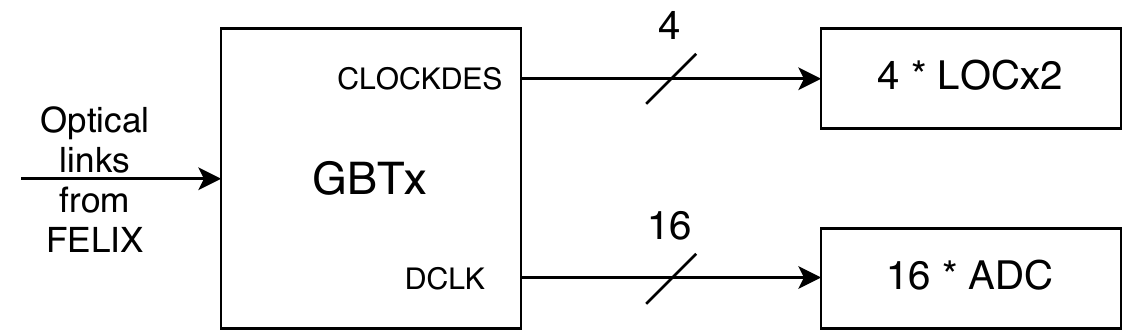}
	\caption{Block diagram of clock distribution on the LTDB. The
	DCLKs of the GBTx are used for the ADCs, and the CLOCKDES clocks
	of the GBTx are used for the LOCx2 chips.}
	\label{fig:ltdb_clkdistribution}
\end{figure}

Timing, control, and monitoring are implemented through five GBTx
links which operate at \SI{4.8}{Gbps} in both directions to communicate with
the FELIX system. The GBTx chips receive timing and control information
through the down-link from the FELIX, then fan-out the clock signals
to each ADC, and distribute the clock and BCR signals to each
LOCx2 as illustrated in Figure~\ref{fig:ltdb_clkdistribution}. The
control commands are transferred to GBT-SCA (Slow Control Adapter)~\cite{sca} chips, then
to be used to configure the MTRx, MTx, ADC and LOCx2 ASICs 
through the I$^2$C bus, and reset the ASICs through the General-Purpose Input/Output (GPIO) ports.

The monitoring information, including voltages, currents, temperatures
and power module status are transferred to the FELIX through the LTDB 
up-link. There are four input power rails drawn from the power bus of the FEC that are used to provide
power to the LTDB. A shunt resistor installed in each power rail allows 
the monitoring of both current and voltage. 
There are also ten thermistors to measure the board
temperature in different positions on the LTDB. All this monitoring
information is collected through the ADC of the GBT-SCA. The power
modules on the PDB have ``Power-Good'' signals that are used to
indicate their operational status. These Power-Good signals are
collected through the GPIO or the ADC of the GBT-SCA.

\begin{figure}[!h]
	\centering
	\includegraphics[width=0.8\textwidth]{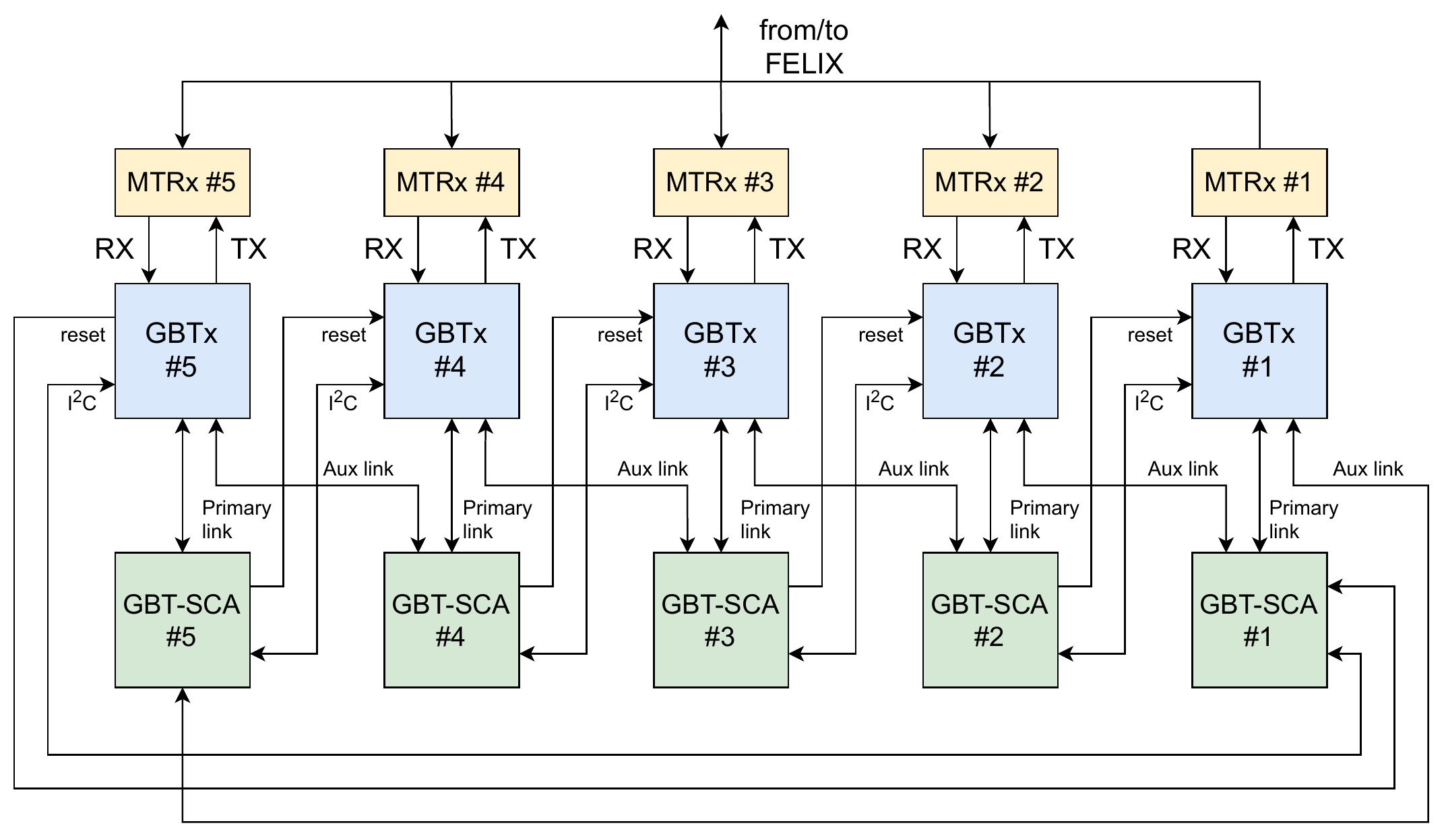}
	\caption{Redundancy design on the LTDB: the redundancy is
	implemented through the auxiliary port of the GBT-SCA. This
	full chain design can recover any GBTx if it gets stuck during
	data taking (when power cycling is not allowed), provided at minimum one GBTx is operating correctly. If all GBTx
	cannot work except GBTx \#5, the GBTx \#4 will be recovered
	through GBTx \#5 and GBT-SCA \#5, then the GBTx \#3 will be
	recovered through GBTx \#4 and GBT-SCA \#4, etc.}
	\label{fig:ltdb_redundancy}
\end{figure}

The LTDBs operate in a high-radiation environment, which may cause
SEEs in the electronics during data taking periods. As some of the
on-board devices will need a reset or power cycle to recover
functionality, a redundant design of GBTx and GBT-SCA is implemented
to mitigate this effect. Typically, the GBT-SCA ASIC connects via an
electrical serial link (e-link)~\cite{GBT2,elink} to the special purpose slow-control e-port of the GBTx
ASIC. This dedicated e-port runs at \SI{40}{MHz} Double Data Rate (DDR) mode,
giving an effective data rate of  \SI{80}{Mbps}. It is also possible to
connect the GBT-SCA ASIC to any of the other GBTx e-ports as long as
its data transfer mode is properly configured for \SI{40}{MHz} DDR
operation. Thus, a possible redundancy scheme can be implemented by
using both of these connections where only one of the GBTx e-ports is
active at any moment. On the LTDB, a daisy-chain is designed to
implement a redundancy scheme, as shown in
Figure~\ref{fig:ltdb_redundancy}. With this scheme, all the GBTx can
be recovered by GBT-SCA even if only one GBTx is operational. 
The GBT-SCA in return can be reset through the GBTx.

\begin{figure}[!h]
	\centering
	\includegraphics[width=0.8\textwidth]{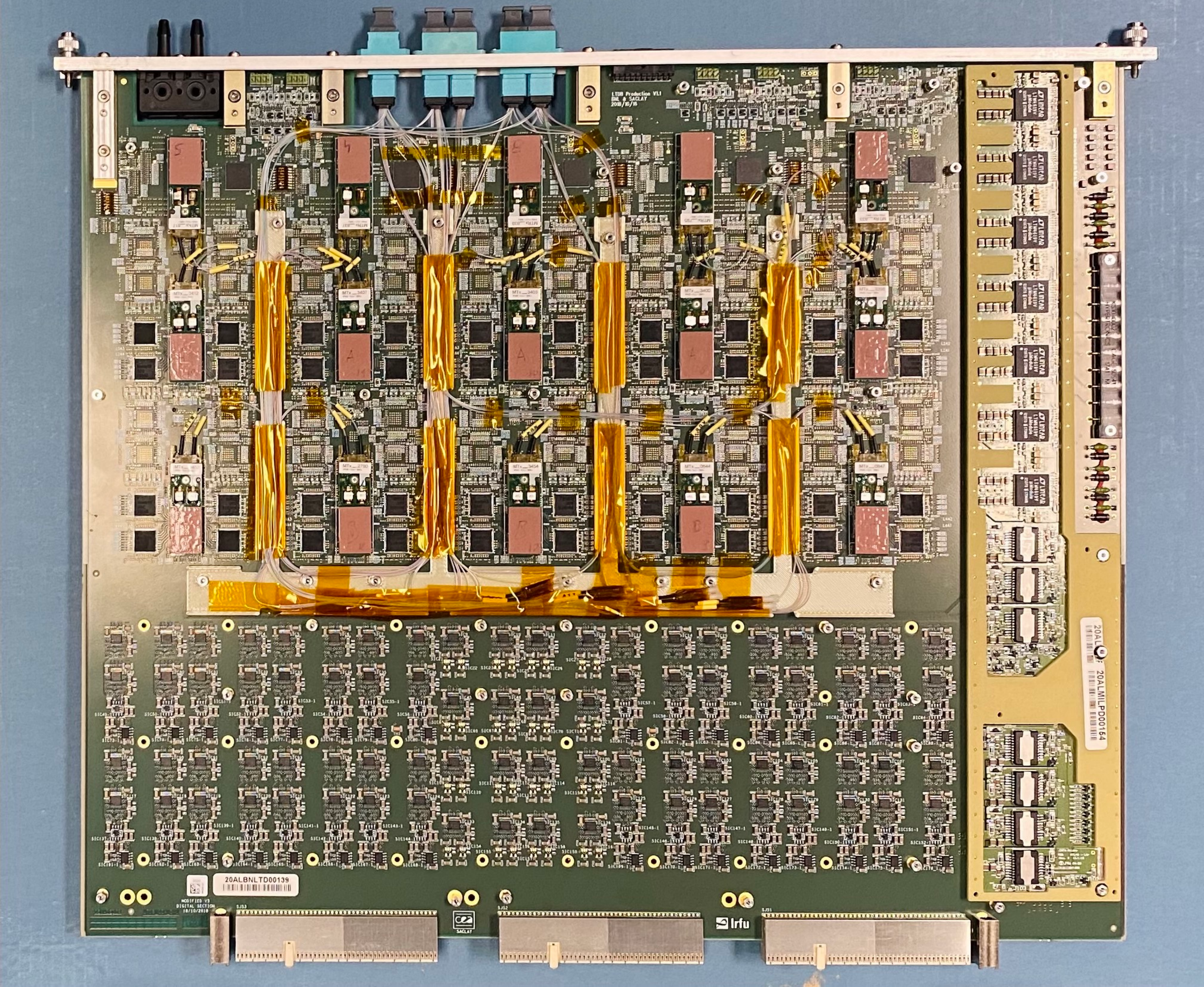}
	\caption{Photograph of an assembled LTDB with PDB installed.}
	\label{fig:ltdb_assembled}
\end{figure}

The LTDB PCB dimensions are \SI{410.7}{mm} x \SI{488.9}{mm}, with a 
thickness of \SI{2.2}{mm} comprising 24 layers\footnote{The manufacturer 
is TTM Technologies, 355 Turtle Creek Ct San Jose, CA  95125 USA}.
A picture of an assembled LTDB is shown in
Figure~\ref{fig:ltdb_assembled}. The board design has been carefully
considered to meet the low noise requirement. The analog amplifier
circuits, which are located on the bottom half of the board, are
separated from the digital circuits. The PDB is placed on the right
side of the board, and separate regulators are used to provide power
supplies for the analog circuitry. To minimize the trace length between
the ADCs and LOCx2, four ADCs are placed around one LOCx2,
where the data from four ADCs are organized and sent out through one
MTx module with two fibers. The MTx module is placed over the LOCx2
chip to save space.

Special care is taken to adapt the analog signals at the input of the LTDB.
The analog output dynamics of the Linear Mixer has to be matched to the
input dynamics of the LTDB ADC. In addition, the single-ended
signal coming from the LSBs located on the FEBs
has to be converted to differential at the ADC input. This conversion and the
dynamic range matching are done by a very low-power (\SI{1.14}{mA/channel})
 fully-differential amplifier (THS4522), equipped with an appropriate passive network to
set the gain and the pedestal at the ADC input. 

In the case of the main
readout, the current pulse coming from the cells is shaped by a CR-RC$^2$ filter.
This filter is optimal in terms of signal over noise optimization. In the case of the trigger path,
the cell signals are first summed in groups of up to four cells by the Linear Mixer, 
and shaped by a CR-RC active filter.
A CR-RC filtering has been chosen to minimize the attenuation
over the trigger path, and a passive RC stage
with a time constant of \SI{15}{ns} is implemented on the LTDB to get the same optimal CR-RC$^2$ as for the main readout. This passive RC stage also acts as
an anti-aliasing low-pass filter in front of the LTDB ADC. 
Super Cell signals corresponding exactly to legacy analog trigger signals are configured as a high 
impedance pick up on the LTDB, since they are directly sent via the backplane to the legacy analog trigger
boards (TBB and TDB), whose inputs are \SI{50}{$\Omega$} adapted. Super Cell signals that need further
summing on the LTDB before being sent back to the backplane for the legacy analog
trigger, are \SI{50}{$\Omega$} adapted on the LTDB.

The Linear Mixer is no longer in its linear dynamic range if its output amplitude goes
above \SI{3.3}{V}. This voltage threshold has to match the \SI{2.4}{V} saturation voltage of the ADC differential input.
Since each ADC has 3600 to 3800 available digital codes, the least significant bit of the LTDB ADC
corresponds approximately to \SI{1}{mV} at the Linear Mixer output everywhere on the detector.

In addition, the pulses from the Super Cells do not all have the same shape,
since the drift time of the charges produced by ionization of the Liquid Argon by the shower particles changes as a function of $\eta$. 
Because a bipolar shaping circuit is used, this translates into a change of the amplitude of the undershoot of the pulse. Since the undershoot
has to be digitized and should not saturate the ADC, the gain and pedestal of each
channel have been adjusted so that signals close to the Linear Mixer
saturation amplitude use the full dynamics of the ADC input. Over the whole
acceptance of the calorimeter, this leads to gain differences up to 30\%.
 
\subsubsection{Power distribution board}
\label{sssec:pdb}
\begin{figure}[tb]
  \centering
  \includegraphics[width=.45\textwidth]{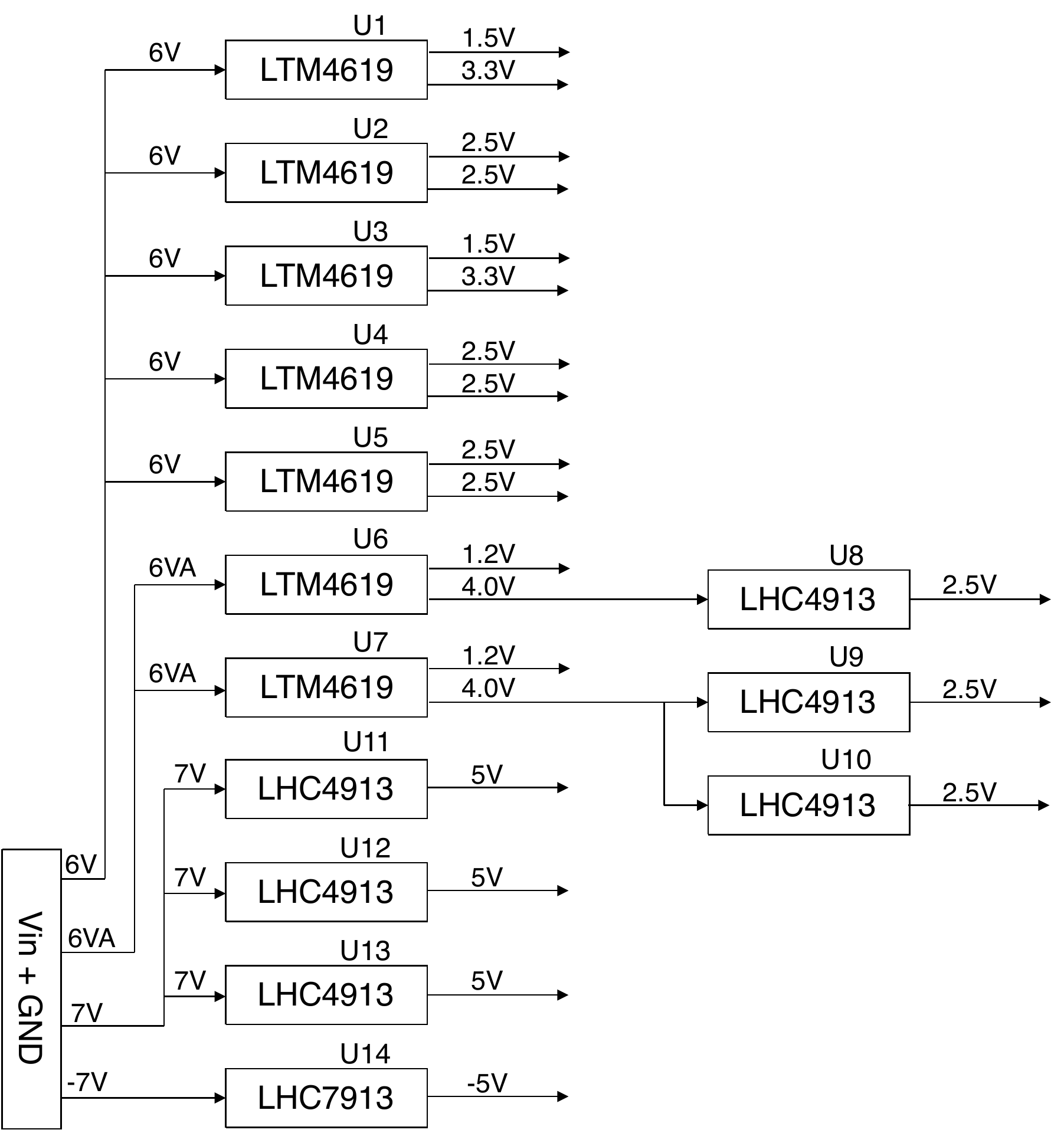}
    \caption{Power scheme of the PDB. The board integrates 
    LTM4619 DC/DC converters (U1-U7) and LHCx913 linear voltage regulators (U8-U14) to produce the voltages needed by the LTDB
    (the number of devices has been established based on the budget of the output currents). From
  the two $+6$ V lines available on the input power bus, one (labelled "6V") is used to generate the digital voltages and
  the other one ("6VA") is used to generate the analog voltages (see text).}
  \label{fig:PDB_fig1}
\end{figure}

Power distribution on the LTDB is provided by the PDB mezzanine card.
The choice of developing a separate board for the generation of the supply voltages for the LTDB ensures 
forward compatibility in the future \phasetwo upgrade, for which a different power distribution scheme for the Front-End 
electronics is planned.
The use of a mezzanine board allows the re-design and replacement of only the PDB, without
modifying the LTDB itself (as explained in Section~\ref{sssec:phase2}).

The PDB consists of digital and analog sections to power the corresponding sections of the LTDB. 
The block diagram of the power scheme is shown in Figure~\ref{fig:PDB_fig1}. 
From the $+6$ V line taken from the FEC power bus, the following 
voltages are created for the digital part of the LTDB: $+1.2$ V, $+1.5$ V, $+2.5$ V and $+3.3$ V. This is achieved using LTM4619 
DC-DC converters from Linear Technology Corporation (now Analog Devices, Inc.).

The analog voltages are generated as follows. From the $+7$ V ($-7$ V) of the FEC power bus, the $+5$ V ($-5$ V) is 
generated using the LHC4913 (LHC7913) Low-DropOut (LDO) linear voltage regulators from STMicroelectronics. 
The LHC4913 is also used to generate the analog $+2.5$ V, starting from the $+6$ V of the power
bus (with an intermediate step down at $+4$ V, using an LTM4619, to avoid a large voltage drop in the LDO).

The PDB is radiation tolerant~\cite{PDB:radtest0,PDB:radtest1,PDB:radtest2} and able to operate in presence of the 
maximum magnetic field expected in the LTDB positions (lower than \SI{0.1}{T})~\cite{PDB:magtest}. 
The radiation tolerance requirements for the PDB are less stringent than for the rest of the LTDB components (see Section~\ref{sssec:adc})
since the board will have to operate only for the LHC \runthree and will be replaced before the start of the HL-LHC.

\begin{figure}[htb]
  \centering
  \includegraphics[width=.9\textwidth]{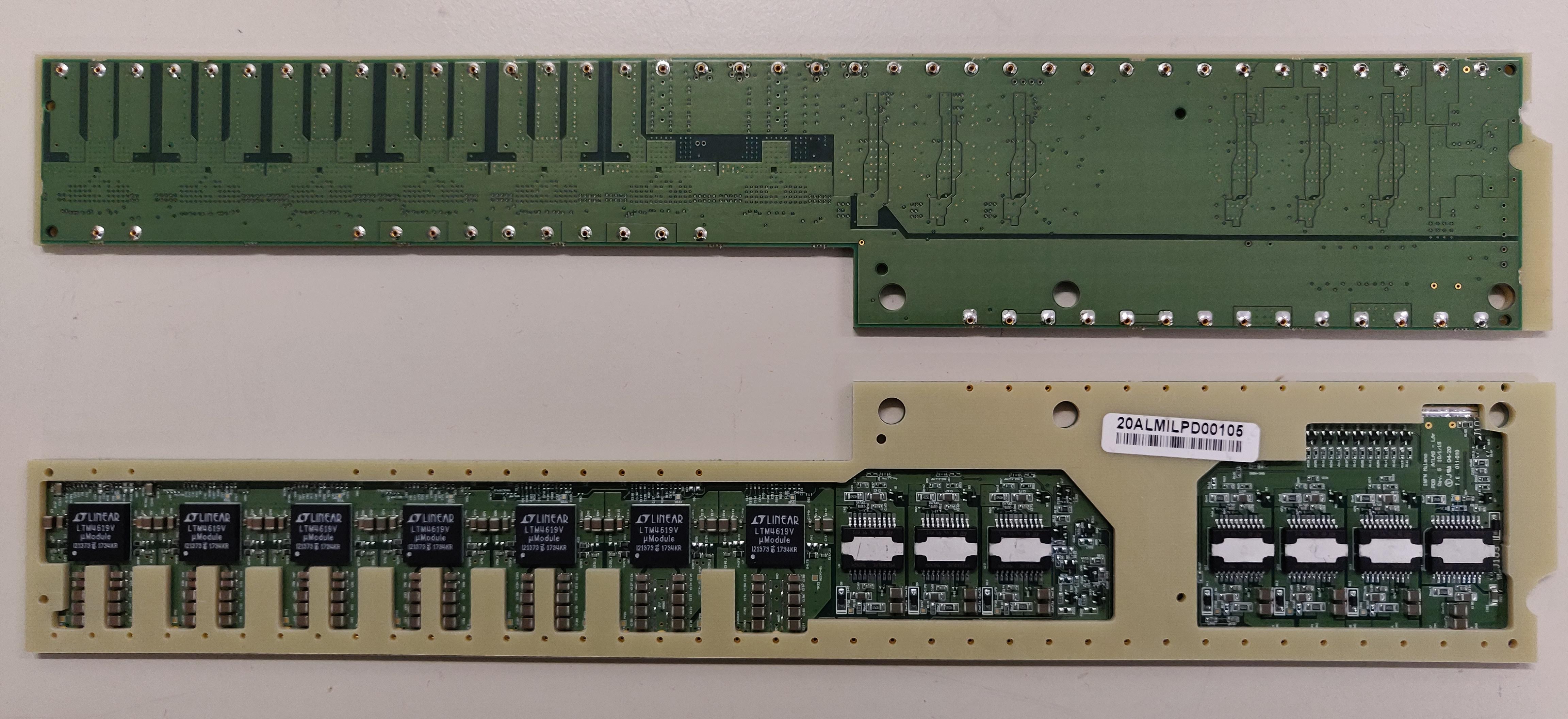}
  \caption{Picture of a
  PDB as seen from the bottom and top. All components are mounted on the top side. On the same side, the G10 reinforcement
  frame is also visible.}  \label{fig:PDB_fig2}
\end{figure}

The board is manufactured as a ten-layer PCB of 1.6 mm thickness\footnote{The PDBs have been produced by ARTEL S.r.l. - 52041 Pieve Al Toppo (AR) - Italy}, reinforced with a fiberglass (G10) frame glued on the top side. 
When mounted on the LTDB motherboard, the total maximum vertical height is about \SI{5.4}{mm}, which fits just below the LTDB cooling plate, 
mounted at \SI{6}{mm} from the motherboard surface. 
A picture of one production PDB
is shown in Figure~\ref{fig:PDB_fig2}.
An automated set-up has been developed to test the correctness of all output voltages of the PDBs at full load
before they were shipped to the LTDB assembly sites.

\subsubsection{Quality assurance and control}
\label{sssec:ltdb-qaqc}

The LTDB quality assurance and control (QA/QC) involves the testing
of three main components: the cooling interface is tested for leaks,
the PDBs are tested to meet the requirements on voltage and current of
the outputs, and each LTDB with a qualified PDB installed is subjected
to functionality testing and -- integrated with the test stand -- to
performance testing as detailed below.

Each PDB module is tested with a standalone test stand to verify basic
functionality before it is cleared to be installed on an LTDB
motherboard. The input and output voltages and current are measured to
obtain efficiency information. All test data and analysis results are
examined and logged into a database. A PDB module is accepted if it
passes the following requirements: the output voltage is within $\pm
2.5$\% of the nominal value; the output voltage ripple is within 10~mV
peak-to-peak; and the efficiency is better than 70\% at nominal load.

The PDBs passing the tests above are then subjected to a Highly
Accelerated Stress Screening (HASS) test, where the PDBs are placed in
a chamber and undergo ten thermal cycles between 0~$^{\circ}$C
and 60~$^{\circ}$C for about 12 hours. The goal of the HASS test is to
find any failures due to component infant mortality, cold solder
joints, etc. After HASS testing, each PDB is installed on an LTDB for
re-testing, where the voltage of all outputs is measured, control and
monitoring signals are tested, and the power rail ramp-up time (from 10\% to 90\% of the rising edge) for the GBTx
is measured as well. If the PDB passes this re-testing, it can be used
on the LTDB for integration testing.

Each LTDB motherboard is initially tested with a standalone test stand to verify
basic functionality. If it passes this test, a HASS test as described
for the PDBs above is performed as well. After the HASS test, the PDB,
MTx, MTRx and fibers are installed on the LTDB. Then, digital
functionality and power distribution as well as analog performance are
verified, followed by an integration test, all performed using the
integration test stand (see Section~\ref{ssec:fe-integration}), before
the LTDB is qualified to be installed on the detector.

Initially, the control links through GBTx as well as the digital readout chain
through LOCx2 are verified. Each ASIC, including GBT-SCA, MTRx, MTx,
ADC and LOCx2, is configured and checked through the GBT link to
verify the control links. The readout chain is verified through ADC
test patterns before analog signal injection is exercised. To pass these
digital functionality tests, an LTDB must meet the following
requirements: slow control, configuration and remote monitoring logic
are functioning properly; pedestal tests of the digital readout chains
are functioning correctly for all 320 channels; and monitoring
information is correct, with the temperature of the digital part at
$20 \pm 5$~$^{\circ}$C, the temperature of the analog part at $30 \pm
5$~$^{\circ}$C, and the voltages at their nominal values $\pm 10$\%.

The analog performance integration focuses on the detailed
characterization of the full board. Both pedestal tests and pulse
tests are performed on the board, as well as tests to
measure the signal amplitude and peaking time. Additional testing
enables assessment of the INL of each channel
while summing testing verifies the summing output and measures the
non-linearity of each summing output. All test data and analysis
results are examined and logged into a database.  To pass these analog
functionality tests, an LTDB must meet the following requirements:
pedestal tests of analog readout chains are functioning correctly for
all 320 channels, with a noise level better than 0.85 of the least significant bit; pulse tests of both
analog and digital readout chains are functioning correctly for all
320 channels; the INL is better than 0.35\%; all 64 summing channels
are functioning correctly; and the power consumption of the whole board 
is within $95.0 \pm 4.5$~W.

After the above tests are passed, the cooling interface is installed.
Each cooling interface including the cooling blocks and cooling plates
is leak-checked in a dedicated test station prior to assembly on the
LTDB, and again after installation on the LTDB. If no leak is
observed, the boards passing above performance tests are shipped to
CERN, where another leak check with pressurized gas is performed in a
dedicated test station before installation. Only the boards passing
this additional leak check are finally installed on the detector.
 
\subsubsection{Compatibility with \phasetwo upgrade}
\label{sssec:phase2}
After the \phaseone upgrade and data taking in \runthree, the LTDB boards will still be used in the \phasetwo upgrade for \runfour and beyond
as part of a level-0 trigger system~\cite{TDAQphase2tdr}.
To fully exploit the functionalities of the LTDB in the \phasetwo configuration
of the LAr trigger and readout system, some design work is needed on the side
of the future FEBs. Correct matching between the analog dynamic
range of the differential trigger sums that will be provided by the Front-End
preamplifier/shaper ASICs and the single-ended analog dynamic range of the LTDB
inputs has to be ensured.
These design aspects will not be covered here, since the \phasetwo FEBs
are currently under development and the details of how to ensure compatibility
are not yet fully defined.

On the LTDB side, it is critical to ensure easy adaptation to
the configuration of the LAr Front-End system during \phasetwo. Seen from the
LTDB side, the main differences between \phaseone and \phasetwo configurations
are:
\begin{itemize}
\item In \phasetwo, the legacy analog trigger system, based on the analog
TBB, will have been decommissioned. The consequence
for the LTDB is that the adders that are present to sum groups of four channels
will no longer be needed. On the LTDB, these adders build the
analog layer sum signals needed by the TBB from the Super Cell signals.
\item The PDB will have to be redesigned to adapt to the \phasetwo power rail 
configuration which will provide only one single positive voltage; negative power 
supply rails will no longer be available. The absence of negative voltages means 
that it will not be possible anymore to power the (then no longer needed) legacy analog adders.
Appropriate measures have been taken so that this situation does not degrade the performance
on the Super Cell signal path.
\end{itemize}

\begin{figure}[htb]
\centering
\includegraphics[width=0.75\textwidth]{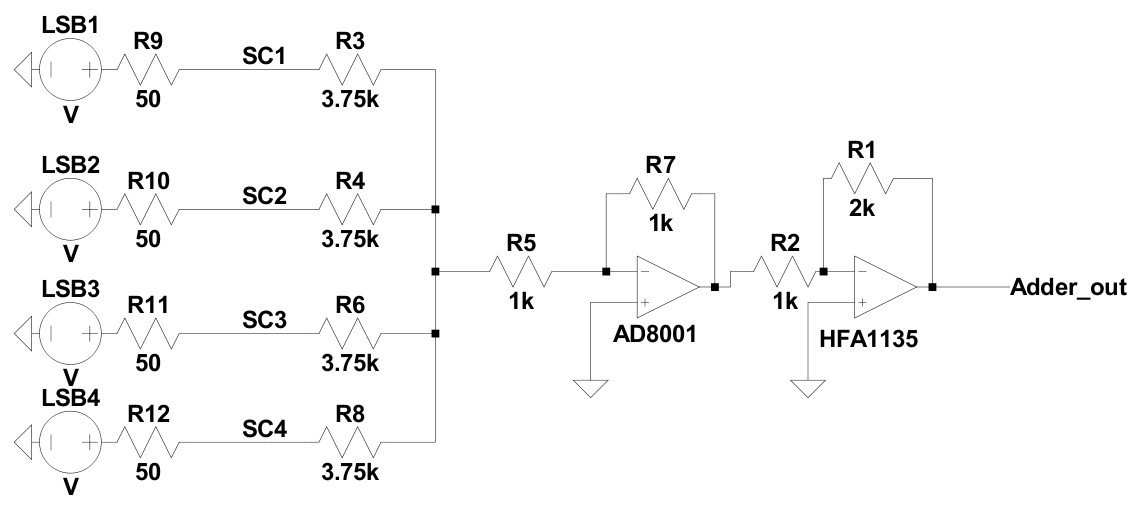}
\caption{Simplified schematic of the adder implemented on the LTDB for compatibility with the legacy analog trigger system.}
\label{adder}
\end{figure}

Figure~\ref{adder} shows a simplified schematic of the LTDB analog
adder circuitry. The Super Cell signals are modelled as a
voltage source in series with a \SI{50}{$\Omega$} resistor. It can be seen that each of
the four Super Cells (SC1 to SC4
in Figure~\ref{adder}) that are
summed by the legacy adders are connected through a \SI{3.75}{k$\Omega$} resistor 
to the inverting input of the first of
the two inverting amplifiers (AD8001) that make up the complete adder.
As long as this amplifier is powered on, its negative input is a virtual ground, and
there is no induced crosstalk between Super Cells. On the
other hand, once the amplifier is powered off, the Super Cells that are summed
together are connected through a resistive network that
introduces some resistive crosstalk. Consequently, the resistor values
have been chosen high enough to keep the crosstalk below 0.2\%.
Taking into account the actual resistor values used, the crosstalk between
any two Super Cells connected to the same adder is equal to 0.17\%.
This number has been checked
on a test board that has been built for this purpose, featuring a group
of four Super Cells and their adder, with the possibility to power off 
specifically the adder circuit.
This shows that the crosstalk induced during \phasetwo by the absence of the negative 
power rail is acceptable.

\section{New Back-End electronics}
\label{sec:be}

\subsection{\lar carrier}
\label{ssec:larc}
\begin{figure}[tb]
  \centering
  \includegraphics[width=\textwidth]{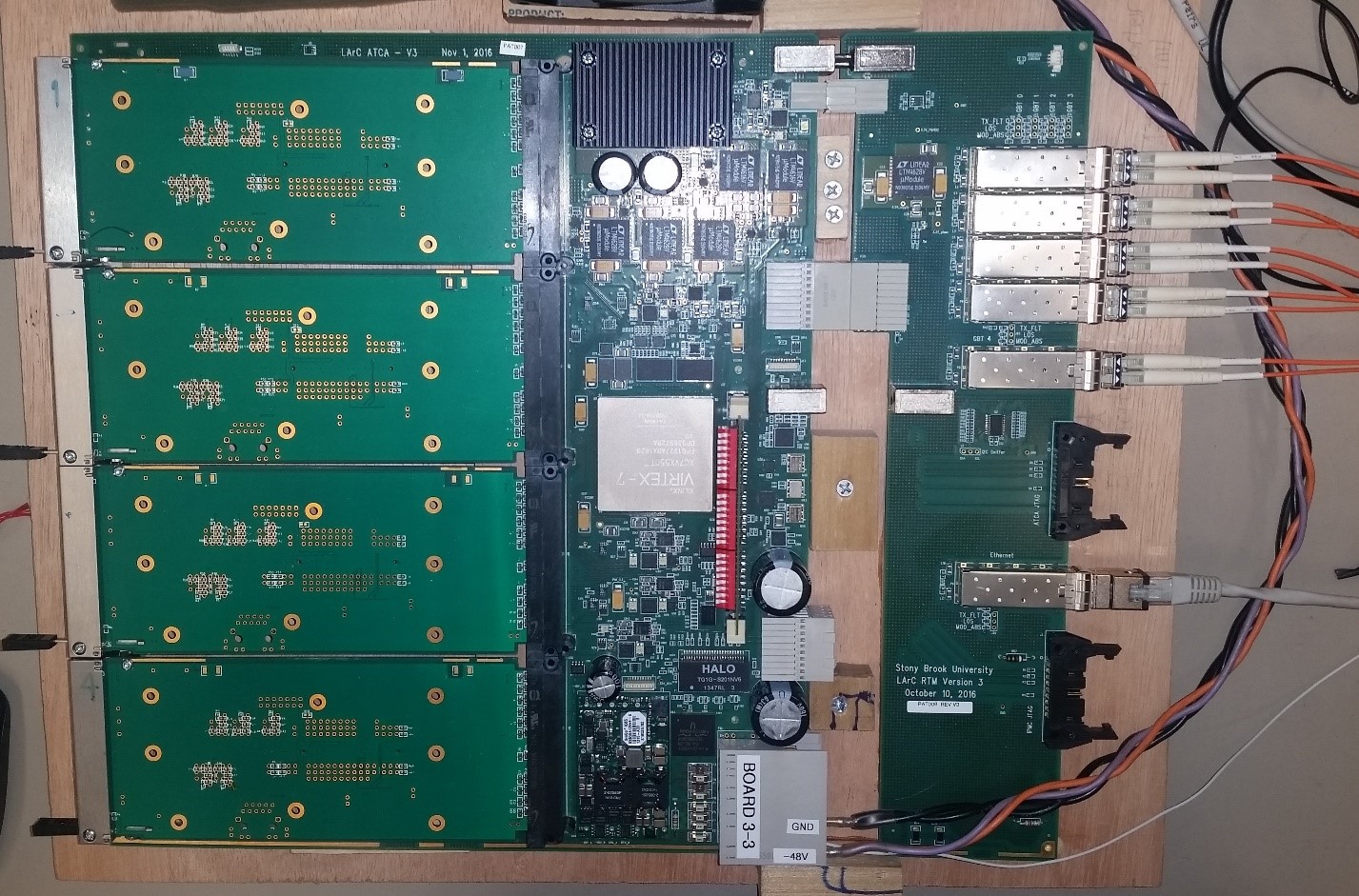}
  \caption{A photograph of the LArC and RTM, with the LArC side panels removed for clarity and testing cards in the AMC sites.\label{f-larc-photo}}
\end{figure}

The LAr carrier (LArC) is an ATCA standard cut-out blade with a Rear Transition Module (RTM).
The LArC hosts the LATOME processor daughtercards (Section~\ref{ssec:latome}),
sends and receives data from the read out system, and distributes clocks and
trigger signals synchronized to the LHC beam clock. Data connectivity includes
the ATLAS Trigger and Data Acquisition (TDAQ) system for the global monitoring
 and standard Ethernet \SI{1}{GbE} and \SI{10}{GbE} links.
 The latter are based on the 10 Gigabit Attachment Unit Interface (XAUI). 
 A photograph of the LArC is shown in Figure~\ref{f-larc-photo}. 

Data are received and/or
transmitted on the LArC using five paths: (1) serial Multi-Gigabit Transceiver 
(MGT) connections of \SI{9.6}{Gbps} 
between the LArC and LATOME cards for readout of data on L1A (TDAQ path),
(2) readout of \SI{10}{Gbps} local monitoring data between the LArC and LATOME cards
upon user request (local monitoring path), (3) Low Voltage
Differential Signaling (LVDS) connections between the carrier and LATOME cards
used for clock and trigger distribution, (4) serial transceiver links to
optical fibers connected to Small Form-factor Pluggable (SFP+)
modules\footnote{SFP+ modules are an industry standard hot-swappable
electro-optical converter module.} on the RTM which provide the external
connection for the TDAQ path and (5) serial transceiver links between the
carrier and other ATCA boards using the ATCA shelf backplane which provide the
external connection for the local monitoring path. The master clock signal and trigger
data are recovered from a dedicated link of the TDAQ data path from the ATLAS
trigger distribution system to the LArC and distributed to all of the
LATOMEs. 

\begin{figure}[tb]
  \centering\includegraphics[width=\textwidth]{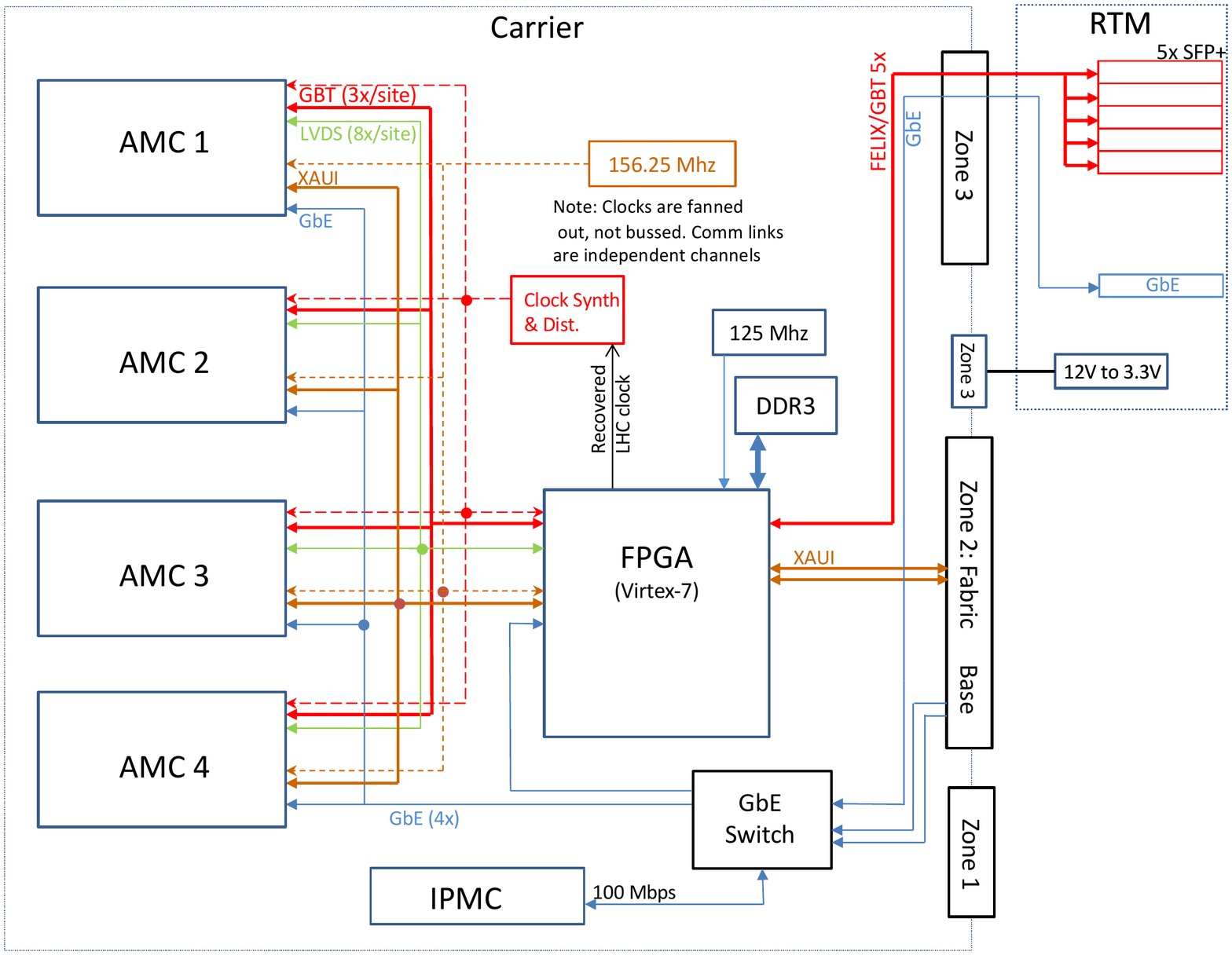}
  \caption{A block diagram of the data and clock connections on the LArC. The GbE connectivity is provided using a BROADCOM BCM5396. The switch is automatically initialized at power on without programming.
  The switch to the AMC, switch to FPGA and switch to RTM SFP connections are standard 1000Base-X connections. The connections between the switch and the Zone 2 base layer are 1000Base-TX as required by the ATCA standard,
  and the connections between the switch and the IPMC are \SI{100}{Mbps} PHY-based links.
  All XAUI and TDAQ connections are routed through the FPGA. I$^2$C based monitoring and configuration connections are not shown.\label{f-larc-block}}
\end{figure}

A block diagram of the carrier data and clock connectivity is shown in
Figure~\ref{f-larc-block}, The main component on the board is a \xilinx 
\virtexseven FPGA, part number XC7VX550T-FFG1927-2-E, 
through which all of the TDAQ and local monitoring data passes. The FPGA boots from
onboard flash and has 256 MB of external DDR3 memory. There are four clock
domains on the LArC for the TDAQ path, the local monitoring path, the 1 GbE path and a
system clock. The TDAQ clock is recovered from the LHC master clock and
provides all trigger-related timing. 

Along with the dedicated purpose TDAQ and local monitoring links, the carrier provides GbE connections. All GbE links are routed through a BROADCOM 
switch on the LArC. The GbE links are connected from the switch to the LArC FPGA, each of the 4 LATOME sites, the IPMC module (at \SI{100}{Mbps}), an SFP cage on the RTM and 2 GbE connections to the ATCA shelf backplane. The redundant backplane connections are required by the ATCA standard.

The LArC uses an ATLAS standard IPMC (Section~\ref{ssec:ipmc}) to provide all
board power and sensor management functions. The ATCA standard also specifies
electrostatic shielding and status Light Emitting Diodes (LEDs) all of which
are included in the carrier. The FPGA can be rebooted without a LArC or shelf
power cycle by using dedicated commands sent to the IPMC over the 1 GbE
network.

The RTM is used to provide connectivity for 5 TDAQ channels via SFP+ modules.
It also has Joint Test Action Group (JTAG) connectors on the back panel providing access to the
onboard JTAG chain and an IPMC JTAG interface. An insertion indicator switch is
integrated into the RTM insertion/removal handle, and by sensing the switch,
RTM power is controlled by user code in the IPMC. This allows hot swapping
control of the RTM DC-to-DC converter. The RTM data connection to the carrier
is a standard ATCA connector located in carrier zone 3. A second zone 3
connector provides \SI{3.3}{V} management power and \SI{12}{V} payload power to
the RTM.

The production LArC's and RTM's undergo multistep testing. The
first step involves checking the power systems, including a simple resistance
test before initial powering on. The second step then tests the infrastructure
including JTAG, I2C, IPMC, DDR3, clock generation, and the 1 GbE
infrastructure. The next step is testing of all transceiver TDAQ and local 
monitoring links to
$10^{14}$ and $10^{12}$ bits respectively, using the industry standard 
PRBS31 (PseudoRandom Binary Sequence of length $2^{31}-1$). The LVDS links are
also tested at 160 MHz using PRBS31. The tests for the links which connect to LATOME's
use special passive loopback cards inserted in the
AMC slots during this check. In the final step a boot
flash memory is programmed with low-level board management code, and the
one-time-programmable FPGA identification eFuse is set. At this point all
components and functionality are verified, and the carriers are ready to have
LATOMEs installed.
 
\subsection{\lar trigger processing mezzanine}
\label{ssec:latome}

\begin{figure}[htb]
	\centering
	\includegraphics[width=0.8\linewidth, height=0.2\textheight]{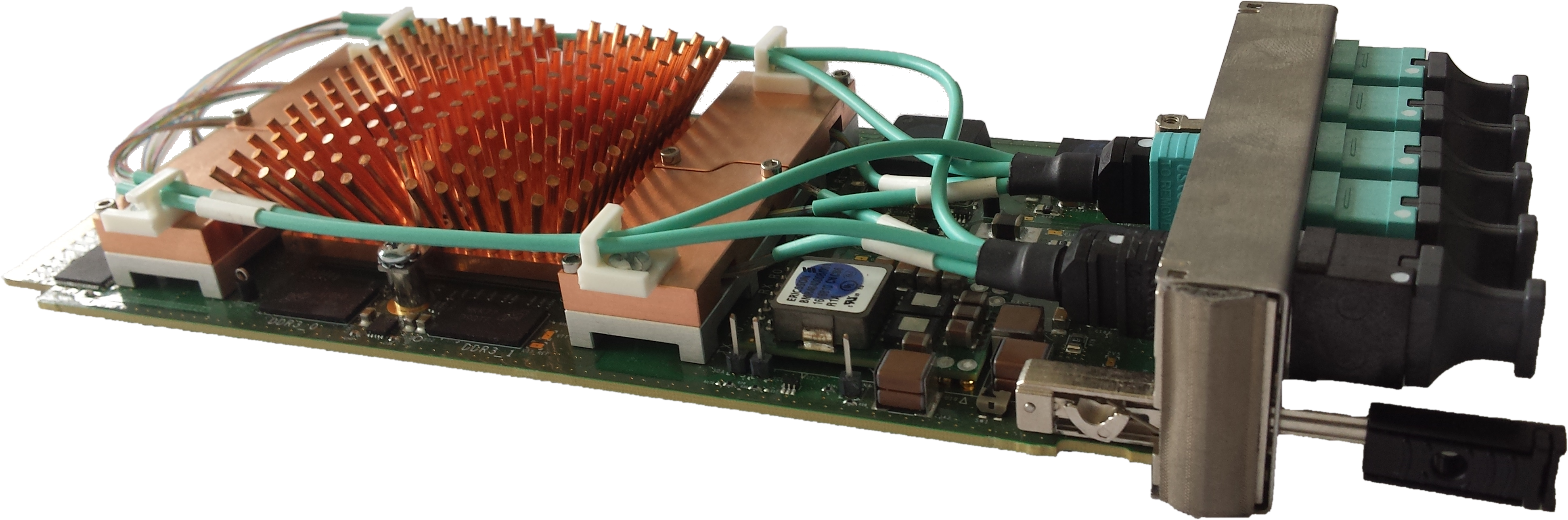}
	\caption[LATOME equipped with optical fibers.]{LATOME board equipped with optical fibers.}
	\label{fig:latome_picture}
\end{figure}

The LATOME board reads incoming data from 48 optical links, processes these data in a high-performance FPGA and sends the results to 48 optical links. The board is also interlinked to the standard Ethernet networks (\SI{1}{GbE} and \SI{10}{GbE}) and the specific ATLAS network (TDAQ) for monitoring and control purposes. This board fulfills the AMC standard and is plugged into the LArC which provides connectivity as discussed above. Figure~\ref{fig:latome_picture} shows the LATOME equipped with optical fibers: the black front panel connectors gather the 48 transmitting fibers and the blue ones the 48 receiving fibers. The FPGA and the optical transceivers are cooled down with large copper heatsinks.

The board receives LTDB data at \SI{5.12}{Gbps} and transmits new FEX system data at \SI{11.2}{Gbps} through four twelve-channel BROADCOM MicroPOD\texttrademark~receivers and four twelve-channel BROADCOM MicroPOD\texttrademark~ transmitters. These MicroPOD\texttrademark~ modules make the optical to electrical and the electrical to optical conversion for the 2$\times$48 FPGA high speed links. All these links can be synchronized either by a local oscillator with a fixed frequency of \SI{160.316}{MHz} or by the external clock provided by the LArC. The links to the standard Ethernet (\SI{1}{GbE} and \SI{10}{GbE}) and the TDAQ (\SI{9.6}{Gbps}) networks are routed through the AMC connector. This connector carries also the common TTC commands over four LVDS links.

\begin{figure}[htb]
	\centering
	\includegraphics[width=1\linewidth, height=0.4\textheight]{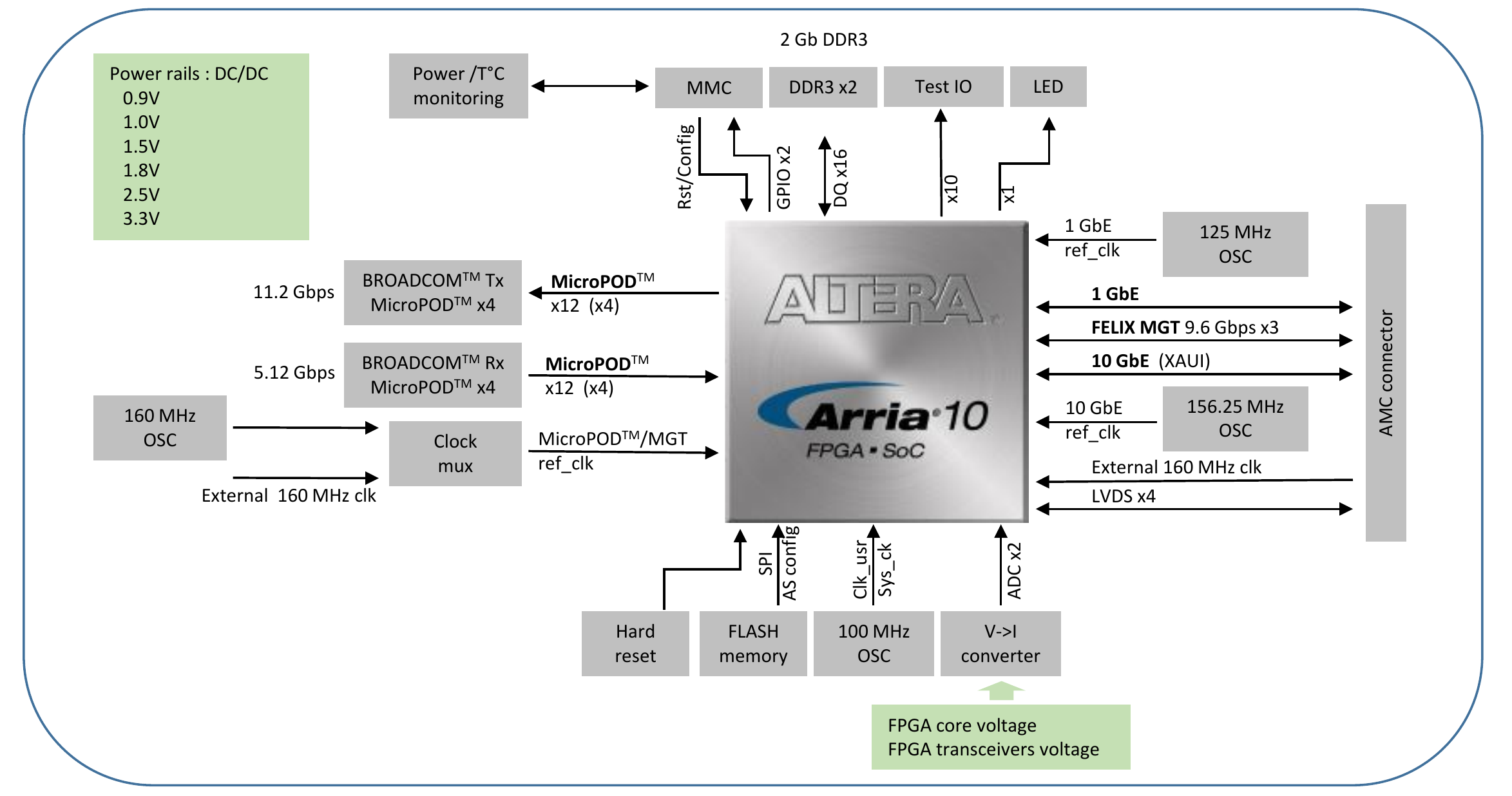}
	\caption{Block diagram of the LATOME board.}
	\label{fig:latome_synoptic}
\end{figure}

On the board, the main component is the \intel{} \arria~10 FPGA, part number 10AX115R3F40E2SG (Figure~\ref{fig:latome_synoptic}). Four DC/DC converters provide the different voltage supplies to this FPGA. They are enabled by a sequencer chip in order to generate the power on and off sequence required by the \arria~10. This sequence is triggered by the Modular Management Controller (MMC) chip which manages the AMC hotswap, the onboard sensors (temperature, voltage and current) and the IPMC connection (see Section~\ref{ssec:ipmc} for the IPMC description).

The FPGA firmware boots from a \SI{1}{Gb} serial flash memory. Two \SI{2}{Gb} DDR3 external memories are also available to exchange data with the FPGA. Four oscillators (125/156.25/160.316 and \SI{100}{MHz}) provide all the clocks needed to synchronize the internal FPGA logic.

The LTDB and FEX optical cables are plugged into the LATOME front panel through high density Multi-fiber Push On (MPO) optical connectors. 
On this front panel there is also access to the FPGA JTAG and test outputs and the MMC JTAG port with external connectors. A reset push button and status LEDs are available as well.

There are two types of LATOME according to the combination of input channels on the MPO connectors and pigtail fibers to the MicroPOD\texttrademark~ modules.
The standard LATOME has four MPO connectors of 12 channels (MPO12) while the special LATOME has two MPO connectors of 24 channels (MPO24) and one MPO12 for data 
reception (Rx). Both LATOME types have one MPO connector of 48 channels for data transmission (Tx).

The LATOME boards were validated after their production in several stages.
Once the LATOME is received from the assembly house, the board is stamped with its unique identifier by setting a ten-bit vector with pull down resistors connected to the FPGA input pins. This identifier is used to track the board all along the process. 
The first stage of validation consists of checking the powering sequence. The board is powered on and the power-up sequence and the DC/DC converters' voltage values are checked. Then the FPGA JTAG connection is verified and all the ports and components interlinked to the FPGA are sequentially tested: oscillators, DDR3, flash memory and ATCA carrier links up to $10^{13}$ bits with PRBS31 vectors.

Once all the FPGA interfaces are validated, the MMC access to the DC/DC voltage
and current values and the FPGA temperature is tested in a stand-alone setup and on an ATCA carrier board to verify the IPMC communication. The final step is to mount all the MicroPOD\texttrademark~ modules and the optical pigtails on the board. An optical loopback is plugged into the front panel to check all the optical links at \SI{11.2}{Gbps} up to $10^{13}$ bits with PRBS31 for approximately 15 minutes. At this stage all the components and links of the board are fully validated. 
\subsection{Intelligent platform management controller}
\label{ssec:ipmc}
\begin{figure}[htb]
\centering
\includegraphics[width=\textwidth]{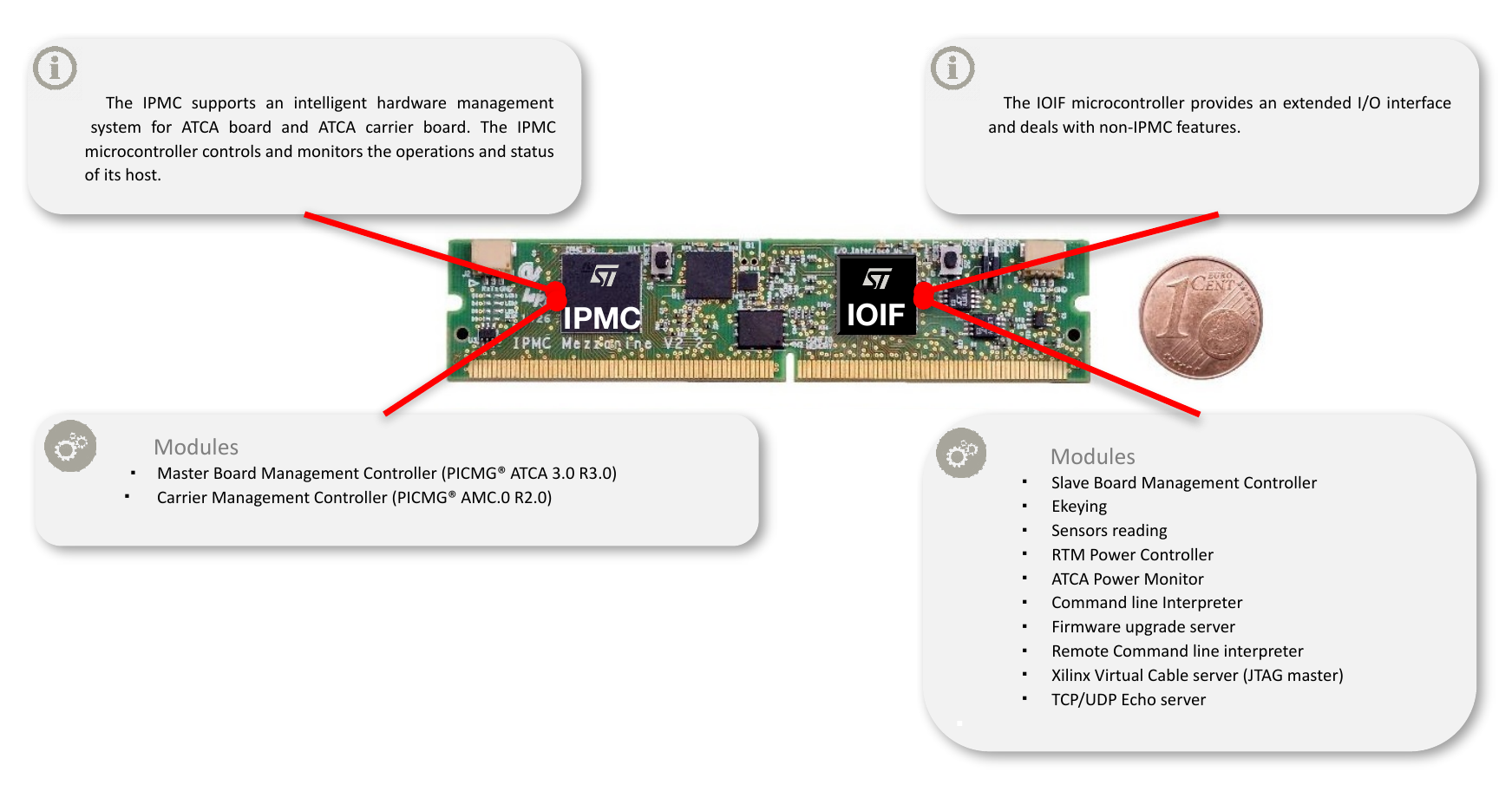}
\caption{A photograph of the IPMC board, with diagrams detailing the functions of each microcontroller.}
\label{fig:ipmc}
\end{figure}

The IPMC supports an intelligent hardware management system for ATCA boards and ATCA carrier boards
(see Figure~\ref{fig:ipmc}) which provides the ability to manage the power, cooling, and interconnect
needs of intelligent devices: monitoring of events created by electronic on-board devices, logging of
such events to a central repository and management of the mezzanine modules according to user’s
implementation, as well as the communication with the Shelf Manager.
Figure~\ref{fig:ipmc} presents a photograph of the IPMC in its mini Dual In-line Memory Module (mini-DIMM) form factor
accompanied by the functions of the main elements of the board. In addition the IPMC mezzanine, via
local IPMB, communicates with the MMC installed on the LATOME AMC.
The dimensions of the IPMC mezzanine are $82\times17.9$~mm\textsuperscript{2}.
The board is equipped with two microcontrollers called IPMC and IOIF (for Input/Output interface) 
that can be seen on Figure~\ref{fig:ipmc}.
The IPMC controller controls and monitor the operations and the status
of the ATCA carrier board, while the IOIF controller provides an extended I/O interface and deals
with the non-IPMC features. The main ATCA functions are fulfilled via IPMC, the master board management
controller: it controls the AMC and RTM power and monitors the ATCA power.
The IOIF, the slave board management controller, ensures dialog
with the sensors located on the carrier. The microcontrollers are loaded with the IPMC software,
based on Intelligent platform management Controller softwARE (ICARE) framework~\cite{icare}
which is fully compliant with the ATCA specifications~\footnote{IPMI v1.5
(document revision 1.1) and some relevant subset of IPMI v2.0 (document revision 1.0); PICMG 3.0 R3.0
(AdvancedTCA\texttrademark~ base specification) and AMC.0 R2.0 (AdvancedMC\texttrademark~ base specification).}.
ICARE is an open-source software which manages ATCA requirements, provides features to build user code
and tools to generate some Field Replaceable Units (FRU) and Sensor Data Records (SDR).
ICARE provides command line and remote command lines interpretation, acts as the firmware upgrade server and
Transmission Control Protocol (TCP) and User Datagram Protocol (UDP) echo server. In addition it provides a \xilinx Virtual Cable server.

\subsection{Back-End firmware}
\label{ssec:firmware}
\subsubsection{LAr carrier firmware}
\label{sssec:larc-firmware}
The primary tasks of the LArC firmware are to provide TTC clock 
and TTC event information to the LATOMEs and to transmit data from the LATOMEs to downstream 
consumers such as FELIX or LAr Monitoring PCs. The FPGA firmware is developed 
using the \xilinx Vivado IDE including SDK for MicroBlaze applications.

\begin{figure}[htb]
  \centering
  \includegraphics[width=\textwidth]{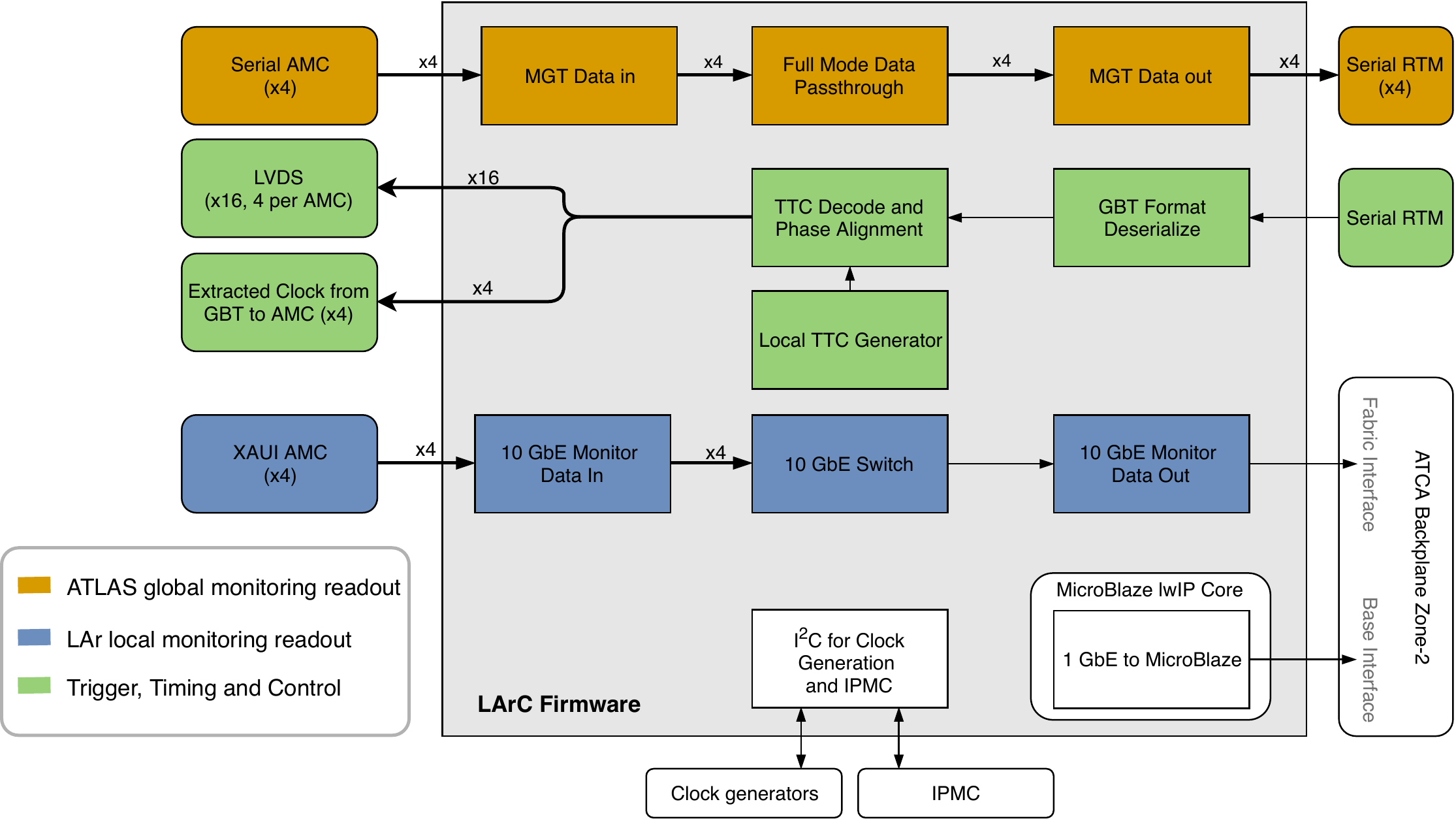}
  \caption{Block diagram of the LArC firmware.\label{t-larc-fw-block}}
\end{figure}

A simplified block diagram of the firmware is shown in Figure~\ref{t-larc-fw-block}.
The firmware framework uses a 
MicroBlaze processor as an embedded system and lightweight IP (lwIP) for network connectivity.
The system clock is a crystal-generated \SI{125}{MHz}.

The main functionalities of the LArC firmware are described in the following paragraphs.

\paragraph{GbE Communication}

GbE is connected to the FPGA via a Broadcom GbE switch on the LArC. This switch is also 
connected to the Zone 2 Base Interface of the ATCA shelf. Ethernet connectivity is implemented 
using the lwIP software running on the MicroBlaze processor.  
GbE data are received and transmitted as Ethernet packets. The protocol used is either raw UDP 
packets or IPbus~\cite{IPBusProtocolSpecification} packets embedded in the UDP packets. The latter is used in normal operation while
the former is useful for sanity tests. The two protocols are assigned different
ports in the LArC software. The handling of these protocols is done through C code 
running on the MicroBlaze processor.

\paragraph{I$^2$C Communication}

There are two I$^2$C buses on the LArC. One is mastered by the IPMC 
and includes a sensor bus connection to the FPGA in the chain.  
The other connects the FPGA with the clock generators and is used to 
configure them. I$^2$C communication with these clock chips is done through C code running 
on the MicroBlaze.

\paragraph{LHC Clock Extraction and TTC LVDS Data}

One of the MGT links to the LArC carries data in GBT~\cite{GBT1,GBT2} format 
at \SI{4.8}{Gbps}.  
The LArC firmware decodes the GBT format data and recovers the \SI{40}{MHz} LHC clock.  
The transceiver IP is one modified from an example provided by the CERN GBT group. The TTC information carried in 
the GBT format data consists of four bits:  L1A, BCR, ECR and Trigger Type data. This TTC information is subsequently 
transmitted to each of the four LATOMEs on four LVDS lines each at \SI{160}{MHz}. Other separate LVDS lines carry a
\SI{160}{MHz} TTC clock signal to each LATOME.

\paragraph{TTC Generator}

The four bits of TTC data can also be generated locally rather than via the GBT link. A multiplexer selects whether
data from the GBT transmitter or TTC generator are used.  The \SI{40}{MHz} clock used to generate the 
\SI{160}{MHz} TTC clock can either be the GBT recovered clock as described above, or a local \SI{40}{MHz} clock.
In either case, a jitter cleaner provides the \SI{160}{MHz} TTC clock and \SI{120}{MHz} MGT clocks.  

\paragraph{\SI{10}{GbE} Switch}

Monitor stream data are generated by each LATOME destined for monitoring computers installed in ATLAS.  
These data are sent as the payload of \SI{10}{GbE} UDP packets from each of the four LATOME's over four 
sets of XAUI lanes to the LArC.  
The LArC reassembles the data from each set of 
XAUI lanes back into a \SI{10}{GbE} packet. Each packet is then received by a dedicated \SI{10}{GbE} MAC IP.  
The packet header is examined by the MAC.  
If the packet is malformed, it will be dropped. If not, the packet enters a \xilinx multiplexer IP.  
This multiplexer receives packets from the 
four \SI{10}{GbE} MACs using a round-robin algorithm and sends them to a bidirectional switch engine connected to
a \xilinx demultiplexer IP. This demultiplexer sends packets to \SI{10}{GbE} output MACs.  
MAC and XAUI IP cores are used to send the switch 
output to the Zone 2 Fabric Interface.
The transmitted data can be standard or jumbo UDP packets.  
The current maximum size of the packet is 16 384 bytes, which matches the capabilities of the network switch 
residing in the ATCA shelf.

\paragraph{MGT Data Transmission}

Four sets of MGT serial lines between the RTM and LArC FPGA are used to carry LATOME data to be recorded by 
ATLAS via the FELIX system.  These data are sent on receipt of an L1A. The bandwidth is \SI{9.6}{Gbps} in
Full Mode~\cite{felix} data format.  It is essentially 8b/10b encoded but with additional control specifications for 
start and end of the packet and 
BUSY. TDAQ stream data are transmitted from each LATOME to the LArC. The data are received and deserialized 
on one transceiver
pair and subsequently serialized and transmitted on a second transceiver pair. The data are transmitted to optical 
modules on the RTM and then to FELIX.  
Plain \xilinx GTH transceiver IP is used for transceiver instantiation in the firmware.  

\paragraph{Resource Usage}

The resource usage with a current LArC project is shown in Table~\ref{t-larcfw-resource}
and comfortably fits in the device.

\begin{table}[htb]
 \caption{Resource utilization in the LArC firmware project.
 Each memory block corresponds to \SI{36}{Kbits} of data.
 \label{t-larcfw-resource}}
 \begin{center}  
  \begin{tabular}{lcc@{}}
  \toprule
     \textbf{Resource} & \textbf{Used blocks} & \textbf{Utilization [\%]} \\
     \midrule
     Logic                & 138343 & 11.4 \\
     Memory               &    452 & 38.3 \\
     DSP                  &      5 & \hphantom{0}0.2 \\
     Pins                 &    128 & 21.3 \\
     Gigabit Transceivers &     34 & 42.5 \\
     Global Clock Buffers &     29 & 90.6 \\
     Clock Managers       &      6 & 15.0\\
   \bottomrule
  \end{tabular}
 \end{center}
\end{table}

\paragraph{Validation and verification}

A limited number of logic modules were simulated via
Questa\textsuperscript{\textregistered} from Mentor Graphics\textsuperscript{\textregistered} and
Xcelium\textsuperscript{\textregistered} from Cadence\textsuperscript{\textregistered}.
A dedicated validation procedure was used to generate and check the expected data.
In general, Vivado projects for individual modules
were developed and tested on the LArC hardware using the \xilinx Integrated Logic Analyzer (ILA).  The 
individual modules were subsequently integrated into a complete project.  The complete project was tested
using a LArC, a number of LATOME modules, and external connections to FELIX and monitoring PCs.  In this
case the integrity of the project was validated by testing carried out at CERN and other laboratories.  
The tests involved stimulation by known inputs and comparison between actual and expected outputs. 
 
\subsubsection{LAr trigger processing mezzanine firmware}
\label{sssec:latome-firmware}
The LATOME firmware performs four main functions:
\begin{itemize}
    \item Handling of up to 48 high speed input links from one or more LTDBs, corresponding to up to 320 Super Cells
    \item Application of a digital filtering algorithm to reconstruct the Super Cell transverse energy every \SI{25}{ns} and to identify the bunch crossing when the deposited energy was initiated (BCID determination)
    \item Outputs results every \SI{25}{ns} to the Level-1 calorimeter trigger system (FEX)
    \item Processes and buffers data to be delivered to the TDAQ readout chain and to the local monitoring, upon request (L1A or other)
\end{itemize}
The FPGA processing of the \SI{40}{MHz} path from the LTDBs and directed to the FEX system has to be performed 
with a fixed latency, in order to correctly work within the ATLAS trigger readout system.
This processing is divided in several steps.
First, the incoming LTDB data are deserialized and descrambled.
The 12-bit ADC data at \SI{40}{MHz} for each channel are aligned accordingly in preparation for the application of the filtering algorithm.
Next, to ease the computations made by the FEX system later on, data words are organized following the detector geometry in a configurable way. The Super Cell transverse energies are then calculated with filtering algorithms.
Finally, the transverse energies are encoded without losing the benefit of the LAr fine energy granularity for cluster reconstruction in the FEX system. 
An important point to be considered is the case of saturated inputs which have different shapes than nominal pulses and for which the BCID determination may fail.
The correct bunch crossing has to be identified with a dedicated filter on the signal in this
case.
The implementation of these steps in the LATOME firmware design is 
detailed in the following paragraphs.

\begin{figure}[htb]
\begin{center}
\includegraphics[width=0.9\textwidth]{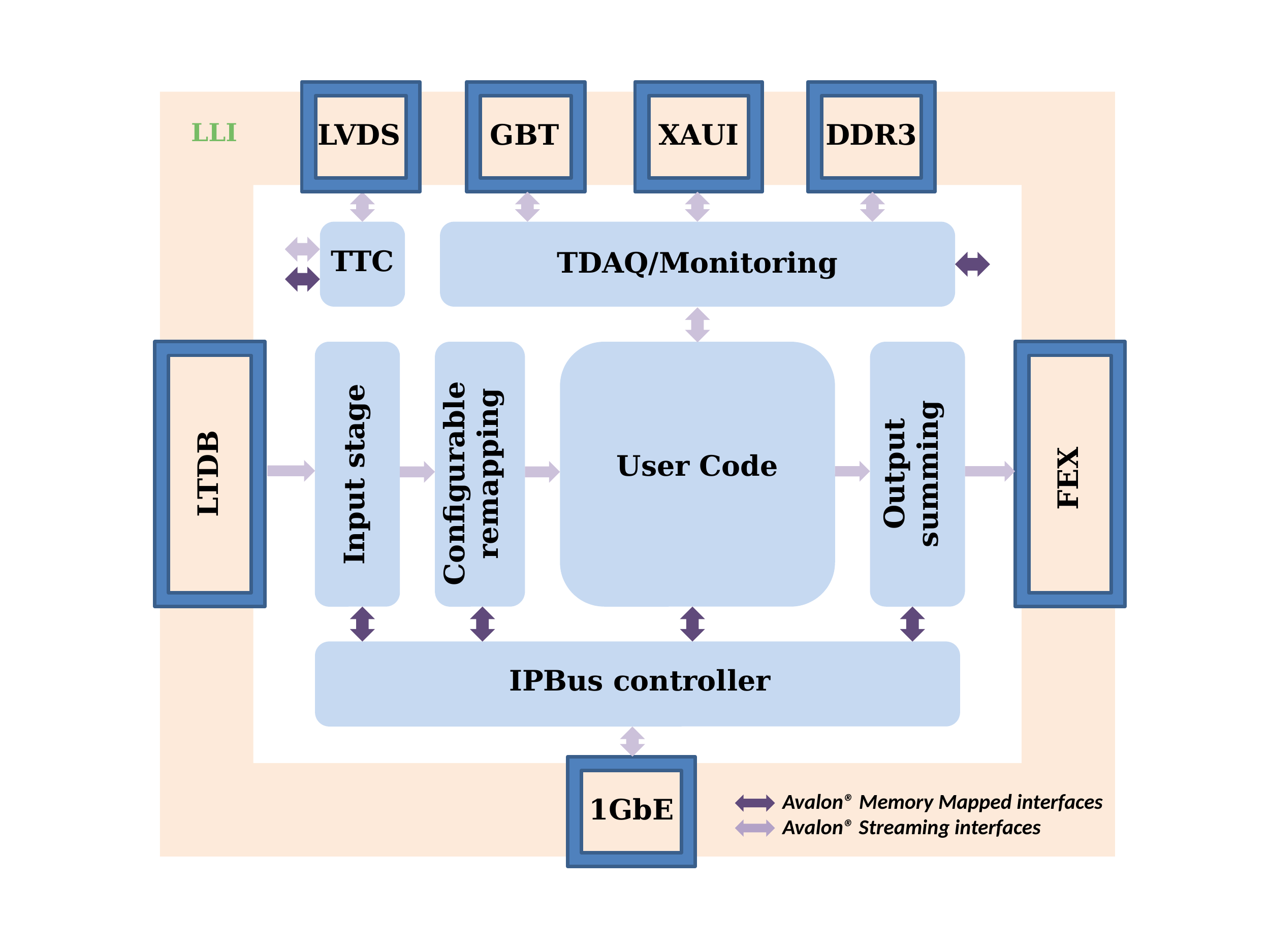}
\caption{LATOME firmware block diagram. The pale brown frame corresponds to the interface with the
hardware while the blue boxes correspond to the higher level functional blocks.
 \label{fig:latome:fw:diagram}}
\end{center}
\end{figure}

Figure~\ref{fig:latome:fw:diagram} presents a block diagram of the LATOME firmware.
The firmware is built around the Low-Level Interface (LLI) which controls the hardware components of the LATOME.
The firmware consists of a main data path (from LTDB to FEX) which has been organized in four blocks, input stage (IS), configurable remapping (CR), user code (UC) and output summing (OS).
Three other functions are included in the firmware: (1) TDAQ/monitoring (MON) which organizes the transfer of data to TDAQ and to the local monitoring processes,
(2) the IPbus~\cite{IPBusProtocolSpecification} controller (IPCTRL) which is the interface for the LATOME slow-control system and (3) the TTC which decodes and provides signals from the TTC system.

\begin{figure}[htb]
\begin{center}
\includegraphics[width=0.9\textwidth]{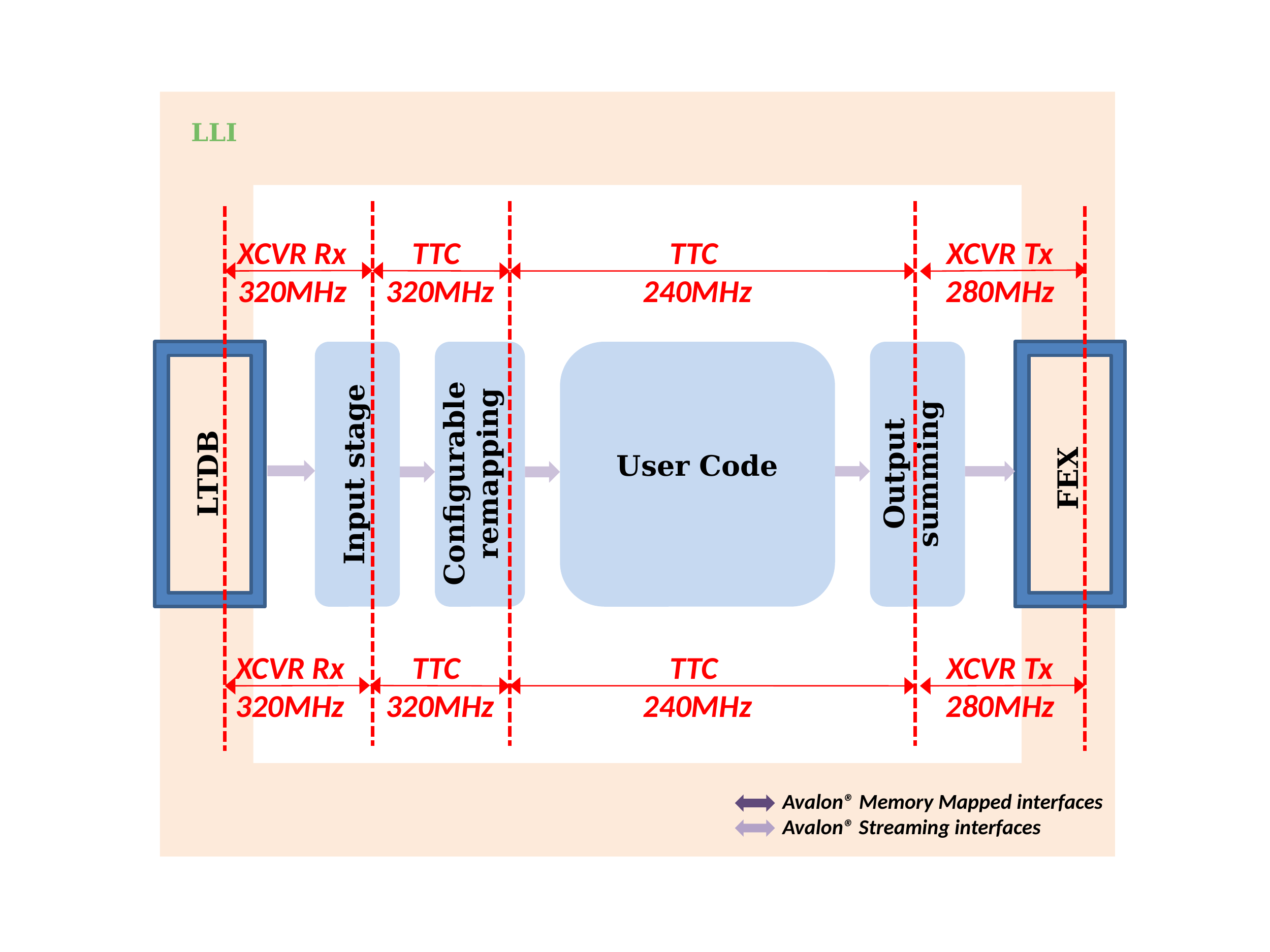}
\caption{LATOME firmware clock domains block diagram. The reference clocks are taken either from the
transceivers (XCVR) connected to the LTDBs or the FEX system, or from the LHC reference clock provided 
by the TTC block.  
\label{fig:latome:fw:clock:diagram}}
\end{center}
\end{figure}

Figure~\ref{fig:latome:fw:clock:diagram} presents a block diagram of the various clocks organisation inside the LATOME firmware.
Three major clock domains are defined. A \SI{320}{MHz} clock is used from the input LTDB to the configurable remapping.
The configurable remapping transfers data to the user code, which receives, computes and transmits, at the frequency of \SI{240}{MHz}.
The clock frequency is reduced to ease the routing and timing of the signals.
This increases the size of the logic needed since it needs to go from 48 input streams to 62 streams at the user code level.
The output summing transfers FEX data at \SI{280}{MHz}.
This frequency is required to handle the output data protocol at \SI{11.2}{Gbps}, comprising 7 LHC cycles per data frame (one LHC cycle corresponds to \SI{40}{MHz}).

The latencies of each processing step have been estimated by simulations and then confirmed by measurements in the system as-built. 
The measured latencies are listed in Table~\ref{tb:latency} in Section~\ref{ssec:system-latency}.  
From the reception of the data to their transmission to the FEX system, the LATOME firmware latency is approximately 16.6 LHC clock cycles.
In the following, the different firmware blocks are briefly detailed.

\paragraph{Low-Level Interface}

The LLI implements all the FPGA device specific IPs configured in the Quartus\textsuperscript{\textregistered} tool from \intel.
The aim of the LLI is to have device agnostic interfaces between the inputs/outputs of the FPGA and the internal blocks.
This is most commonly known as board support package (BSP).
All the interfaces between the LLI and the FPGA core layers are designed with standard Avalon Streaming and Avalon Memory Mapped interfaces \cite{AvalonInterfaceSpecification}.
This allows for well-defined and documented standard interfaces between each one
of the blocks.

The LLI implements the following IP blocks:
\begin{itemize}
    \item Clock PLLs and synchronous resets for all various clock domains 
    \item Flash interface for remote programming 
    \item Gigabit Ethernet for slow control 
    \item 4 LVDS lines at \SI{160}{\MHz} for TTC data input 
    \item 48 \SI{5.12}{Gbps} transceivers for LTDB inputs 
    \item 48 \SI{11.2}{Gbps} transceivers for FEX outputs 
    \item XAUI \SI{10}{Gbps} ethernet (\SI{10}{GbE}) for local monitoring output 
    \item DDR3 interface for local memory
\end{itemize}

\paragraph{Input Stage}

The LTDB data are sent to the LATOME on optical fibers that carry the data stream at \SI{5.12}{Gbps} using the LOCic protocol \cite{LOCICSpecification}.
The role of the input stage is to decode this protocol and align all input streams to the same BCID. It is split into three parts:
\begin{itemize}
\item Reception and processing of the LTDB data streams where decoding of LOCic protocol is handled 
\item Fiber to fiber alignment where all input streams are aligned to the same BCID 
\item Test pattern generator which can be used internally to generate test data
\end{itemize}

\paragraph{Configurable Remapping}

The configurable remapping block is used to reorder data coming from the input stage following the detector geometry for each bunch crossing.
This block groups data from Super Cells belonging to the same Trigger Tower and then sends them to one of the User Code block instances.
This prepares data for some part of the FEX system where the unit is a Trigger Tower.
As the data arrangement and the geometry are not uniform across the detector and are therefore different for each LATOME, the block is configured specifically at startup.

\paragraph{User Code}

\begin{figure}[htb]
  \centering
  \includegraphics[width=.9\textwidth]{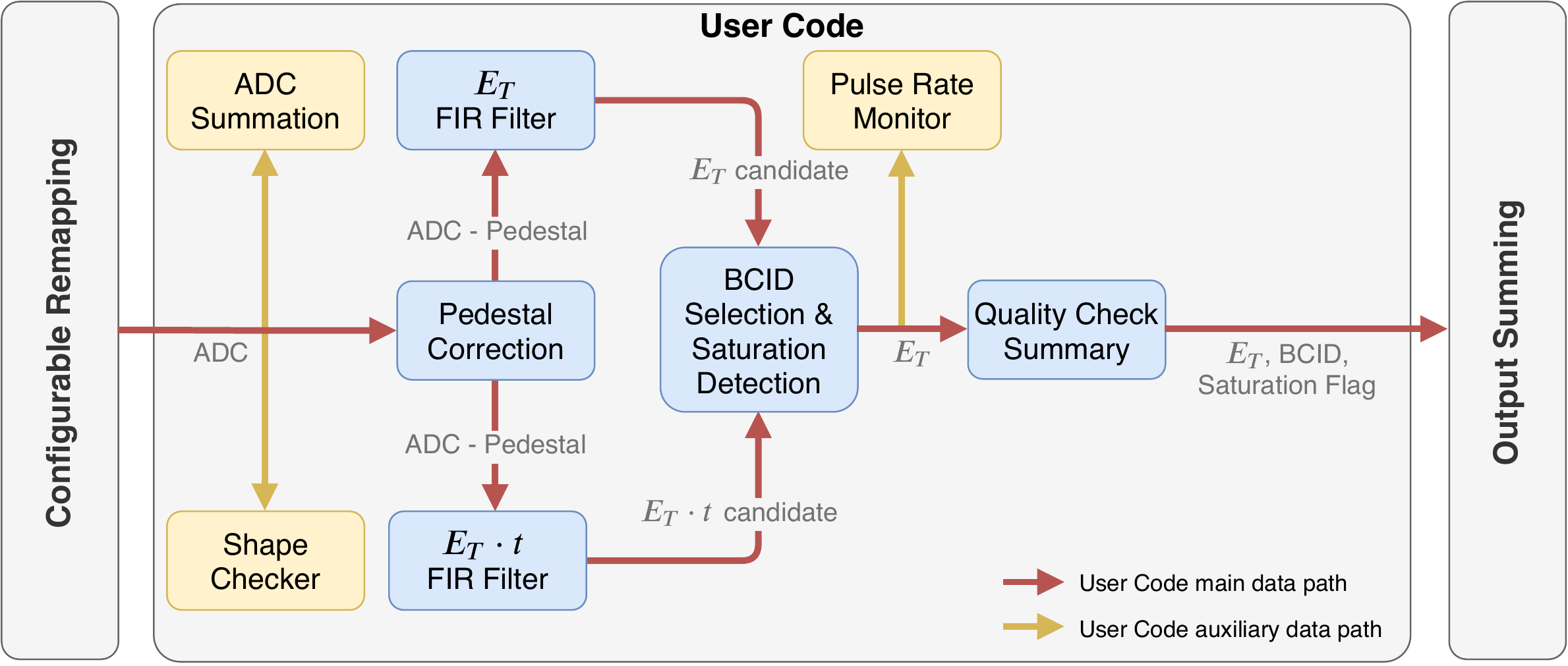}
  \caption{Block diagram of the User Code module in the LATOME firmware. The red arrows represent the data flow
  corresponding to the digital trigger readout, while the yellow arrows represent auxiliary paths used for basic
  tests (more details in the text).}
  \label{fig:UserCodeBlockDiagram}
\end{figure}

The User Code block receives the Super Cell data from the Configurable Remapping block and outputs synchronously the 
reconstructed transverse energy and data quality bits to the Output Summing block, at the corresponding bunch crossing with 
a fixed latency.
The different operations implemented in the User Code module are illustrated in Figure~\ref{fig:UserCodeBlockDiagram}.

First, a BCID-dependent pedestal correction is applied to the Super Cell ADC value, in order to take into account the
imperfect pile-up effect cancellation with the bipolar pulse shape from events occurring at the beginning
of an LHC bunch train: 
With long trains and for distances from the beginning of the trains
smaller than the typical drift time in LAr gaps (\SI{450}{\ns} in the EM
barrel calorimeter, less in the FCAL) an average positive energy shift from
pile-up is expected, as the contributions from out-of-time pile-up does not
cancel the in-time pile-up contribution. After 20 BCIDs (\SI{500}{\ns}) the
average shift is alleviated thanks to the bipolar shaping applied in the
readout, since the total contribution from out-of-time pile-up compensates
here for the in-time pile-up contribution.
This BCID-dependent pedestal correction depends on the LHC bunch scheme, the 
instantaneous luminosity of each bunch and the pulse shape of each Super Cell.
Consequently, the correction is computed directly on the LATOME FPGA for every BC 
position and for every Super Cell and updated approximately every 10 seconds with 
a round-robin scheme. 
It is calculated from Super Cell ADC counts averaged over 1024 LHC cycles,
processing only 156 BC positions at a time, with a $1/8$ ADC count precision and a maximal absolute value of $32$ ADC counts. 
Two Finite Impulse Response (FIR) filters, based on an optimal filtering algorithm~\cite{OFC},
then compute the \ET and the product of \ET and time ($\ET\cdot t$) 
from a fixed number of ADC input values (four in the actual implementation). 
The coefficients used in the optimal filtering algorithm include a geometrical factor
to compute directly the \ET from the ADC values rather than the Super Cell energy.
From the \ET and $\ET\cdot t$ candidate values computed every bunch crossing a BCID selection algorithm determines the correct \ET
and BCID of the Super Cell signal. The \ET and BCID are considered correct
if at a given BCID $b$ the following condition is fulfilled:
\begin{equation}
  \begin{cases}
    \phantom{-}8 \ET(b) < \ET\cdot t\ (b) < -8 \ET(b),& \text{if~}\  -1 <\ET(b) \leq \SI{0}{\GeV}\\
    -8 \ET(b) < \ET\cdot t\ (b) < \phantom{-}8 \ET(b),& \text{if~}\  0 <\ET(b) \leq \SI{10}{\GeV}\\ 
    -8 \ET(b) < \ET\cdot t\ (b) < 16 \ET(b),& \text{if~}\ \ET(b) > \SI{10}{\GeV}. 
  \end{cases}
  \label{eq:bcidSel}
\end{equation}
The threshold values utilize powers of two to ease the FPGA computation. Because of resolution effects,
some thresholds are negative ($-8\ET$). For \ET values above \SI{10}{\GeV} the selection window is
asymmetric, with a looser boundary towards high $\ET\cdot t$ values, to avoid wrong BCID 
assignments to delayed signals coming from hypothetical long-lived particles. 

\begin{figure}[tb]
  \centering
  \includegraphics[width=.7\textwidth]{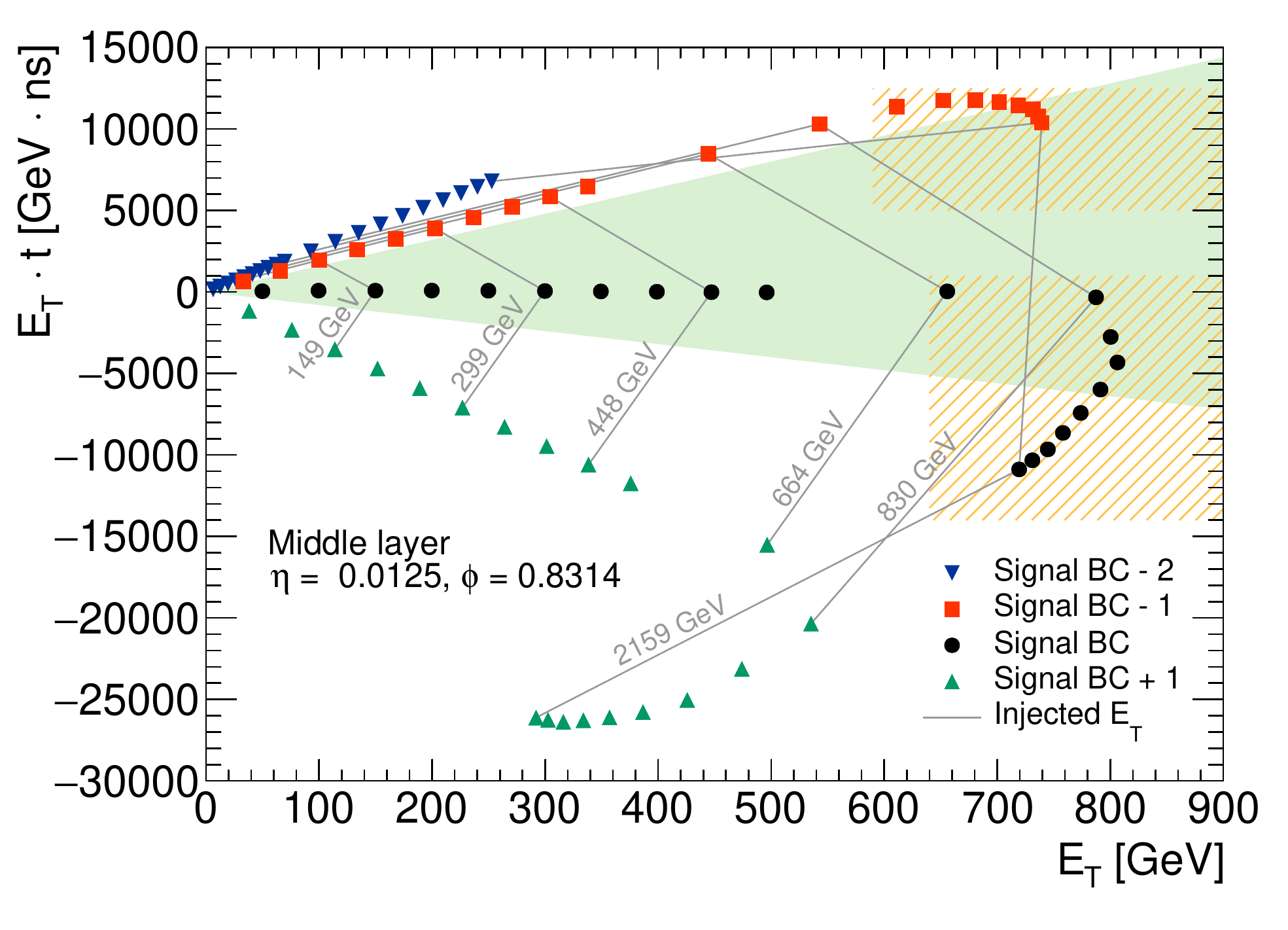}
  \caption{$\ET\cdot t$ as a function of \ET measured in calibration runs with injected signals of known \ET
  and bunch crossing. Values are either computed with the four ADC inputs leading to a correct BCID assignment (Signal BC) 
  or with four ADC inputs in neighboring time windows leading to incorrect BCID assignments of $+1$, $-1$ or $-2$ BC (Signal BC $\pm1$ 
  or $-2$). For some injected \ET values, grey contours link the (\ET, $\ET\cdot t$) points measured from the same injected 
  signal but with ADC inputs from different time windows.
  The green shaded area corresponds to the BCID selection criterion in the case of non-saturated 
  inputs. The orange hatched areas represent the saturation detection conditions.
  The data correspond to the middle layer Super Cells in the $\eta=0.0125$ and $\phi=0.8314$ calorimeter region, where saturation
  occurs approximately above \SI{750}{\gev}.
  \label{fig:ettau_vs_et}}
\end{figure}

In the case of saturated inputs, the proper bunch crossing is identified with a dedicated selection criterion
based on the expected shape deformation and a flag is raised on the passed data. 
Figure~\ref{fig:ettau_vs_et} shows the $\ET\cdot t$ as a function of \ET measured in calibration runs with injected signals of known \ET and bunch crossing (BC),
 for some middle layer Super Cells in the EMB calorimeter.
Values are either computed with the four ADC inputs leading to a correct BCID assignment (circles) or with four ADC inputs in 
neighboring time windows leading to incorrect BCID assignments of $+1$, $-1$ or $-2$ BC (triangles and squares).  
The BCID selection criterion of Equation~(\ref{eq:bcidSel}) is represented in the figure by the triangular
green shaded area.
The data correspond to special calibration runs with injected \ET values spanning the linear and saturated regimes.
The saturation effect can be observed with the $\ET\cdot t$ versus \ET dependence becoming non-linear at high \ET values.
In this case, $\ET\cdot t$ is biased towards lower values because the saturated pulses have a shorter rise time.

The saturation flag is raised in case the \ET and $\ET\cdot t$ candidate values
for the two consecutive BCIDs $b-1$ and $b$ follow simultaneously the conditions
\begin{equation}
  \begin{cases}
  \phantom{0} \ET(b-1) > \ET^{\text{min previous}}& \text{and~}\ \xi^{\text{min previous}} < \ET\cdot t\ (b-1) < \xi^{\text{max previous}}\\
  \phantom{0} \ET(b) > \ET^{\text{min}}& \text{and~}\ \xi^{\text{min}} < \ET\cdot t\ (b) < \xi^{\text{max}},
  \end{cases}
\end{equation}
where $\ET^{\text{min previous}}$, $\ET^{\text{min}}$ and $\xi^{\text{min previous}}$, $\xi^{\text{min}}$ are configurable thresholds on \ET and $\ET\cdot t$ values for
BCID $b-1$ and $b$, respectively. These detection criteria correspond to the orange hatched areas
in Figure~\ref{fig:ettau_vs_et}. The condition on the previous BCID $b-1$ could in principle be
sufficient to detect the saturation, but enforcing the simultaneous verification on BCID $b$ 
adds more robustness without degrading the computation time.

Once the BCID selection and saturation detection is completed, the corresponding 
quality bits are stored into the input of the Output Summing firmware block together with the 
Super Cell's \ET and BCID.
If the BCID selection is fulfilled neither for the regular nor the saturated inputs processing,
meaning that no physics signal is present, the \ET result is set to zero.
While the ADC values are encoded with 12 bits at the input of the User Code, the calculated \ET at the output is encoded with 18 bits 
with a precision of \SI{12.5}{\mev} per least significant bit.
This is needed to reach the numerical accuracy required by the FEX system
(see the "Output Summing" paragraph).

Three auxiliary functions in the User Code module were developed to ease testing and commissioning of the firmware. The ADC
Summation adds the Super Cell ADC values over a period of time corresponding to one LHC orbit (3564 BCs). This helps to quickly
check the pedestal values of every channel. 
The Shape Checker detects if a pulse is observed in a given Super Cell from the ADC values of six consecutive samples.
The Pulse Rate Monitor provides a similar functionality, but based on the \ET value computed by the FIR Filter. It measures the number of detected pulses above a configurable \ET threshold. 
These two submodules
are useful to check the hardware connectivity, the channel mapping in the firmware and the time alignment of the measured signals.
All of these auxiliary functions are not part of the main data flow and can only be read out via the slow-control interface based 
on IPbus.

\paragraph{Output Summing}

The main task of the Output Summing block is to group the data received from the User Code in order to compute the sums over specific $\eta - \phi$ areas and to send the data to the FEX output fibers.
The characteristics of these sums are compiled in Table~\ref{tab:osum_sumcar}.
The eFEX system aims for identifying isolated energy deposits indicative of electrons, photons
and tau leptons. To achieve an optimal hadronic background rejection with a topological 
discrimination, the full Super Cell granularity is required: 
no sum is performed and the \ET of each Super Cell is sent directly. 
The jFEX system targets jet, large-area tau leptons, missing \ET and total-\ET triggers, while
the gFEX system is designed for large-area jets, missing \ET and sum \ET online detection.
Consequently, a coarser granularity can be used compared to the eFEX system:
for the jFEX (gFEX), Super Cells are summed to
provide the \ET of the Trigger (gFEX) Towers, covering $0.1\times0.1$ ($0.2\times0.2$) $\eta - \phi$ areas in the central part of the detector. In the
forward region a coarser granularity is used. 
In case of the gFEX, a position code is added to the output data to provide some information about the location of the most energetic deposit within the tower.
Flags are attached to the output data for the three subsystems to specify if a Super Cell
passes the saturation condition or if there was a problem in receiving or calculating the Super Cell's 
\ET.

\begin{table}[htb]
  \caption{Characteristics of the sums made in the Output Summing block for the electromagnetic (eFEX), jet (jFEX) and global (gFEX) feature extractors. The quoted granularity is provided for the central part of the detector; a coarser granularity is used
  in the forward region.\label{tab:osum_sumcar}}
  \begin{center}
    \begin{tabular}{@{}lccc@{}}
      \toprule
      \textbf{\ET sum characteristics}         & \textbf{eFEX}        & \textbf{jFEX}       & \textbf{gFEX} \\
      \midrule
      $\Delta\eta\times\Delta\phi$ granularity & $0.025\times0.1$ or $0.1\times0.1$ & $0.1\times0.1$ & $0.2\times0.2$\\
      Longitudinal granularity                 & All layers separated & \multicolumn{2}{c}{All layers summed together}\\
      Total number of sums                     & 34048    & 5760 & 1984 \\
      \bottomrule
    \end{tabular}
  \end{center}
\end{table}

The accuracy (number of bits) for each FEX subsystem is adjusted on multi-linear precision plateaus,
as described in Table~\ref{tab:osum_precision}, and data are encapsulated with headers and trailers
to be sent to the LLI and further to the FEX system.
Another task of the block is to duplicate some FEX outputs several times, according to the number of receivers of the data for specific $\eta - \phi$ areas.
Because of these duplications, the data flow rate at the output of one LATOME directed to the FEX system (about \SI{350}{Gbps}) 
is larger than the data flow rate at the input of one LATOME coming from the LTDBs (about \SI{200}{Gbps}). Taking into account
all LATOMEs, the total data rate sent to the FEX system is slightly above \SI{40}{Tbps}. 

\begin{table}[htb]
  \caption{Energy precision of the sums made in the Output Summing block depending on the energy range considered
    for the electromagnetic, jet and global feature extractors.\label{tab:osum_precision}}
  \begin{center}
    \begin{tabular}{@{}lccc@{}}
      \toprule
\textbf{Energy precision [\mev]} & \multicolumn{3}{c}{\textbf{Energy range [\mev]}} \\
                 & \textbf{eFEX}      & \textbf{jFEX}      & \textbf{gFEX} \\
      \midrule
      25         & $[-750,1600[$      & $[-3150,6400[$     & $[-12800,12800[$ \\
      50         & $[1600,6400[$      & $[6400,25600[$     & $[-50000,-12800[\cup[12800,51200[$\\
      100        & $[6400,25600[$     & $[25600,102400[$   & $[51200,200000[$\\
      200        & $[25600,102400[$   &  $[102400,409600[$ & --\\
      400        & $[102400,200000[$  &  $[409600,800000[$ &  --\\
      25600      &   --               &  --                & $[200000,1019200[$\\
      102400     & $[200000,1019200[$ &  --                &  -- \\
      \midrule
      Encoding   & 10 bits            & 12 bits            & 12 bits\\
      \bottomrule
    \end{tabular}
  \end{center}
\end{table}

\paragraph{TDAQ Readout and Monitoring}

The TDAQ/Monitoring block takes care of sending data to two external paths upon L1A.
This is used for monitoring, debugging and online or offline analysis.
ADC data and transverse energy data from each Super Cell are buffered until the arrival of a L1A signal, and then written to the output links via the transceivers in the LArC as discussed in Sections~\ref{ssec:larc} and~\ref{sssec:larc-firmware}.

\paragraph{Slow Control}

The slow control interface connects the firmware blocks to the outside world to allow the user to load or change all the configuration parameters of the system.
It is also used to monitor its status.
The transport mechanism for the slow control interface is the IPbus protocol~\cite{IPBusProtocolSpecification} over Gigabit Ethernet (\SI{1}{GbE}).

\paragraph{TTC}

The TTC block receives the TTC signals along the LVDS lines, from the LArC board.
It takes care of decoding the relevant TTC information to be provided to the other firmware modules.
The TTC block communicates with the input stage and TDAQ/Monitoring blocks.

\paragraph{Verification and validation}

In order to test the firmware, a Python/C model of the firmware has been developed. Several types of data have been defined and are fed to this model which generates expected data for all stages in the firmware.
These datasets are used in multiple simulation testbenches to test all blocks one by one, bit
by bit.
The same datasets are used to test the functionality on the real hardware using a secondary LATOME that acts as a data generator and also as a data checker.

Simulations and tests on the target are integrated in the GitLab Continuous Integration flow (CI) to be able to detect and track regressions.
   
\section{Integration tests}
\label{sec:integration}
\subsection{Front-End boards integration tests}
\label{ssec:fe-integration}
\begin{figure}[!h]
	\centering
	\includegraphics[width=1\textwidth]{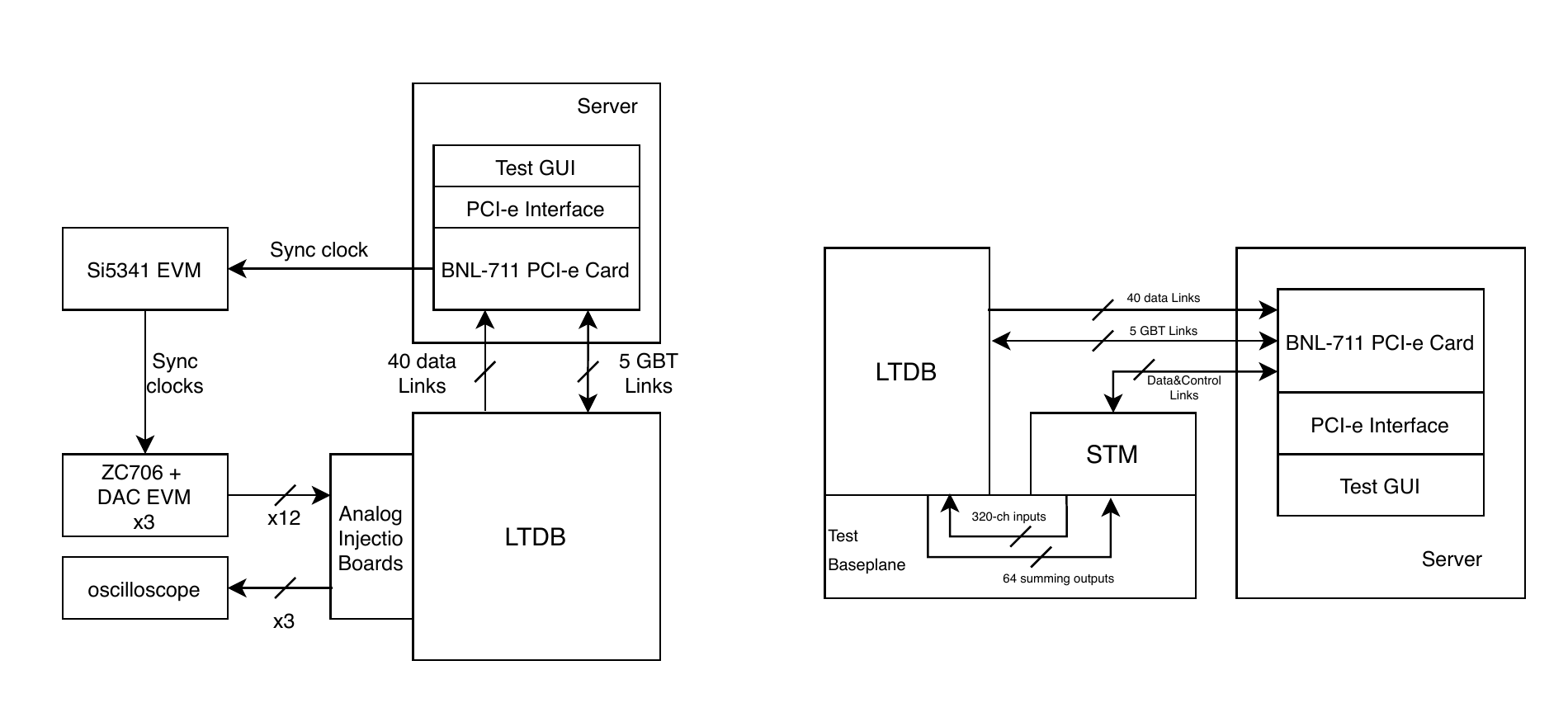}
        \caption{Block diagrams of the LTDB test stands. The left one
	is the test stand with analog injection boards fed by
	commercial evaluation boards as signal generator. The outputs
	of the summing signals are collected and measured on the
	analog injection boards. The right block diagram illustrates
	the integrated test stand with STM and test baseplane, where
	the signals generated on the STM are sent to the LTDB through
	the test baseplane. Outputs of the summing signals are
	collected and measured on the STM as well. Both test stands
	use a FELIX prototype PCI-e card to serve as Back-End
	electronics for receiving data from the LTDB, timing, control
	and monitoring.}
        \label{fig:ltdb_testsetup}
\end{figure}

The LTDB production test stand evolved from initially using analog signal injection
boards based on the commercial FPGA and digital-to-analog converter
evaluation boards \xilinx ZC706 and TI DAC3484EVM, to their eventual
full integration in the Saclay Test Module (STM),
as shown in Figure~\ref{fig:ltdb_testsetup}~\cite{ltdbtns}. All test
results shown below are obtained with the evaluation-boards-based test
stand, where the LAr Super Cell signals are produced
on a signal generator board with 320 DAC channels, and then injected to the
LTDB. The 64 summing signals from the LTDB are sent back to the injection
board for measurement. A FELIX prototype (a PCI-e card)
is used to simulate the Back-End electronics to communicate with the
LTDB through five control links and up to 40 data links. The control links
send the control and timing information to, and collect
monitoring information from the LTDB. The data links receive the ADC
data from the LTDB.

\begin{figure}[!h]
	\centering
        \subfloat{
	\includegraphics[width=0.62\textwidth]{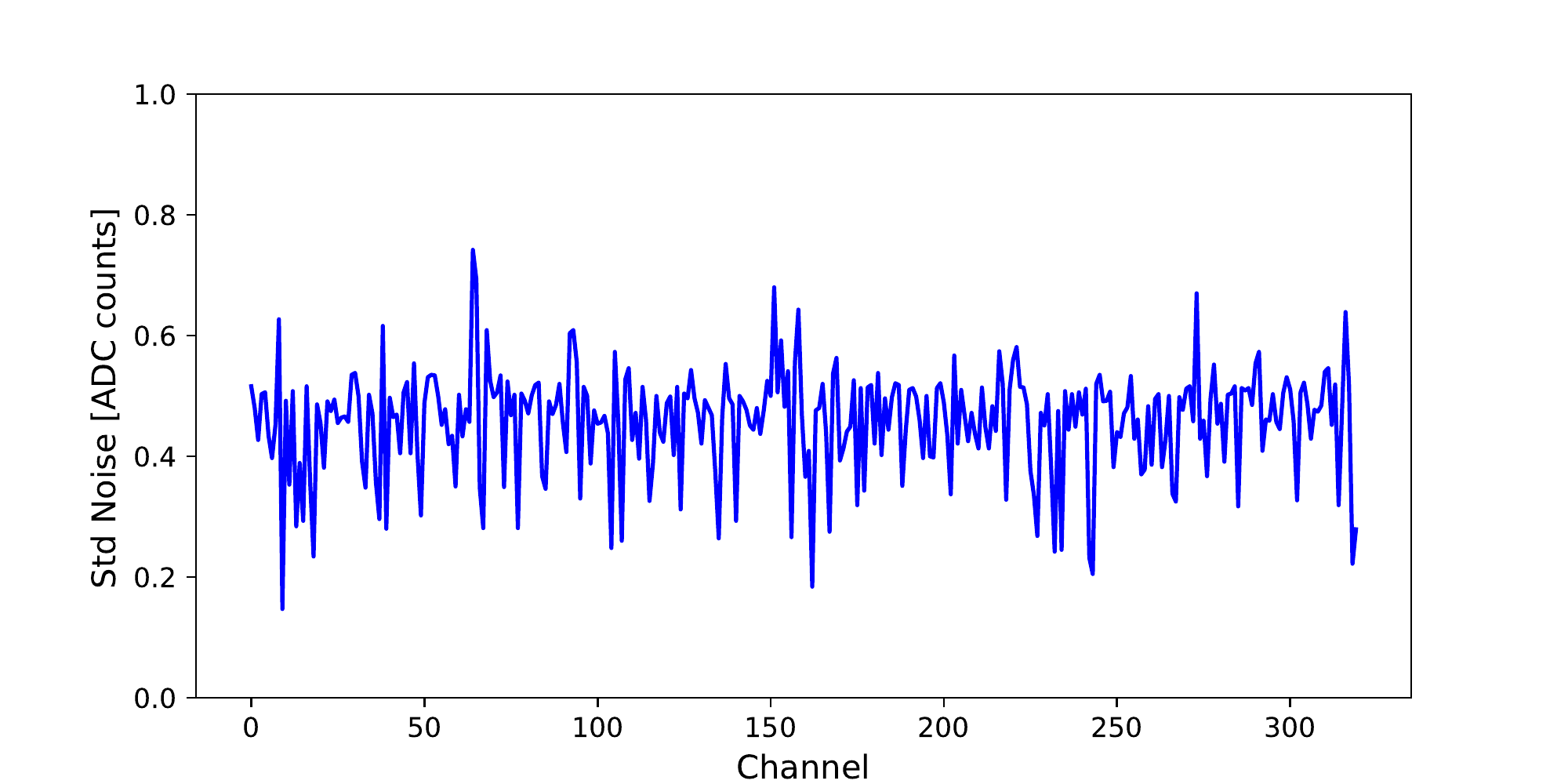}
        }
        \subfloat{
        \includegraphics[width=0.38\textwidth]{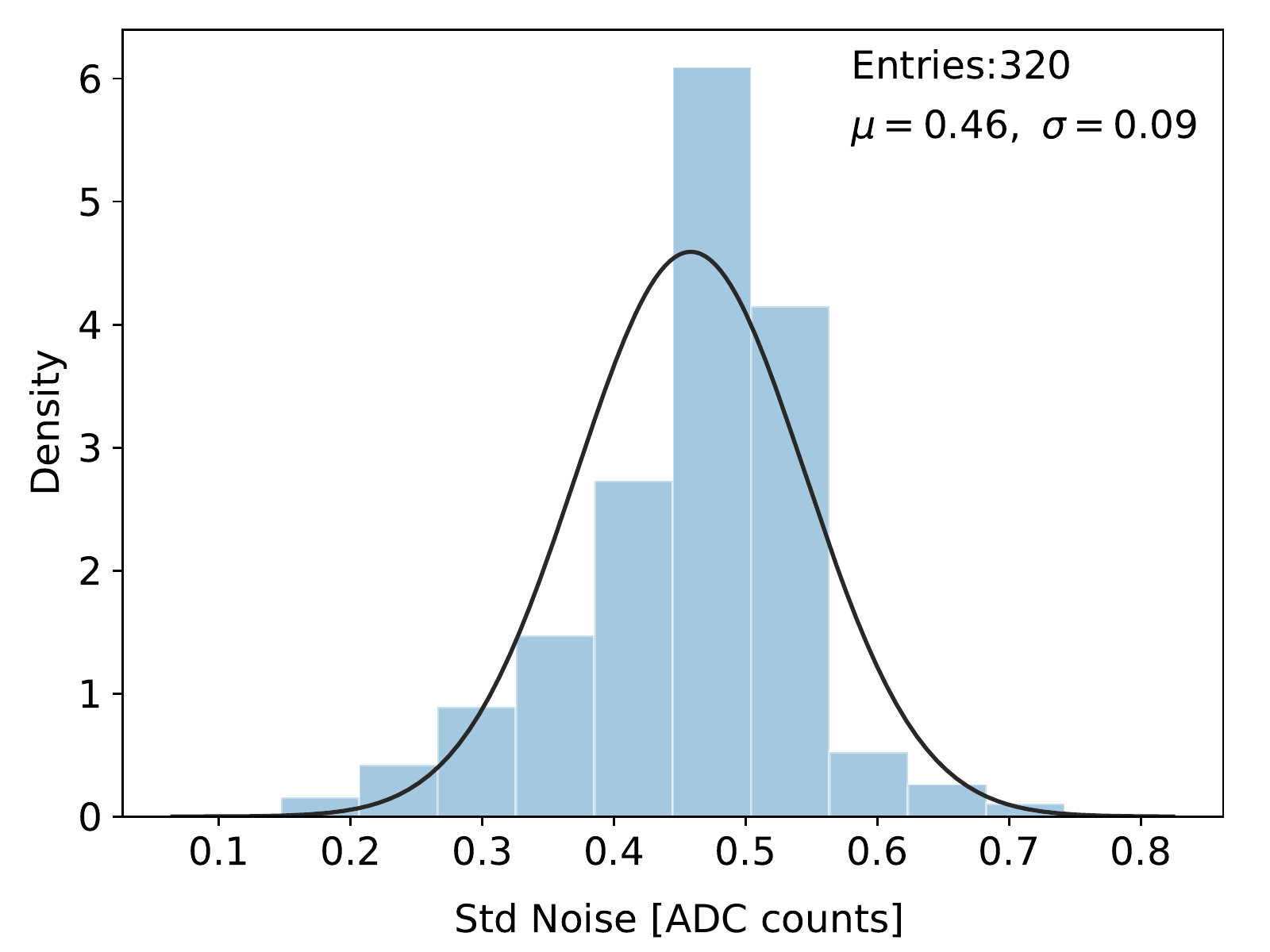}
        }
        \caption{Example noise distribution of an EMB LTDB. The 
        left plot shows the noise of each channel, and the right 
        histogram shows the distribution of the noise values of 
        all the 320 channels with an overlaid Gaussian fit. The average noise is 0.46 ADC counts, 
        which meets the LTDB design requirements.}
        \label{fig:ltdb_noise}
\end{figure}

Pedestals and noise are measured for each LTDB for each channel without any signal
injection. The pedestal values then show the electronics baseline
level, which simplifies the inspection of the full LTDB for dead channels.
The noise distribution of one LTDB board is shown in
Figure~\ref{fig:ltdb_noise}.

\begin{figure}[!h]
	\centering
        \subfloat{
	\includegraphics[width=0.5\textwidth]{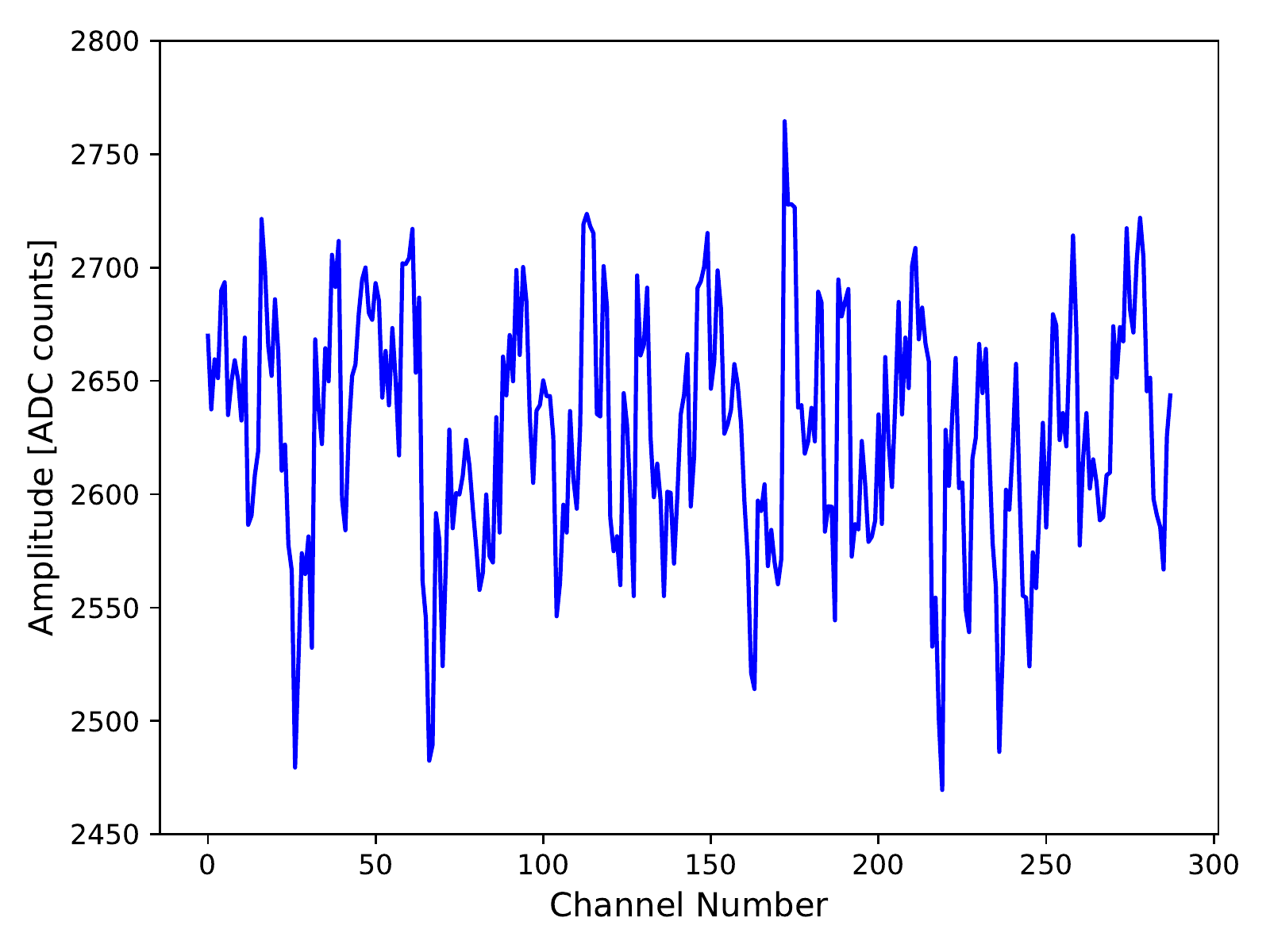}
        }
        \subfloat{
	\includegraphics[width=0.493\textwidth]{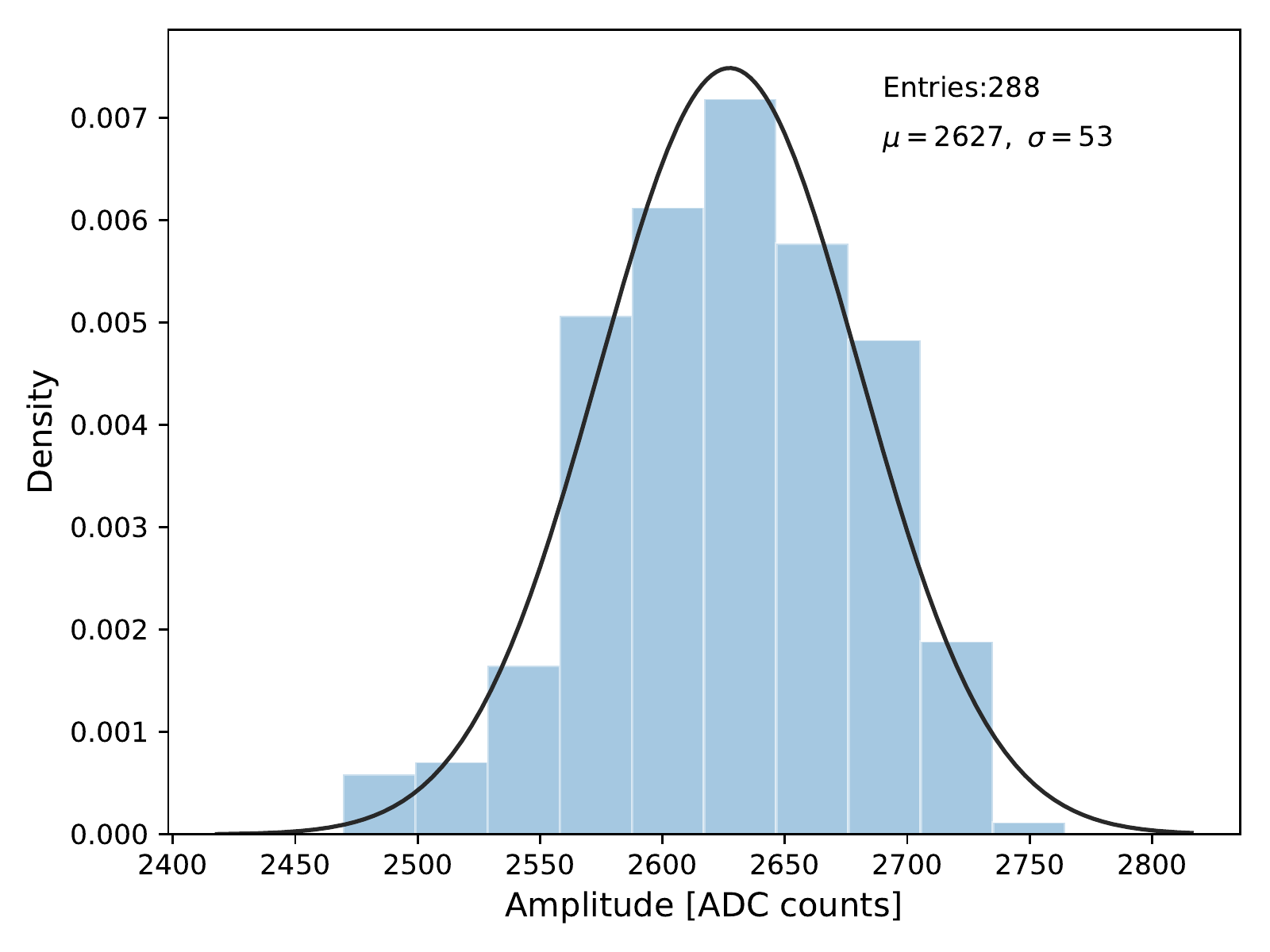}
        }\\
        \subfloat{
	\includegraphics[width=0.5\textwidth]{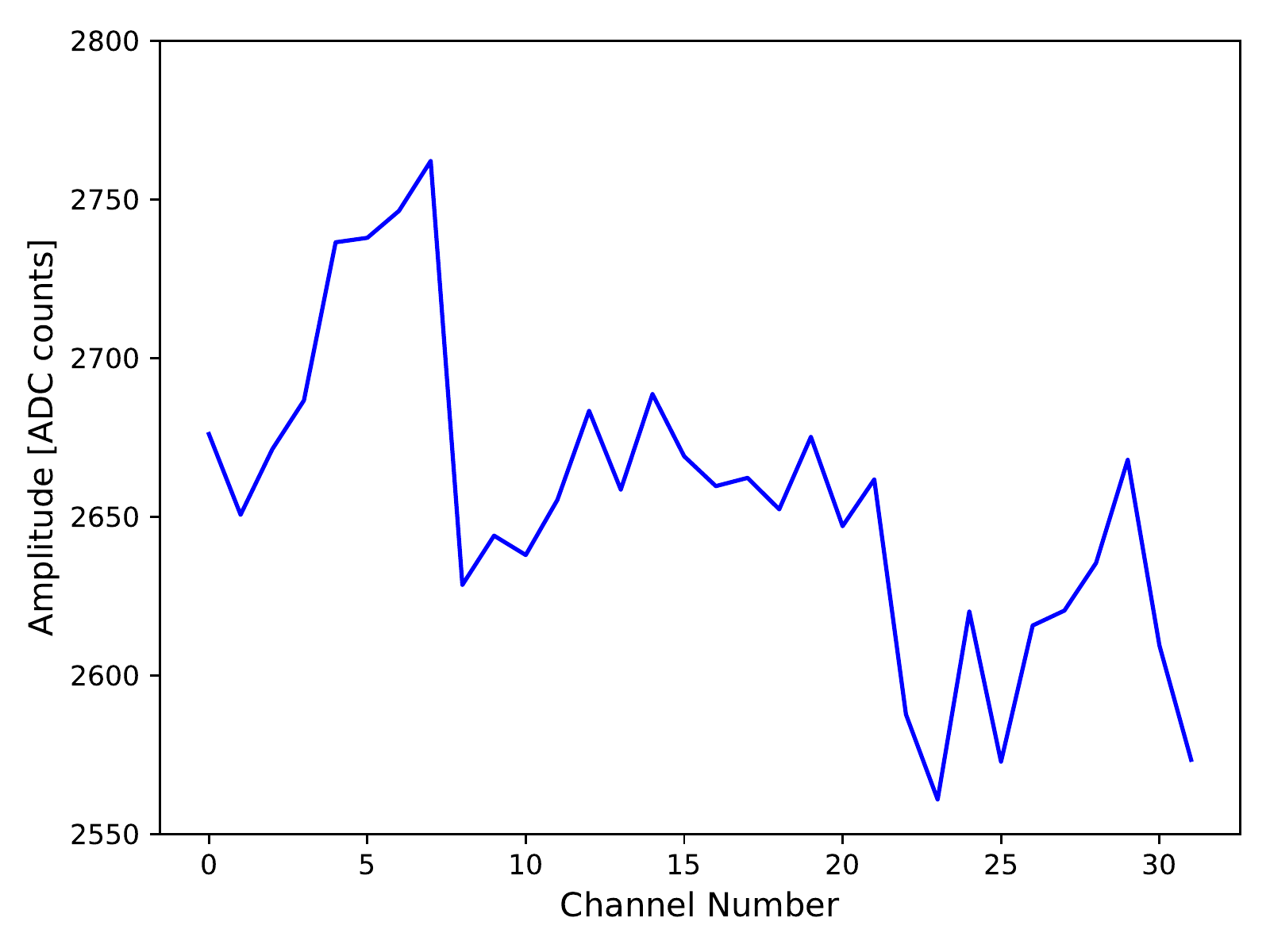}
        }
        \subfloat{
	\includegraphics[width=0.493\textwidth]{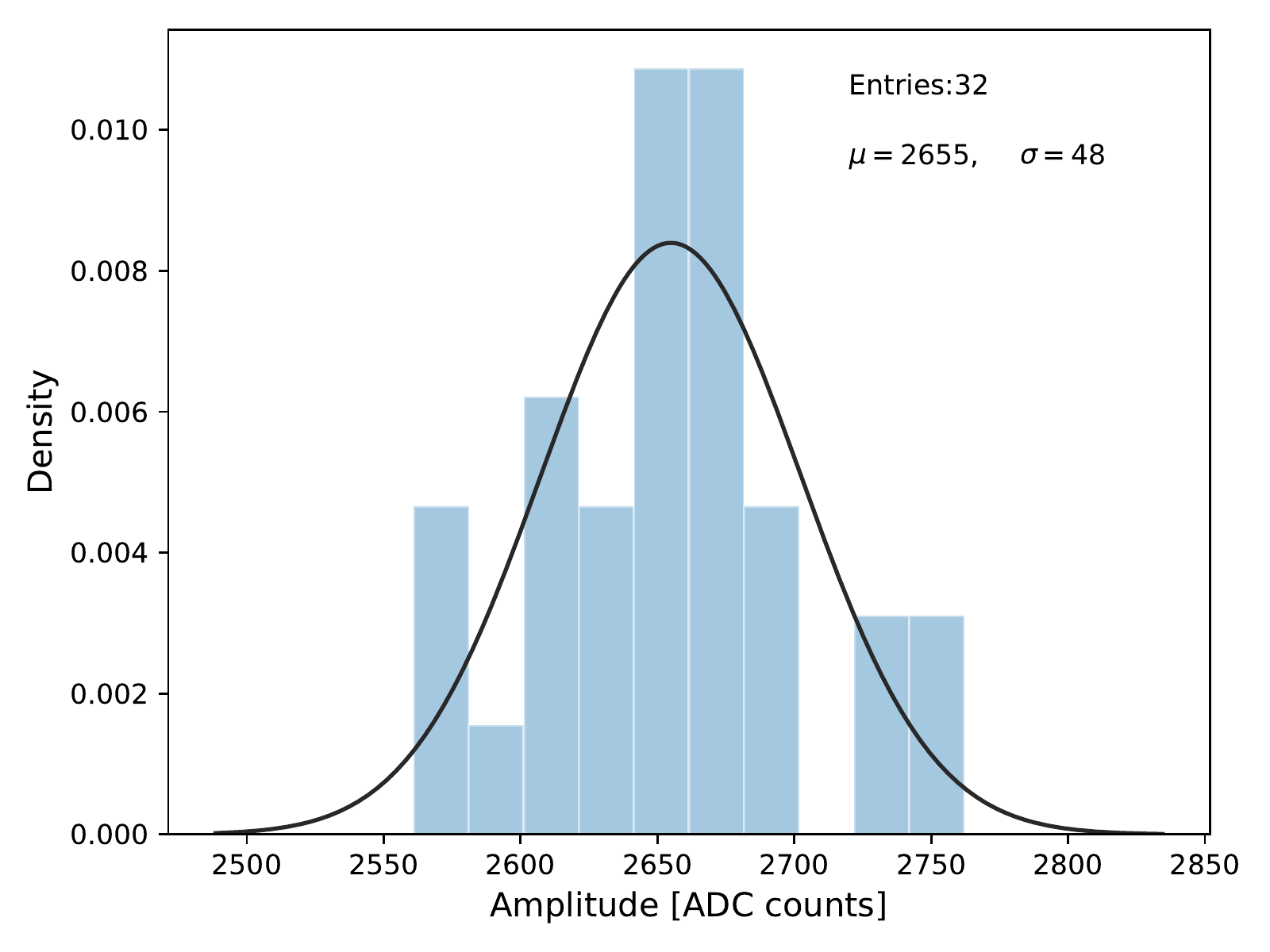}
        }
        \caption{Example amplitude distribution of an EMB LTDB, where 
        a front layer LAr signal with about 1.5~V amplitude is injected 
        into the LTDB. The left plots show the amplitude of each channel, 
        while the right plots show the distribution of the amplitude values with an overlaid Gaussian fit. 
        The top two plots show the amplitude of the 288 50-ohm channels with 
        an average amplitude value of 2627 ADC counts. The bottom two 
        plots show the amplitude of the 32 high-impedance channels with an 
        average amplitude value of 2655 ADC counts.}
        \label{fig:ltdb_amp}
\end{figure}

\begin{figure}[!h]
	\centering
        \subfloat{
	\includegraphics[width=0.5\textwidth]{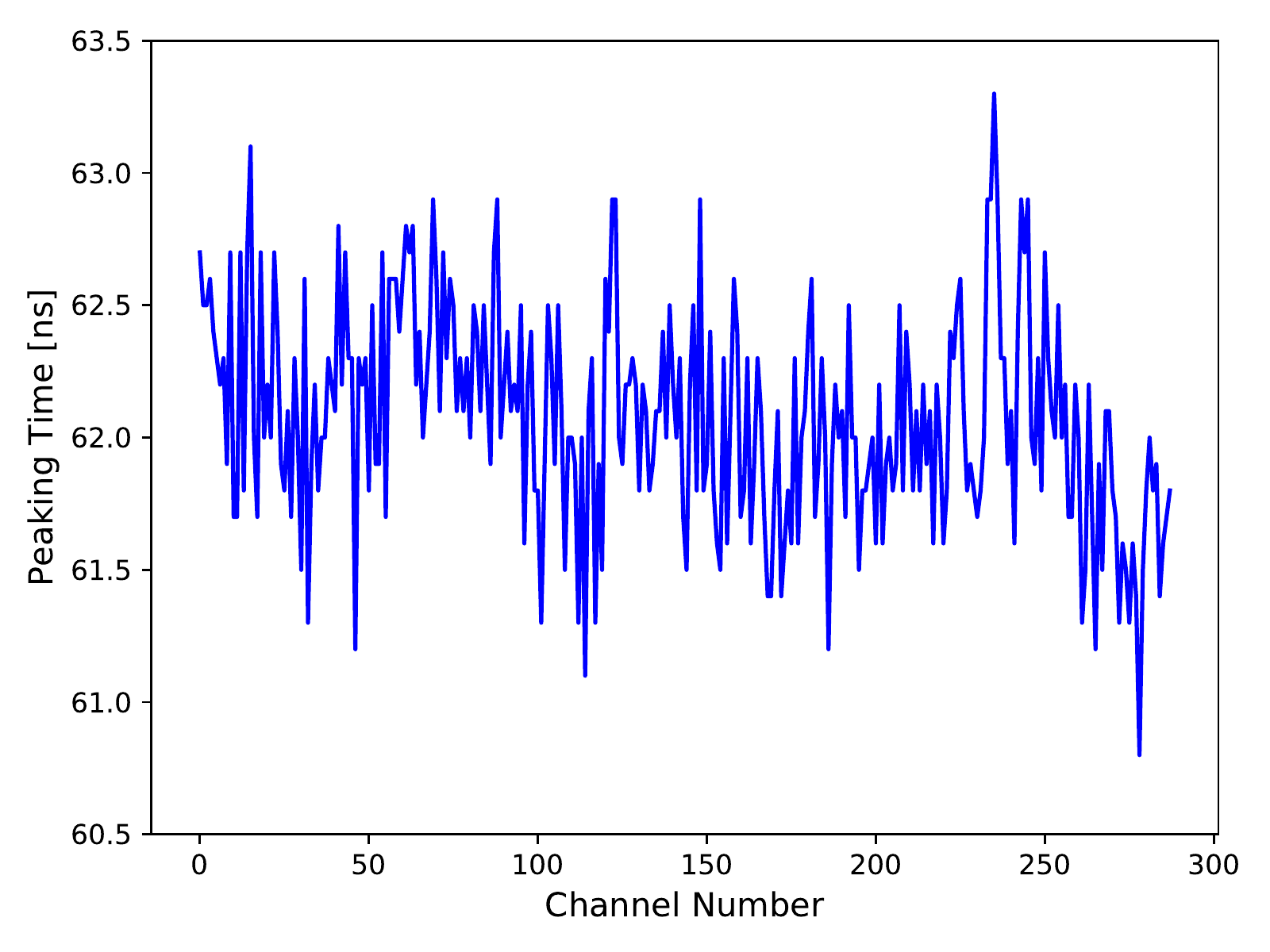}
        }
        \subfloat{
	\includegraphics[width=0.493\textwidth]{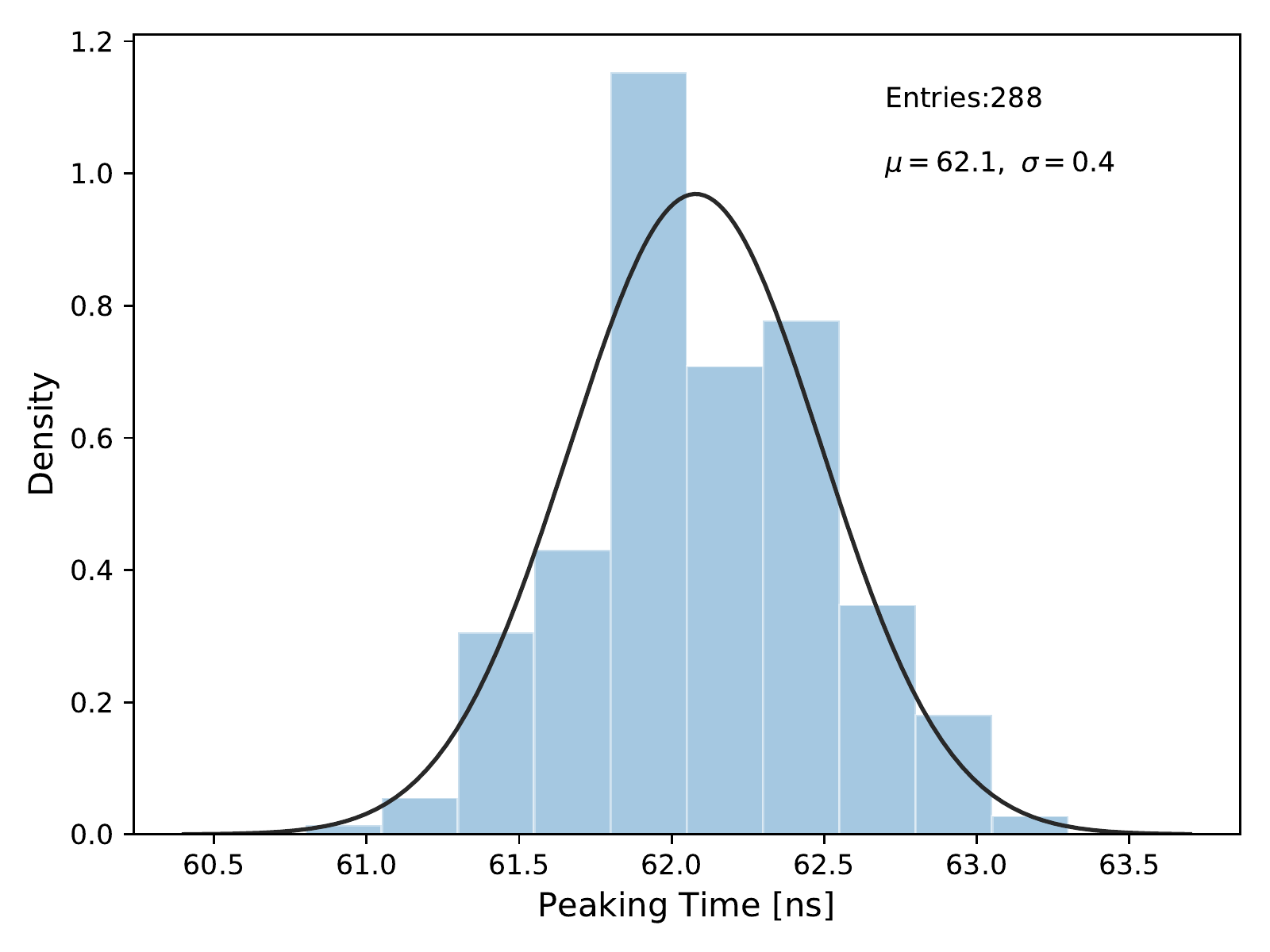}
        }\\
        \subfloat{
	\includegraphics[width=0.5\textwidth]{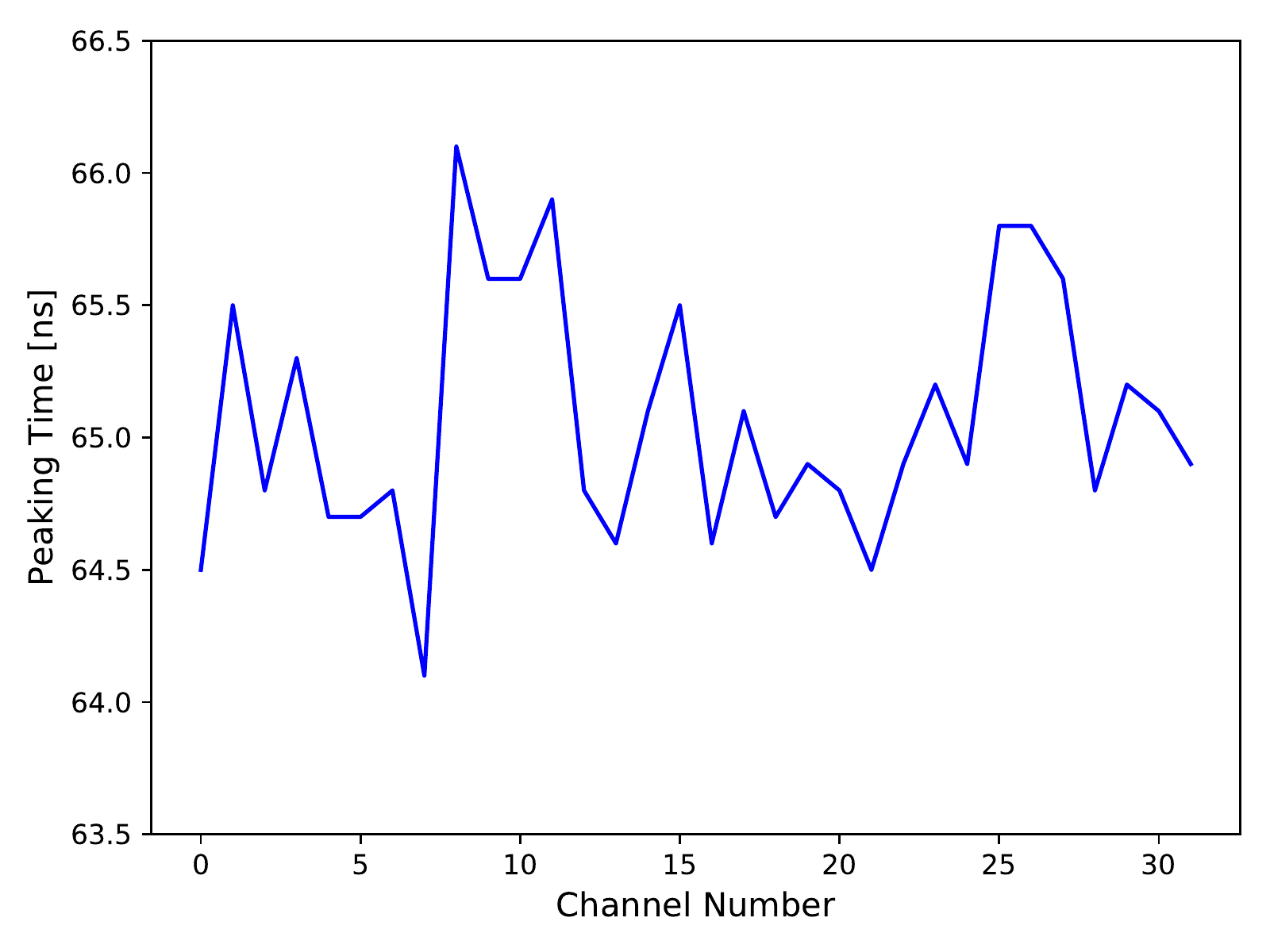}
        }
        \subfloat{
	\includegraphics[width=0.493\textwidth]{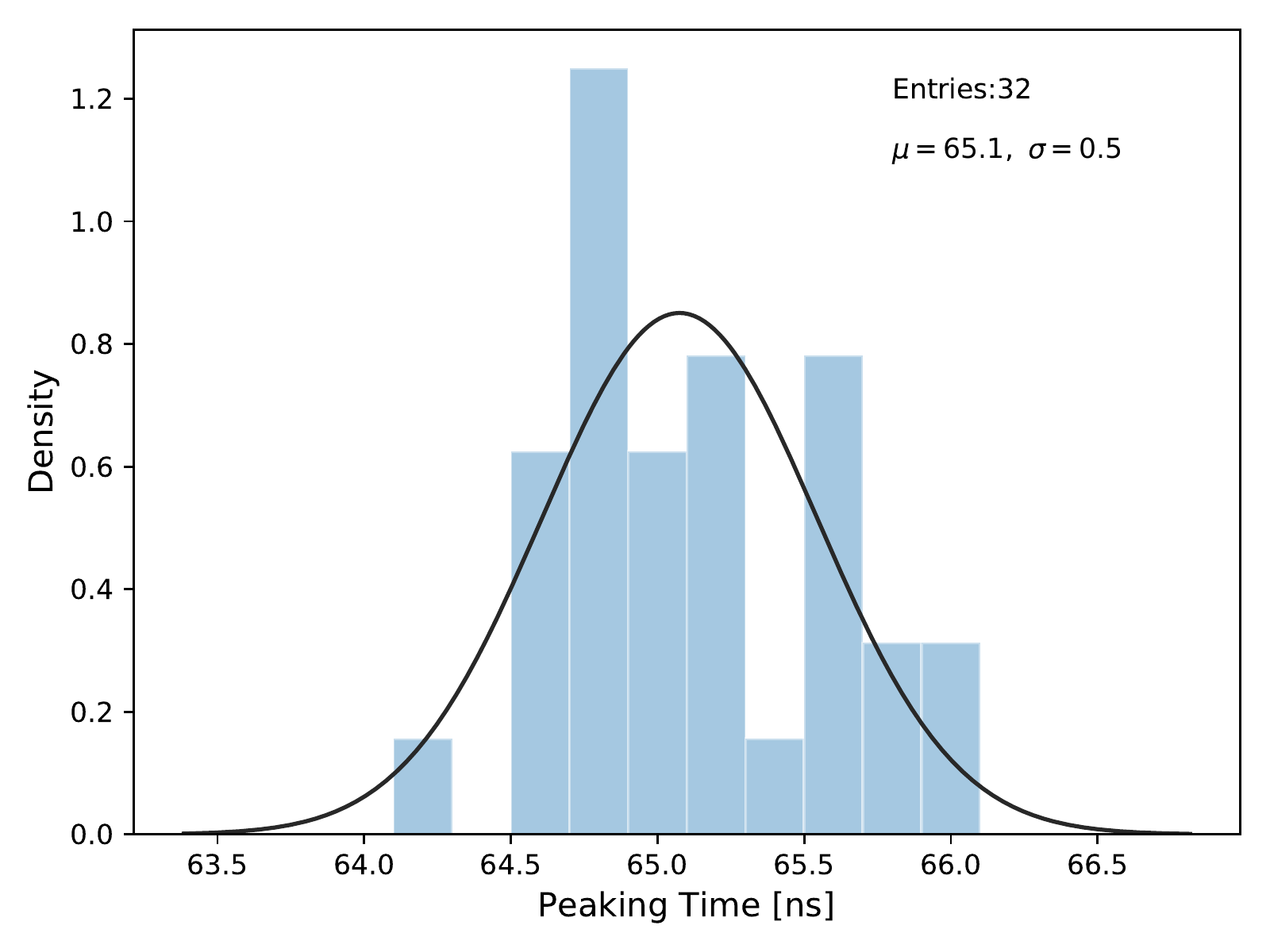}
        }
	\caption{Example peaking time distribution of an EMB LTDB, 
        where a front layer LAr signal with about 1.5~V amplitude is 
        injected into the LTDB. The left plots show the peaking 
        time of each channel, and the right plots show the distribution 
        of the peaking time values with an overlaid Gaussian fit. The top two plots show the peaking 
        time of the 288 50-ohm channels with an average peaking time of 62.1~ns. 
        The bottom two plots show the peaking time of the 32 high-impedance 
        channels with an average peaking time of 65.1~ns.}
        \label{fig:ltdb_peaking}
\end{figure}

\begin{figure}[!h]
	\centering
        \subfloat{
         \includegraphics[width=0.6\textwidth]{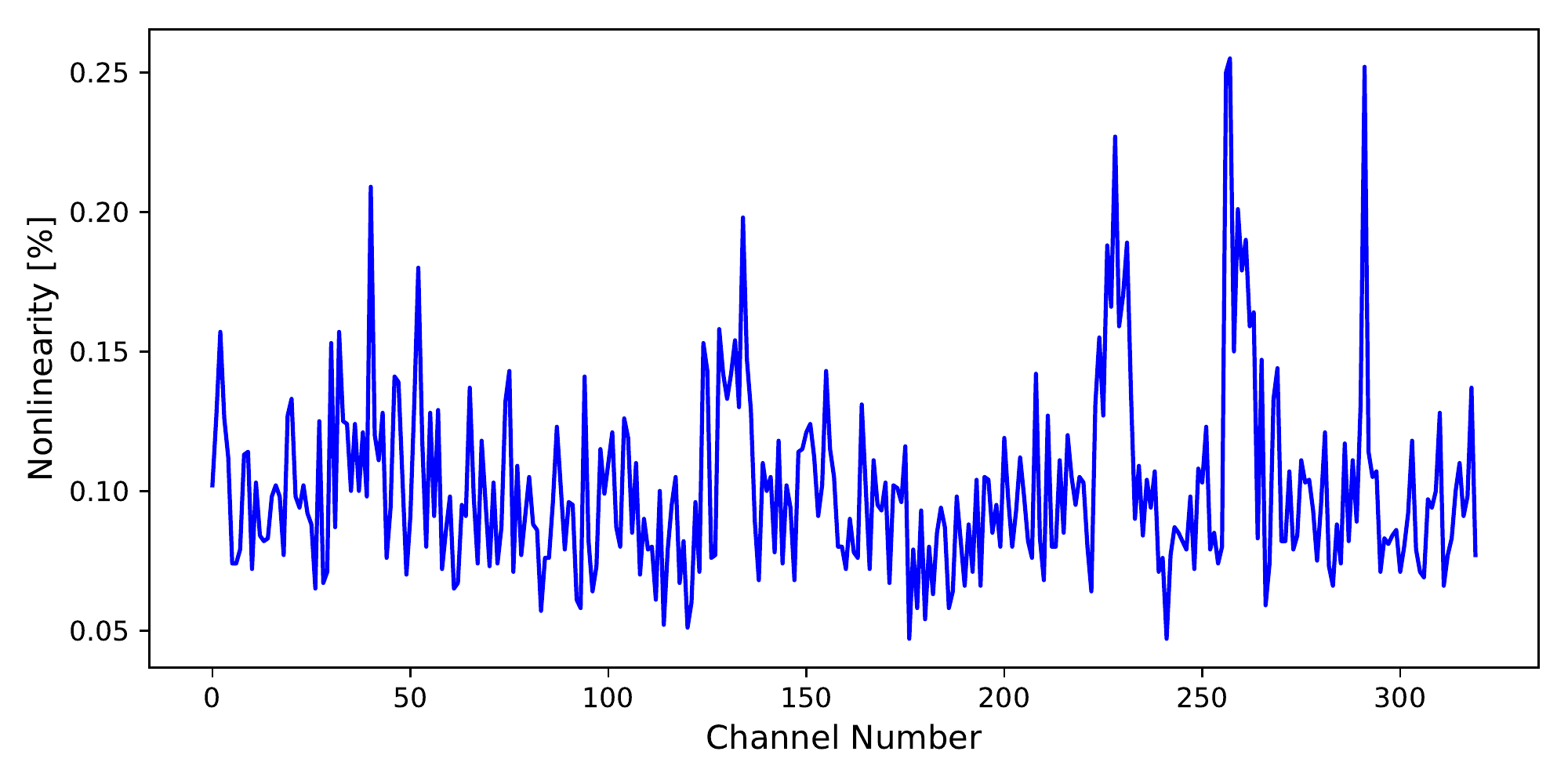}
        }
        \subfloat{
         \includegraphics[width=0.4\textwidth]{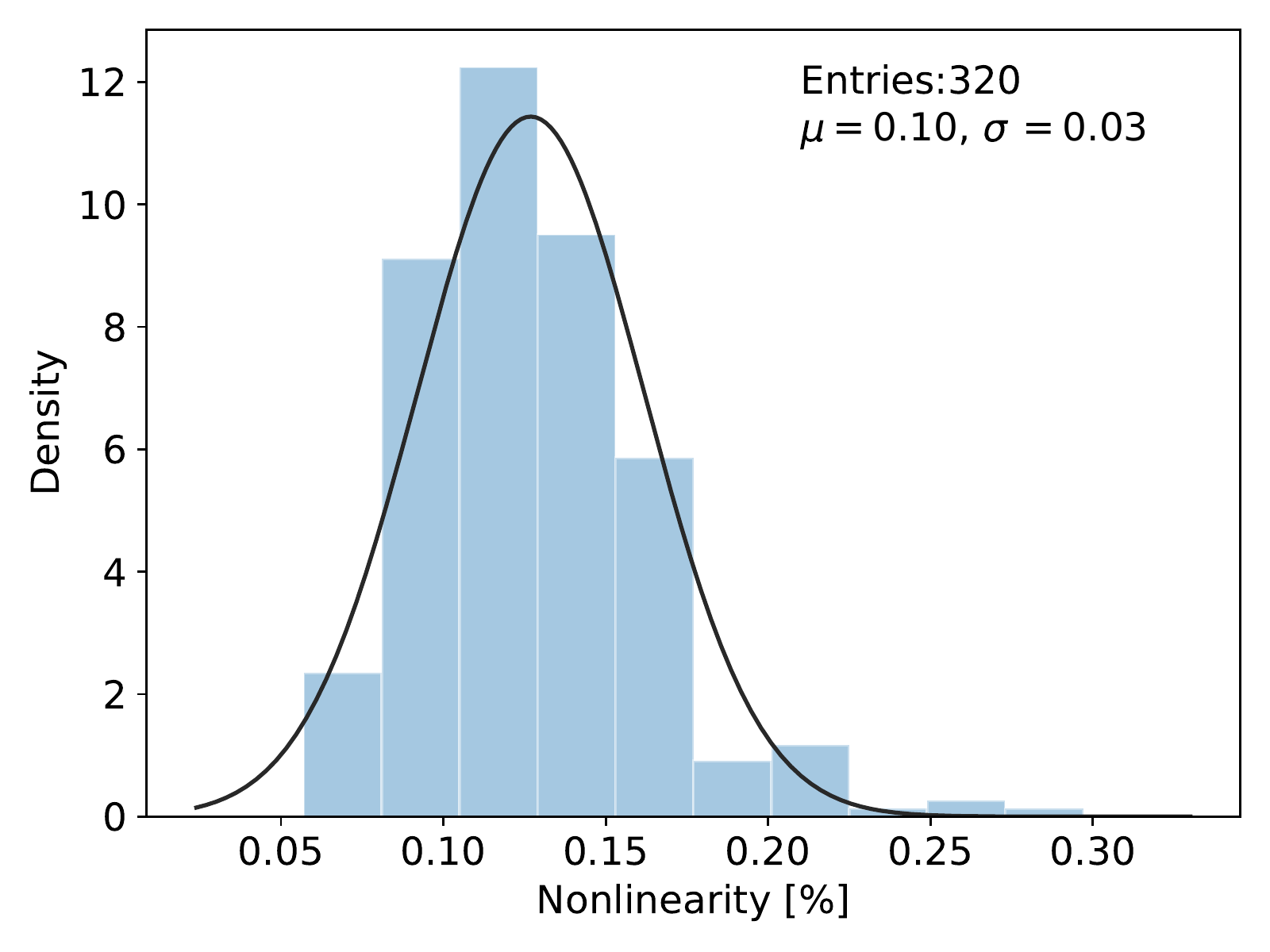}
        }
	\caption{Example INL distribution of an EMB LTDB. The 
        left plot shows the INL of each channel, and the right 
        histogram shows the distribution of the INL values of 
        all the 320 channels with an overlaid Gaussian fit. The average INL is 0.10\%.}
        \label{fig:ltdb_inl}
\end{figure}

Simultaneous LAr signals are injected to each LTDB for the integration
performance tests. Delay runs are used to measure the amplitude and
peaking time of the signals in each channel, using a series of 15 LAr 
pulses, each delayed by \SI{1.67}{ns} with respect to the previous. With this method,
the reconstruction of the LAr signal pulse has an effective
sampling interval of \SI{1.67}{ns}. An example distribution of amplitude and
peaking time is shown in Figures~\ref{fig:ltdb_amp}
and~\ref{fig:ltdb_peaking}. Ramp runs are used to measure the
linearity of each channel, using a series of 16 LAr signals injected 
with different amplitudes, which are equally spaced from the
baseline to the maximum signal close to ADC saturation. An example
distribution of the INL is shown in
Figure~\ref{fig:ltdb_inl}. 

There are 64 summing output channels per
LTDB. The average INL of each channel is around
0.1\%, measured with an oscilloscope.

\subsection{Back-End boards integration tests}
\label{ssec:be-integration}
These integration tests are performed on LArC and LATOME assembled together before their installation. There are two types of LATOME according to the combination of input channels on the MPO connectors and pigtail fibers to the microPODs as already discussed in Section~\ref{ssec:latome}. 
Up to four LATOMEs of the same type are used to assemble one LDPB. In total 22 (8) standard (special) LDPBs are assembled and tested. Spare LDPBs are validated in the same way. 

After further visual inspection\footnote{Already inspected at production stage, the identification number on the sticker and on the front panel, PCB edge cutting, and holding of heat-sinks are verified once more.}
to verify the type of LATOME and the IPMC on the LArC to be mounted properly, the LDPB and the RTM board are plugged into a slot of the ATCA shelf in order to check the ATCA system
connectivity:
\begin{itemize}
    \item The state of the module is checked with LEDs on the front panel. On the ATCA shelf-manager, connectivities to the IPMC on the LArC and MMCs on each LATOME are verified by checking that the topology and the hardware names are correctly recognized.
    \item A JTAG cable is connected to the RTM in order to verify that the JTAG chain has five FPGAs, four \intel \arria~10 FPGAs and one \xilinx \virtexseven with the Quartus and the Vivado software.
    \item  The factory and the application firmware images are loaded to the flash on each LATOME and the following checks are performed: (1) the boot sequence of the factory firmware is verified upon a power cycle of the LATOME, and (2) connectivities to the GbE network are checked with responses of the ping command to IPMC, LArC and all LATOMEs from the PC on the same network. 
\end{itemize}

The next step is an optical data transmission test on the main data path for all links. This is performed with a pair of LATOMEs connected to each other, instead of a simple loop-back, with custom made 48 fiber bundles. These have an MPO48 connector for the LATOME Tx side and four MPO12 (or two MPO24 and one MPO12) connectors for the Rx side. On the LATOME, the application firmware is loaded and configured with a fiber mapping corresponding to the LATOME connector type. On the LArC, the TTC generator is used to provide the \SI{160.2}{MHz} reference clock and BCR to LATOMEs via an LVDS line. The synchronization of the LATOMEs with the reference clock is verified, as well as the reception of the BCR with a known period. Then the data transmission test is performed with the PRBS31 data.
These procedures certify the stability of clock and BCR reception at the LATOMEs and the main optical data transmission capabilities. 

The TDAQ and local monitoring readout paths are both verified by counting the numbers of packets sent from each LATOME and received on the LArC. These numbers are compared to the numbers of trigger signals provided by the TTC generator on the LArC. 

The long-term stability is also checked by keeping power up and monitored for one week. All monitoring sensors related to power load and temperatures are read out. After the long-term stability test, all connectivity tests are repeated.
 
\subsection{Full chain tests}
\label{ssec:be-full-chain}
The full chain of the LAr trigger and data acquisition electronics including the Front-End and the Back-End electronics is tested in situ after installing the parts of the \phaseone upgrade electronics in the ATLAS cavern.
A set of input signal scans is designed to validate the new system: mapping scans to check the connectivity of all channels; timing scans to align various components in time; and calibration scans to validate the pedestal values, the shape of the pulse and the value and linearity of the gains.

\begin{figure}[htb]
  \centering
  \includegraphics[width=\textwidth]{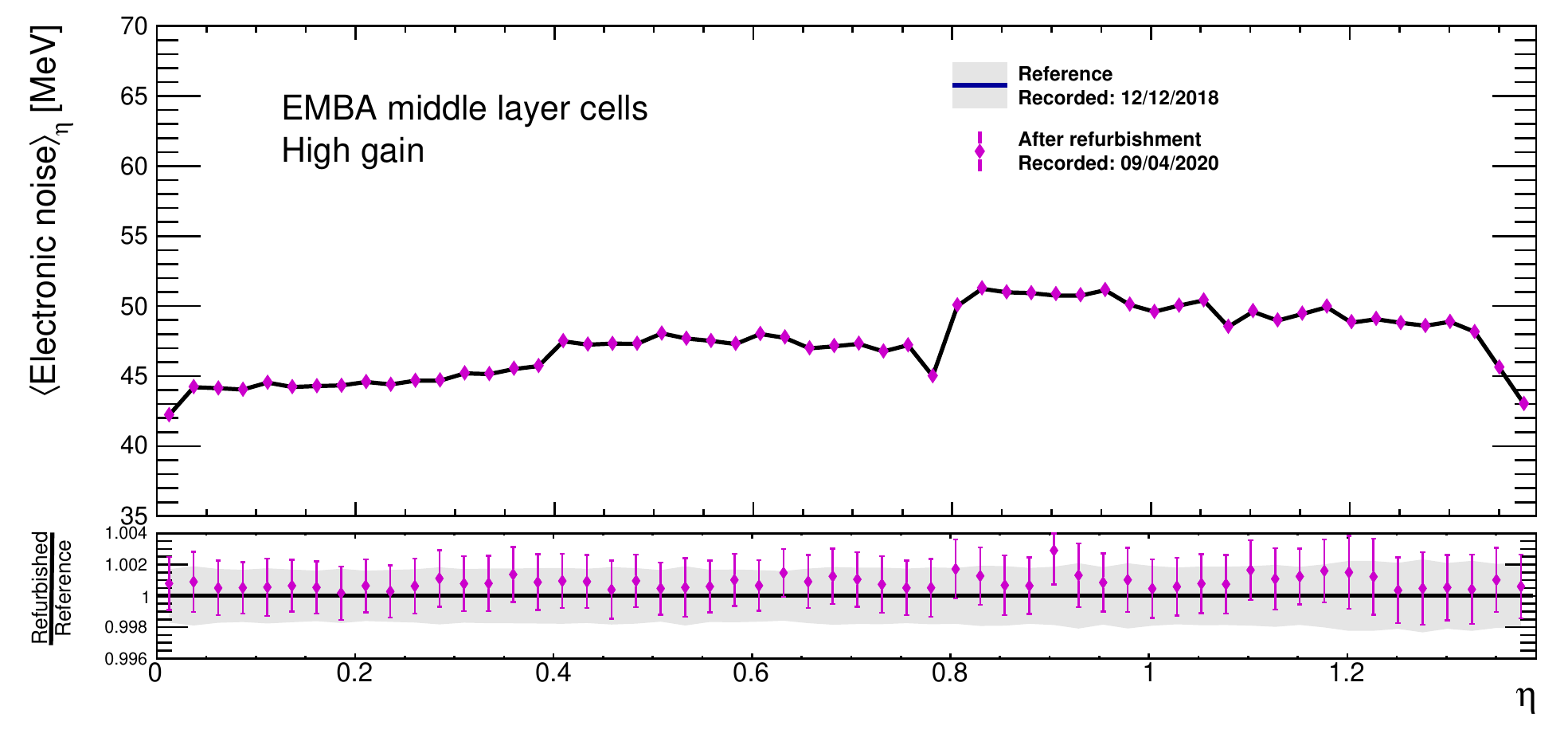}
  \caption{Noise level as function of the pseudorapidity for cells before and after the installation of the new Front-End electronics.
The noise level corresponds to the mean value of the readout electronic noise in MeV over the cells in a given pseudorapidity ($\eta$) range. Only the cells of the middle layer of the EMB on side A are included. The black line and the gray uncertainty band show the values measured at the end of \runtwo. The purple dots show the values measured after the refurbishment of the FECs and FEBs during the LS2. The displayed uncertainties are computed as $\sigma_{\textrm{noise, cells}}/\sqrt{N_{\textrm{cells}}}$, where $\sigma_{\textrm{noise, cells}}$ are the noise values integrated over the cells in the given $\eta$ range and $N_{\textrm{cells}}$ is the number of cells in the given $\eta$ range. The cause of the systematic increase of approximately one per mil over the whole $\eta$ range is under investigation but will have negligible impact on data analyses.
}
  \label{fig:noise_vs_eta_legacyMain}
\end{figure}

\begin{figure}[t]
  \centering
  \includegraphics[width=\textwidth]{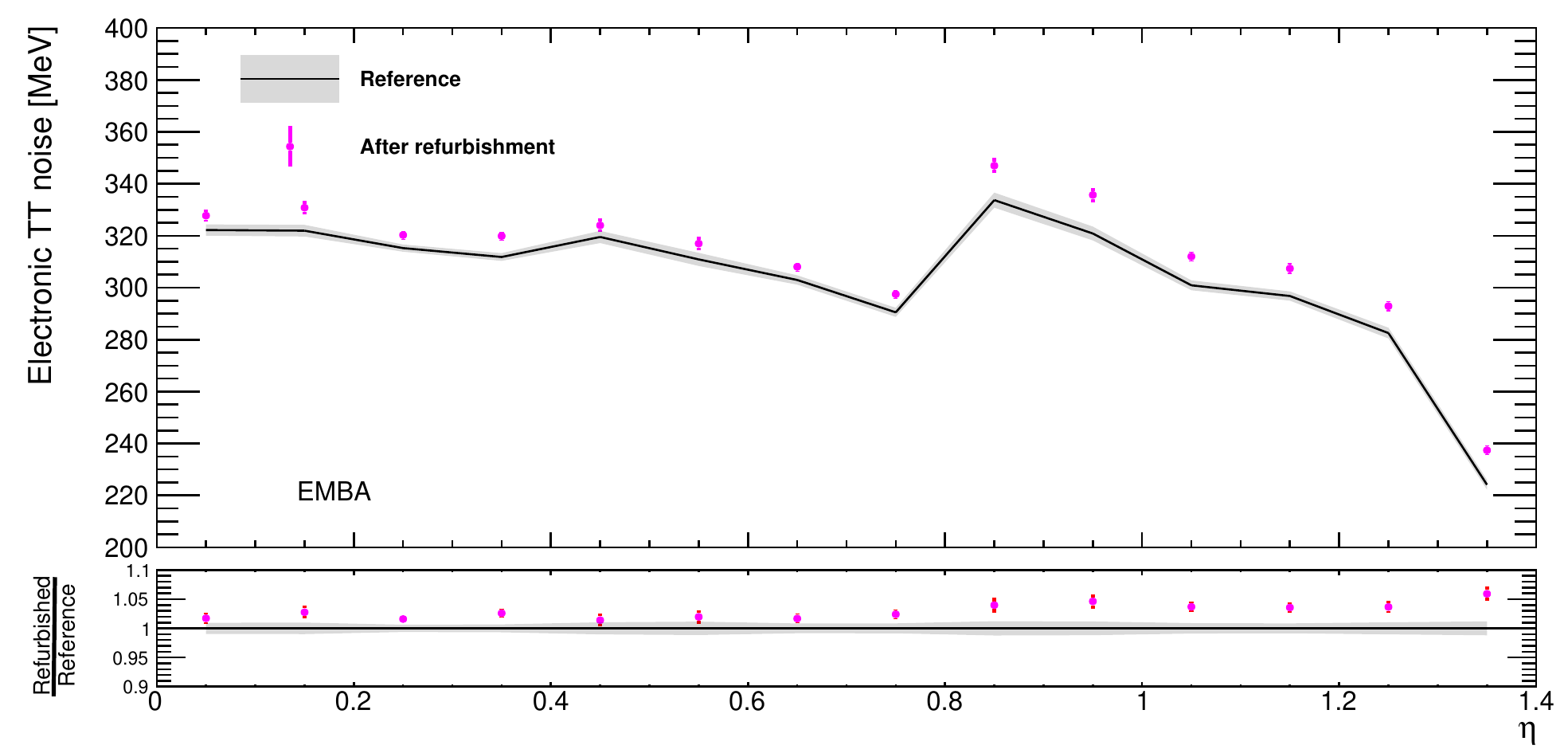}

  \caption{Noise level as function of the pseudorapidity for Trigger Towers (TTs)
  before and after the installation of the new trigger readout electronics.
  The noise level corresponds to the mean value of the readout electronic noise
  in MeV over the TTs in the same pseudorapidity ($\eta$) position. Only
  the TTs of the EMB on side A are included. The black line and the gray uncertainty band
  show the values measured at the end of \runtwo. The purple dots show the values measured
  after the refurbishment of the FECs and FEBs  and the installation of
  the LTDBs during the LS2. The displayed uncertainties are computed as 
  $\sigma_{\textrm{noise, TT}}/\sqrt{N_{\textrm{TT}}}$, where 
  $\sigma_{\textrm{noise, TT}}$ are the noise values integrated over the TTs in the given $\eta$ position and $N_{\textrm{TT}}$ is the number of 
  TTs in the given $\eta$ position. The systematic increase of approximately $2$ to $5\%$ over the whole $\eta$ range is due to a more complex
  analog signal path and has a negligible impact on trigger performance. 
  }
  \label{fig:noise_vs_eta_legacyTrigger}
\end{figure}

Due to the installation of the new baseplanes and the LTDBs, the electronic path for some of the input channels to the legacy trigger towers were modified.
Moreover, the FEBs were removed from the old baseplanes, refurbished with new LSBs and installed on the new baseplanes. These operations might have
affected the electronic legacy main readout.
The legacy trigger system is required to run in parallel to the new digital trigger system during \runthree.
For these reasons, both main and trigger legacy readout paths are validated after the installation of the new Front-End electronics.
The noise levels for the different cells and Trigger Towers before and after the installation of the new baseplanes and LTDBs can be seen in Figure~\ref{fig:noise_vs_eta_legacyMain} and~\ref{fig:noise_vs_eta_legacyTrigger} respectively. 
No significant change in the level of the electronic noise is observed due to
the installation of the new electronic boards.
The jump at $\eta=0.8$ observed in both figures is due to a change in the barrel calorimeter sampling fraction,
that affects the conversion of the signal in $\mu$A to a signal in \MeV.

\begin{figure}[htb]
  \centering
  \subfloat[]{
  \includegraphics[width=0.5\textwidth]{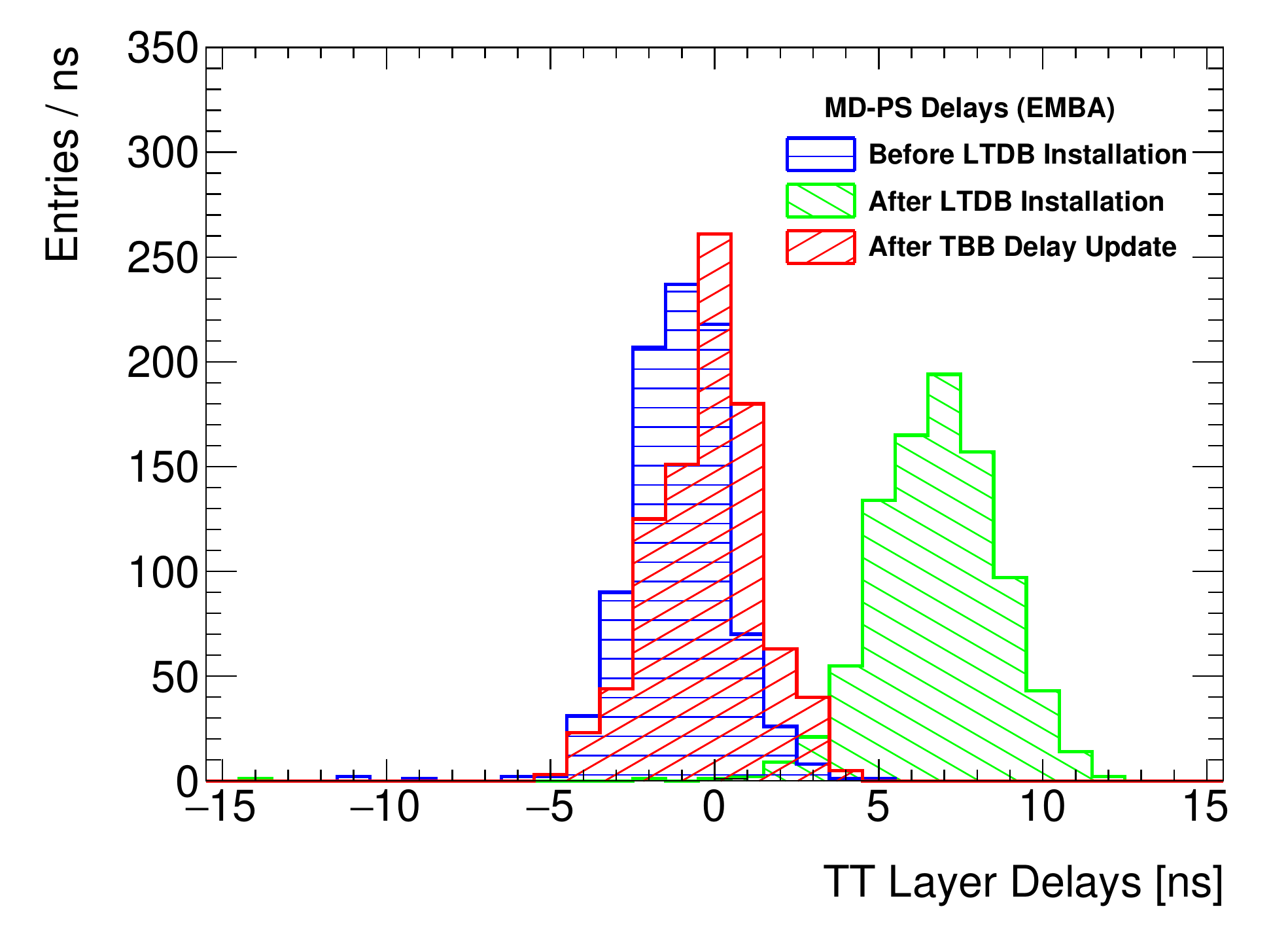}
  }
  \subfloat[]{
  \includegraphics[width=0.5\textwidth]{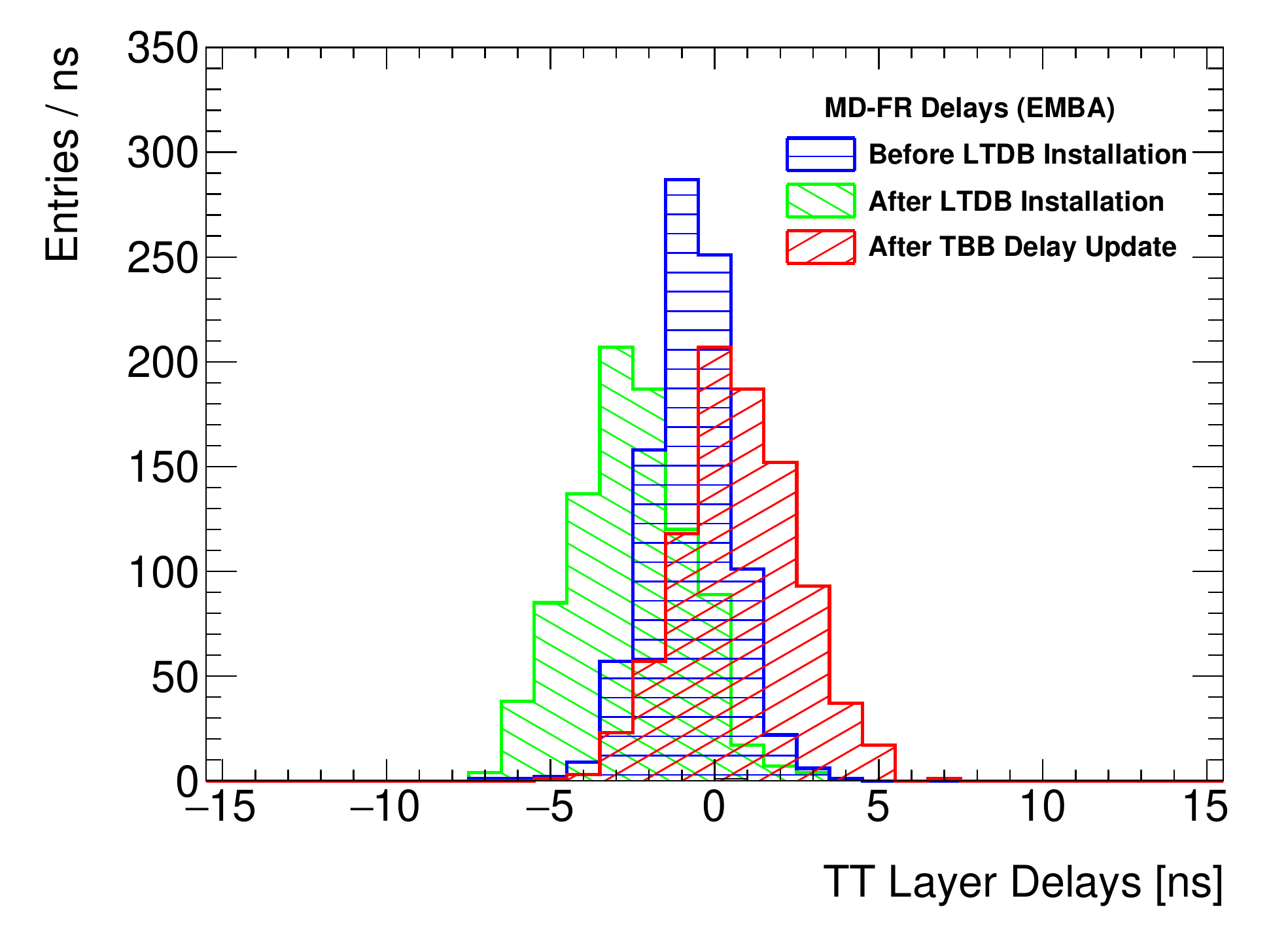}
  }\\
  \subfloat[]{
  \includegraphics[width=0.5\textwidth]{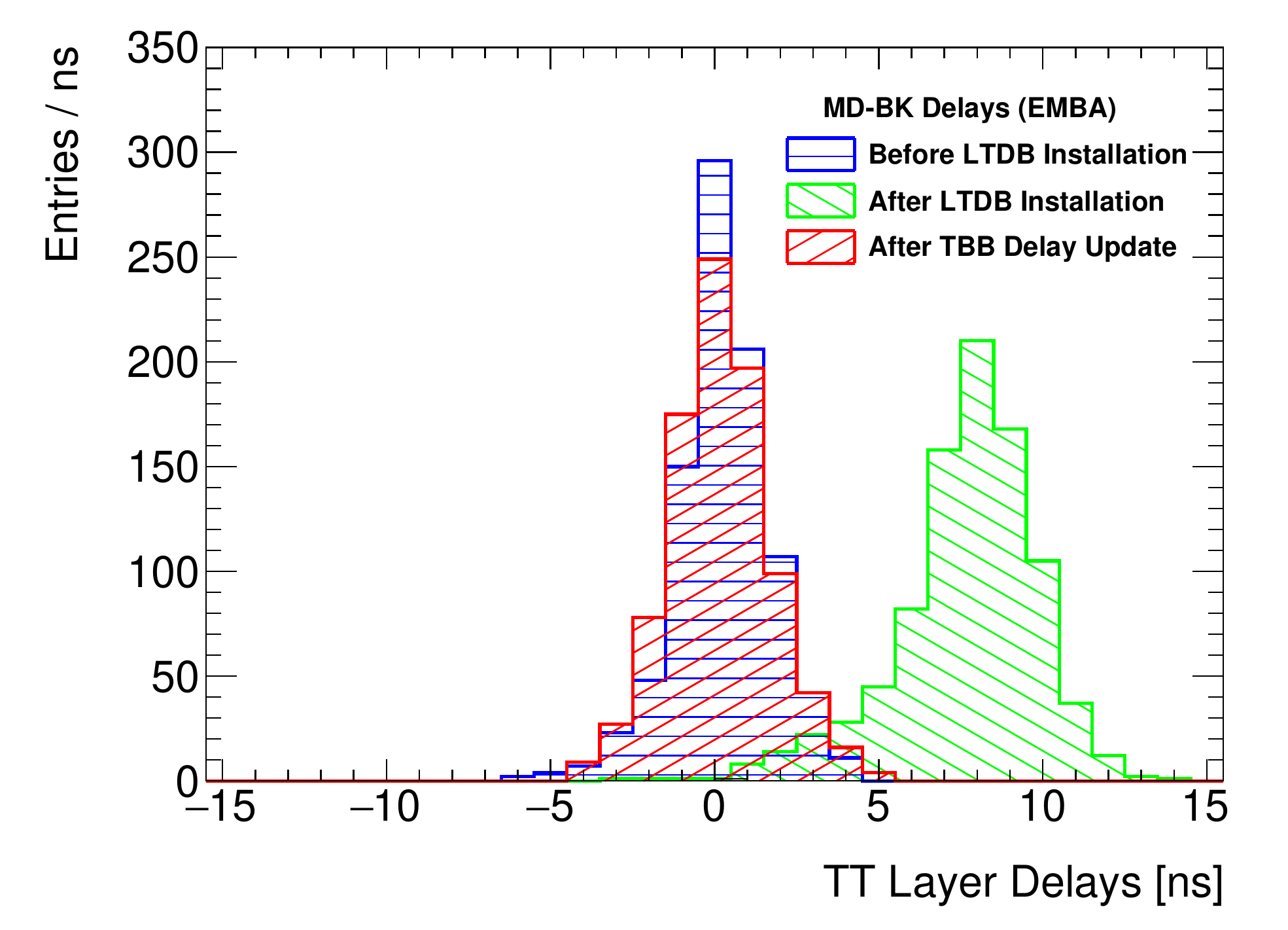}
  }
  \caption{Time difference between (a) middle (MD) layer and presampler (PS) channels, (b) between middle 
  and front (FR) layer channels, and (c) between middle and back (BK) layer channels before and after the 
  baseplane replacement and the LTDB insertion. After the corrections applied to the TBB delays 
  to take into account the new hardware, the distributions match well. Only EMB side A is considered here.}
  \label{fig:tt_layer_timing}
\end{figure}

The timing of the different Trigger Tower channels is also adjusted on the TBBs to take into account the new signal path due to the new boards. The time difference relative to the middle layer channels in the different layers before and after the LTDB installation and before and after the correction is shown in Figure~\ref{fig:tt_layer_timing}.

\begin{figure}[htb]
  \centering
  \includegraphics[width=.75\textwidth]{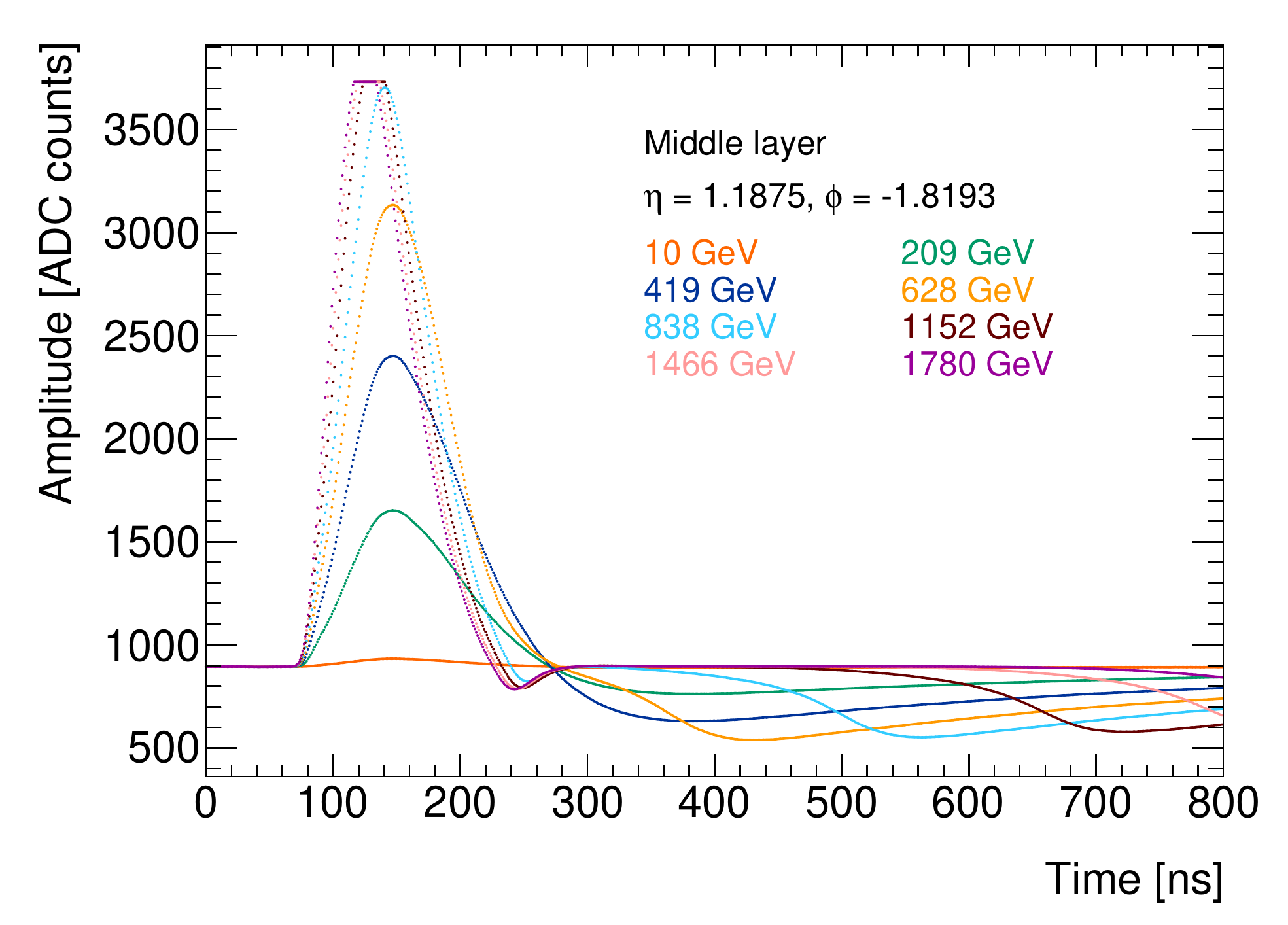}
  \caption{Pulse shape collected by the LATOME board for one Super Cell in the barrel part of the detector at $\eta=1.1875$ and $\phi=-1.8193$ in the middle layer for several injected current pulses corresponding to different \ET values.}
  \label{fig:pulse_shape}
\end{figure}

To validate the new digital trigger path, data from the LTDB are received, time-aligned and processed on the LATOME boards and then sent to the local monitoring path to be collected by a UDP server running on a dedicated machine.
More precisely, 
a calibration pulse is injected in the Front-End electronics synchronously to a delayed L1A signal.
The LTDB data, corresponding to 32 ADC samples, are received at \SI{40}{MHz} and buffered on
the LATOME boards, which send the data to the local monitoring path when the L1A is received. 
To be able to retrieve the 32 ADC samples that correspond to the injected pulse the LATOME
boards are time-aligned. 
With this procedure the shape of the pulse collected by the LATOME can be verified for
different energy regimes, as illustrated for one Super Cell in Figure~\ref{fig:pulse_shape}. To obtain this shape with a fine granularity from a 32 ADC sample
readout, a series of twenty consecutive calibration pulses is used, each one with an
incremental delay of about \SI{1.04}{\ns}.
Distortion in the pulse shape can be seen at high energy due to saturation effects which will 
be discussed in the following.

\begin{figure}[htb]
  \centering
  \subfloat{
  \includegraphics[width=\textwidth]{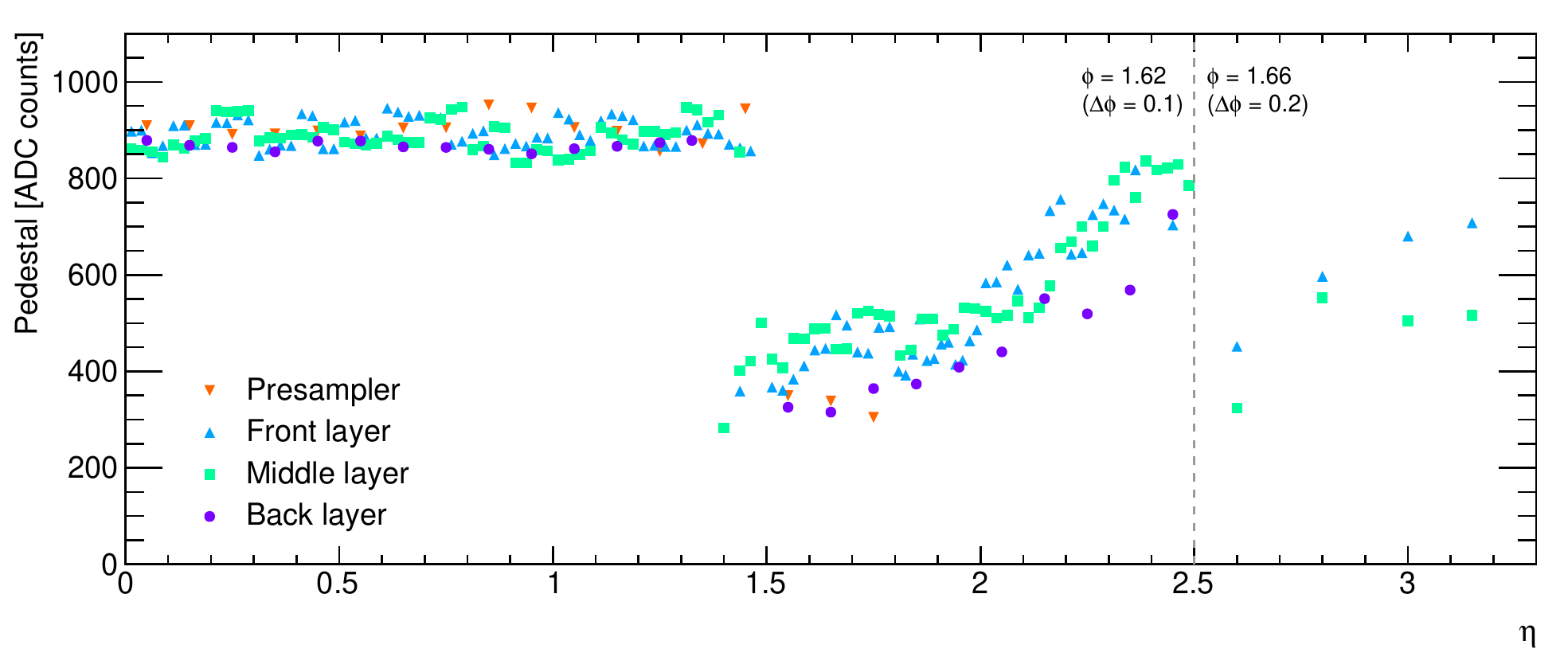}
  }\\
  \subfloat{
  \includegraphics[width=\textwidth]{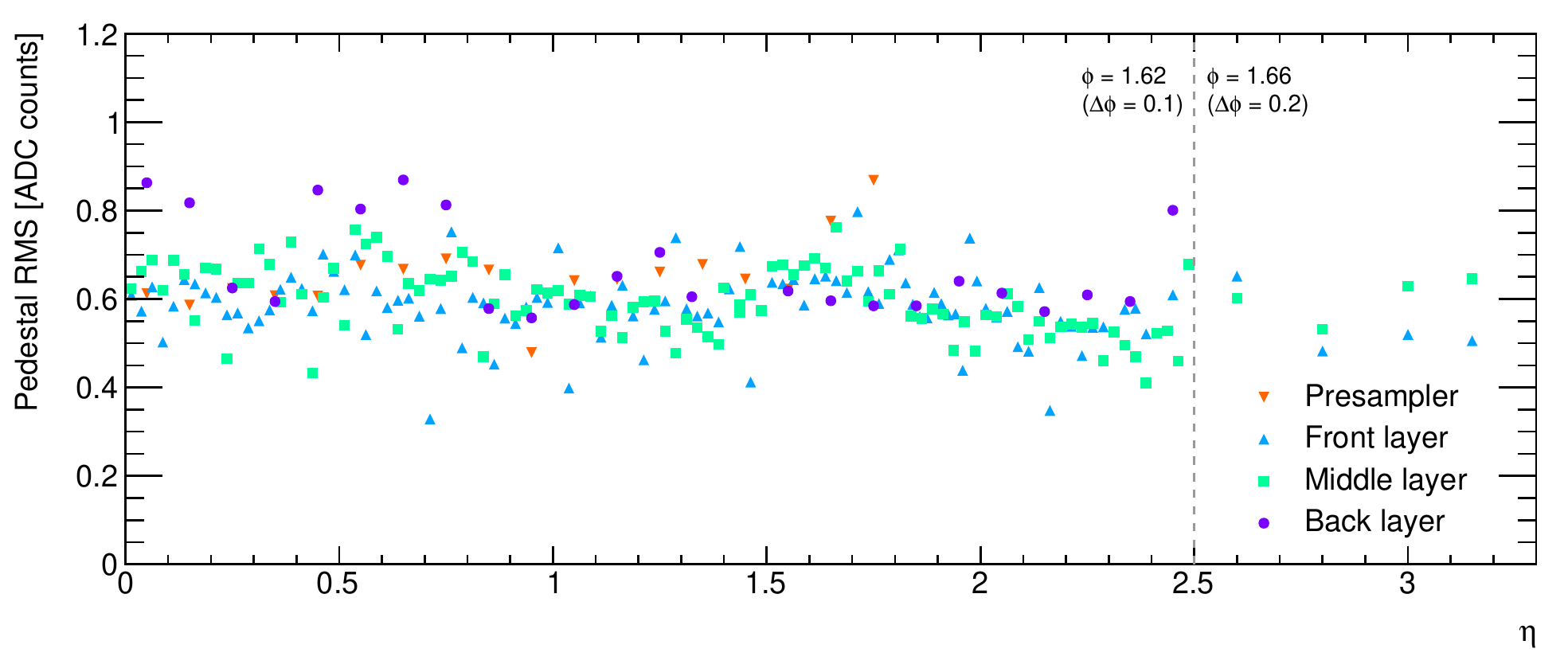}
  }
  \caption{The pedestal value (top) and its root mean square (bottom) as function of the $\eta$ position of the Super Cells at $\phi=1.62$ 
  ($\phi=1.66$) and $0<\eta<2.5$ ($2.5<\eta<3.2$), corresponding to a
   $\Delta\phi$ granularity of $0.1$ ($0.2$) and for the different 
   calorimeter layers.}
  \label{fig:pedestal_vs_channel}
\end{figure}

In order to compute the Super Cell energy from a pulse, the pedestal value
must be measured beforehand. Pedestal data are collected without injecting any pulse in the Front-End electronics. From this data, the electronic noise can also be studied since it 
corresponds to the root mean square of the pedestal value.
In Figure~\ref{fig:pedestal_vs_channel} the pedestal value and its root mean square can be seen as function of the $\eta$ position of the 
Super Cells, in the EMB and EMEC areas and for the different calorimeter 
layers.
They are found to be consistent with the expectation from the design of the LTDB boards, with the electronic noise always smaller than \SI{1}{ADC count}.

\begin{figure}[htb]
  \centering
  \includegraphics[width=.75\textwidth]{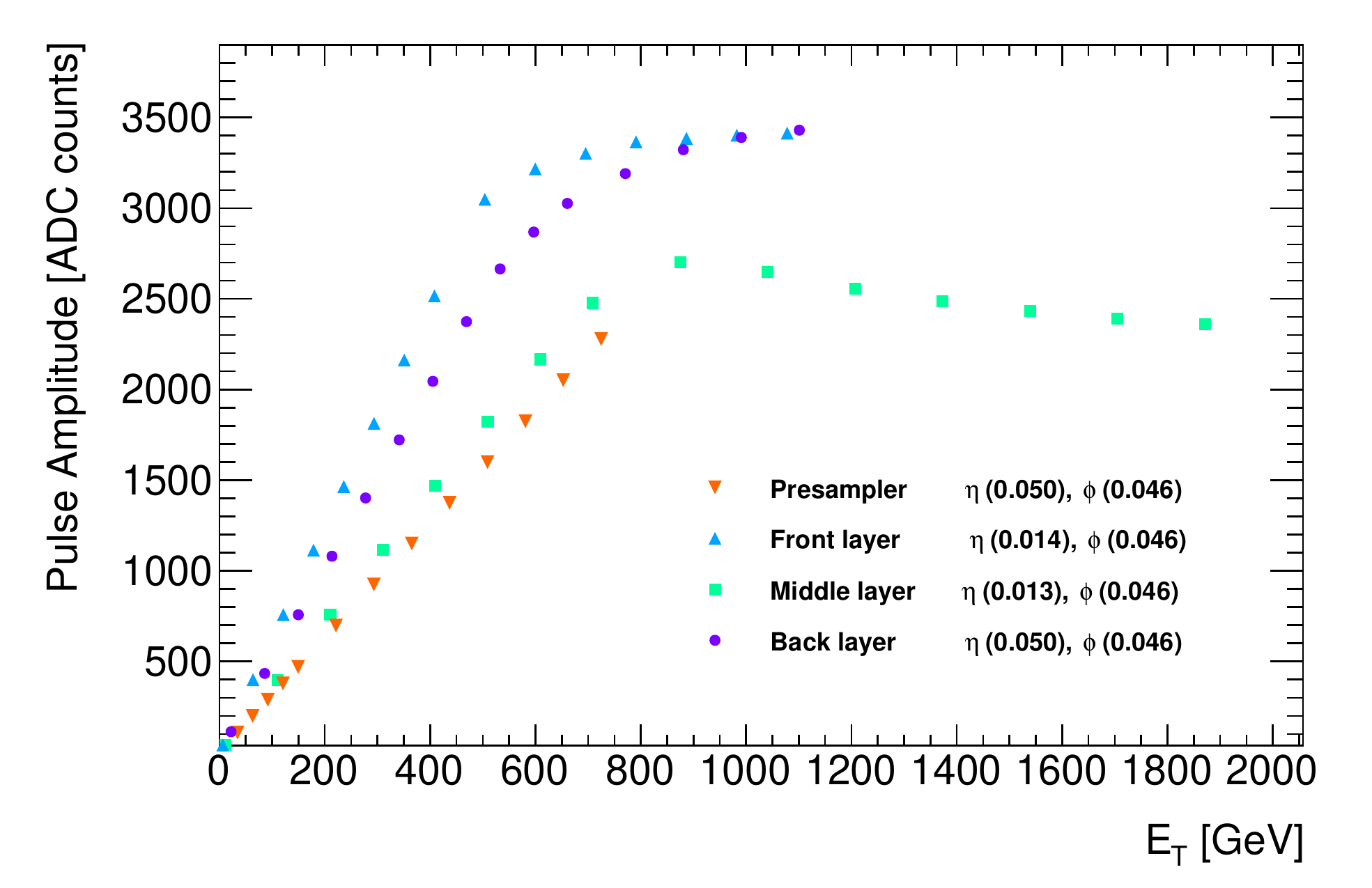}
  \caption{Peak ADC value with respect to the pedestal as function of the \ET corresponding to the injected pulse as seen on a LATOME board by four channels of the different calorimeter layers.}
  \label{fig:peakADC_vs_et}
\end{figure}

\clearpage
The linearity of the new digital trigger chain is measured using a ramp run similar to the one described in Section \ref{ssec:fe-integration}.
The peak ADC value with respect to the pedestal as function of the \ET 
corresponding to the injected pulse as seen by four channels on a LATOME board is presented in Figure~\ref{fig:peakADC_vs_et}. The peak ADC value is extracted with 
an optimal filtering algorithm~\cite{OFC}.
The ADC values are linearly increasing with the deposited \ET up to about \SI{800}{\GeV}, where saturation of the Super Cell pulse occurs.
This saturation is mainly due to the analog electronics (Linear Mixer), however digital saturation can occur for some channels in addition to the analog one at very high energies.

\begin{figure}[htb]
  \centering
  \includegraphics[width=\textwidth]{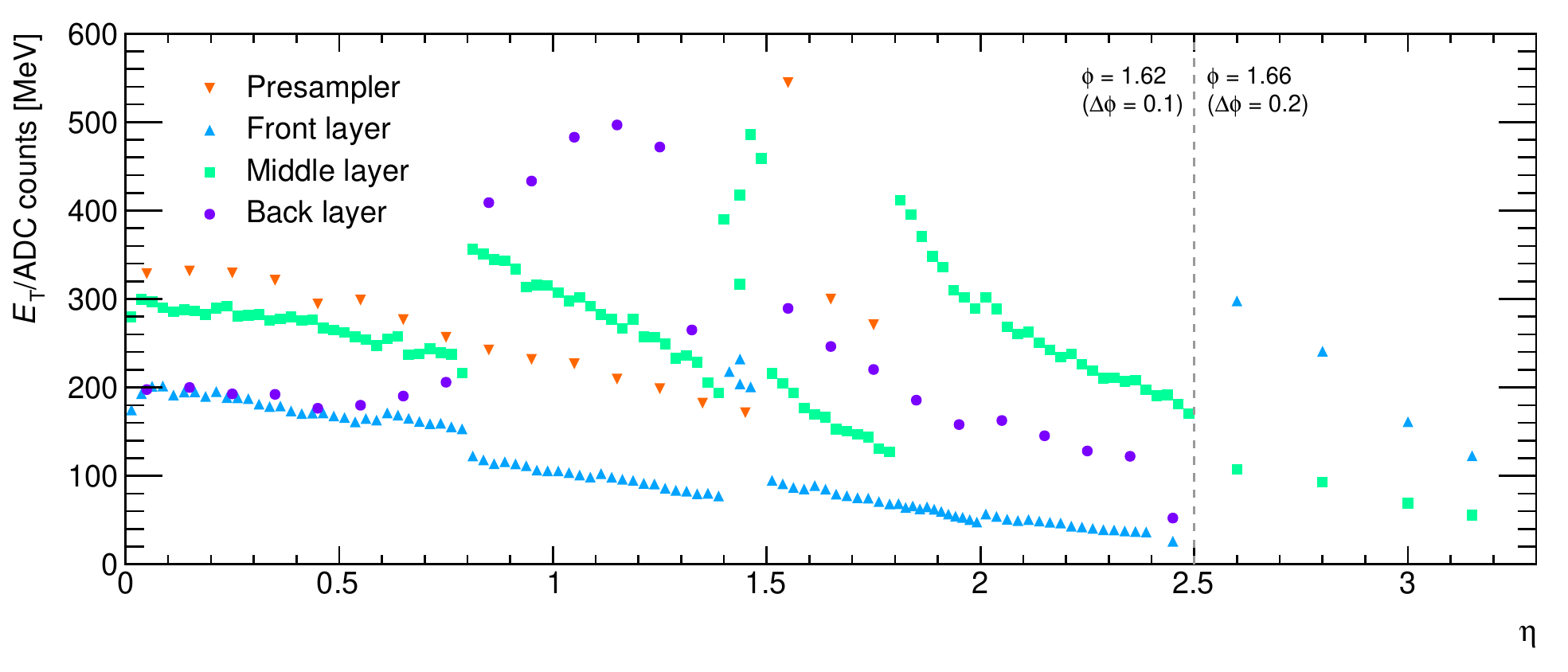}
  \caption{Measured \ET per ADC count as function of the $\eta$ position of the 
  Super Cells at $\phi=1.62$ ($\phi=1.66$) and $0<\eta<2.5$ ($2.5<\eta<3.2$),
  corresponding to a $\Delta\phi$ granularity of $0.1$ ($0.2$) and for the different calorimeter layers.}
  \label{fig:et_vs_eta}
\end{figure}

The \ET per ADC count is computed from the ramp runs.
This value is shown in Figure~\ref{fig:et_vs_eta} as function of the pseudorapidity in the EMB and EMEC areas.
The jump at $\eta=0.8$ is due to a change in the EMB sampling fraction, the one
at $\eta=1.5$ corresponds to the EMB-EMEC transition and the one at $\eta=2.5$ to the boundary between the EMEC
 inner and outer wheels.
The values are consistent with the ones expected from the design of the LTDB boards.

\begin{figure}[htb]
  \centering
  \includegraphics[width=.7\textwidth]{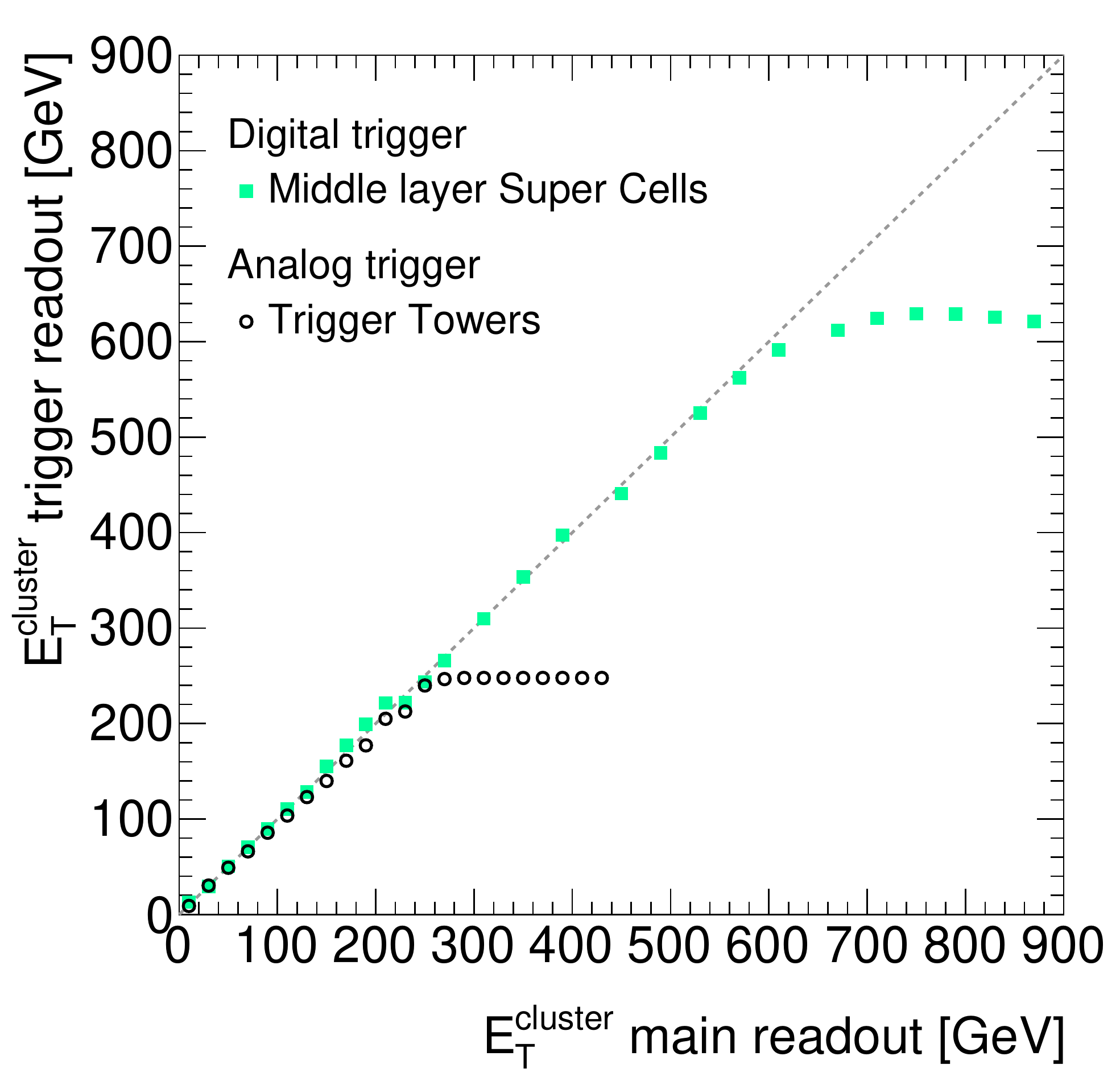}
  \caption{Comparison between the \ET deposited in middle layer Super Cells or Trigger Towers and the sum of the \ET in the corresponding cells of the LAr barrel calorimeter. The Super Cell data are collected by the LATOME boards while the energy deposited in the cells and the Trigger Towers is collected by the legacy main and trigger readout systems for injected signals. The deposited energy measured by the LATOME or the legacy trigger system corresponds well to the one measured by the main readout system up to the level where the signal on the Super Cells or the legacy Trigger Towers is saturated. 
  \label{fig:et_sc_tt_vs_et_cell}}
\end{figure}

The main readout system, which reads individual cells of the LAr calorimeter, is well understood and calibrated based on years of ATLAS operation.
To fully validate the trigger system, the \ET values computed by the LATOME for certain Super Cells are compared to the sum of \ET values computed for the corresponding cells using data collected with the main readout.
The data are collected simultaneously by both readout paths after injecting an electronic pulse in the Front-End electronics. In both cases the \ET values are computed with an optimal filtering algorithm~\cite{OFC}.
Figure~\ref{fig:et_sc_tt_vs_et_cell} shows good agreement between the Super Cell \ET and the sum of
the \ET values in the corresponding cells for the middle layer of the calorimeter in the barrel region.
At high energy the Super Cell pulse shape is saturated and the energy computed using the LATOME data does not increase anymore with the deposited energy.

A similar comparison is performed between the Trigger Tower \ET computed by the legacy trigger system and the sum of the \ET
values in the corresponding cells.
Figure~\ref{fig:et_sc_tt_vs_et_cell} shows good agreement between the two \ET measurements, with a saturation that occurs earlier in energy than with the Super Cell
readout.

\clearpage{}
\subsection{System latency}
\label{ssec:system-latency}
As discussed in the introduction, the scope of the ATLAS \phaseone upgrades
is limited to a few subdetectors. The readout of most of the
systems remains unchanged, limiting the capability of extending
some of the parameters of the Level-1 trigger system, such as the maximum rate and maximum latency. 
This situation will only change with the \phasetwo upgrades. ATLAS maintains a precise bookkeeping, for each 
subdetector contributing to the Level-1 trigger, of the different elements leading to the latency of the ATLAS L1A decision. 
Table~\ref{tb:latency} details the measured latency 
for the LAr trigger path starting from the proton-proton collision time up to the reception of the computed energy by the FEX
system.
The total latency amounts to $43.8$ BCs (one BC corresponds to a \SI{25}{ns} time interval). 
The listed numbers account for the slowest path; the detector readout being heterogeneous not all signals 
have the same latency (e.g. due to optical fiber cables of different lengths) although they are all aligned at the input of the LATOME firmware. 
At its output, corresponding to the encoding and summing firmware block in Table~\ref{tb:latency}, the listed latency is the one 
of the data sent to the jFEX system. The data sent to eFEX and gFEX systems have a lower latency by $0.4$ and $0.1$~BCs, respectively.  
The results in Table~\ref{tb:latency} are consistent with the measurements of the single ASIC components~\cite{latencyLOCx2} of the LTDB 
and with the simulations and on-target measurements of the firmware code operating in a standalone setup.

In addition to the LAr part, the FEX processors require \SI{14}{BCs}~\cite{phase1tdr} to extract the trigger primitives and 
transmit them to the Topological Trigger processors. The overall \SI{57.8}{BCs} latency of the calorimeter trigger system conforms with 
the maximum (\SI{65}{BCs}) value allowed by ATLAS at the input of the Topological Trigger processors where data from both 
the calorimeter and the muon trigger modules are combined.

\begin{table}[htbp]
\caption{Latency measurements for the \phaseone upgrade of the LAr trigger readout electronics. BCs
correspond to \SI{25}{ns} time intervals.}\label{tb:latency}
\begin{center}
\begin{tabular}{lrccc}
    \toprule
    & \multicolumn{2}{c}{\textbf{Latency}} & \multirow{2}{*}{\textbf{Sub-total [BCs]}} & \multirow{2}{*}{\textbf{Total [BCs]}} \\ & \textbf{[ns]} & \textbf{[BCs]}       &  &  \\ \midrule
    Time-of-flight at $\eta=2$              &  \phantom{0}15.0 & \phantom{0}0.6 & & \\
    Cable to pulse preamplifier             &  \phantom{0}30.0 & \phantom{0}1.2 &  & \\
    Preamplifier, shaper, and linear mixer  &  \phantom{0}21.6 & \phantom{0}0.9 &  \textbf{\phantom{0}2.7} & \textbf{\phantom{0}2.7}  \\ 
    \midrule 
    LTDB                                    &            153.7 & \phantom{0}6.1 & \textbf{\phantom{0}6.1} & \textbf{\phantom{0}8.8}  \\ 
    \midrule 
    Optical fiber cable (77 m) LTDB to LDPB &            385.0 &           15.4 &           \textbf{15.4} & \textbf{24.2}\\
    \midrule
    Deserializing and descrambling          &  \phantom{0}81.1 & \phantom{0}3.2 &  &  \\
    BCID aligning                           &  \phantom{0}28.2 & \phantom{0}1.1 &  & \\
    Channel remapping                       &  \phantom{0}67.7 & \phantom{0}2.7 &   & \\
    Optimal filtering                       &            108.2 & \phantom{0}4.3 & & \\
    Encoding and summing                    &  \phantom{0}94.6 & \phantom{0}3.8 &  & \\
    Serializing                             &  \phantom{0}35.7 & \phantom{0}1.4 &            \textbf{16.6} & \textbf{40.8}\\
    \midrule
    Optical fiber cable (15 m) LDPB to FEX  &  \phantom{0}75.0 & \phantom{0}3.0 & \textbf{\phantom{0}3.0} & \textbf{43.8}\\
    \bottomrule
\end{tabular}
\end{center}
\end{table}

\FloatBarrier

\section{Conclusion}
\label{sec:conclusion}
The ATLAS Liquid Argon Calorimeter \phaseone upgrade of its trigger readout
electronics increases the readout granularity by up to a factor of ten, 
enabling shower shape parameter calculation at the Level-1 trigger stage and thereby 
increasing the trigger rejection power while retaining high efficiency. 
The upgrade provides substantial improvement to the Level-1 trigger 
electron, photon, jet and missing energy resolution.
To realize this trigger system upgrade, new electronic boards including dedicated
ASICs were produced and tested to meet specifications and finally installed 
in the experiment during the second long shutdown of the LHC.

The new trigger readout is being commissioned, and initial tests
show that the system is operating according to specifications.
The legacy analog trigger readout is kept operational in parallel, reaching the
same performance as in Runs 1 and 2 in terms of noise and gain stability
without any loss of channels.
Noise level, linearity, radiation tolerance and overall energy measurements of 
the new digital trigger readout meet all expectations.

The next step in the commissioning of the new trigger readout is its full integration
with the new ATLAS Level-1 calorimeter trigger system.

\section*{Acknowledgements}

We thank the support staff from our institutions without whom ATLAS could not be operated efficiently.

We acknowledge the support of CNPq and FAPESP, Brazil; NSERC, NRC, CFI and CMC Microsystems, Canada; CERN; CAS, MOST and NSFC, China; IN2P3-CNRS and CEA-DRF/IRFU, France;  BMBF, HGF and MPG, Germany; INFN, Italy; MEXT and JSPS, Japan; MES of Russia, NRC KI, MSHE, Russian Federation; JINR;  MSSR, Slovakia; STFC, United Kingdom; DOE and NSF, United States of America. In addition, individual groups and members have received support from Compute Canada and CRC, Canada; ERC, ERDF, Horizon 2020, Marie Sk{\l}odowska-Curie Actions and COST, European Union; Investissements d'Avenir Labex, Investissements d'Avenir Idex and ANR, France; DFG and AvH Foundation, Germany; The Royal Society and Leverhulme Trust, United Kingdom.

The crucial computing support from all WLCG partners is acknowledged gratefully, in particular from CERN, the ATLAS Tier-1 facilities at TRIUMF (Canada), NDGF (Denmark, Norway, Sweden), CC-IN2P3 (France), KIT/GridKA (Germany), INFN-CNAF (Italy), NL-T1 (Netherlands), PIC (Spain), ASGC (Taiwan), RAL (UK) and BNL (USA), the Tier-2 facilities worldwide and large non-WLCG resource providers. Major contributors of computing resources are listed in Ref.~\cite{ATL-SOFT-PUB-2021-003}.

We acknowledge, in addition, the excellent work done by employees of the Etablissements Publics pour l’Integration (EPI), Geneva, with the refurbishment of the Front End Boards at the CERN site.

\printbibliography

\end{document}